\documentclass[10pt]{article}
\usepackage{amsmath,amssymb,graphicx}
\usepackage{amsthm}

\batchmode

\newtheorem{lem}{Lemma}
\newtheorem{thm}{Theorem}

\newtheorem{cor}{Corollary}

\newcommand{\pr}{\noindent{\bf Proof}. }
\newcommand{\re}{\noindent{\bf Remark}. }
\newcommand{\res}{\noindent{\bf Remarks}. }

\newcommand{\pa}{\partial}
\newcommand{\one}{\cO(1)}

\newcommand{\hs}{ \hspace{1cm}}

\newcommand{\loc}{ \textrm{loc}}
\newcommand{\Vol}{\textrm{Vol}}
\newcommand{\nat}{\natural}
\newcommand{\tk}{\bbT^{-k}_{\sM +\sN -k}}

\newcommand{\be}{\begin{equation}}
\newcommand{\ee}{\end{equation}}
\newcommand{\bs}{\begin{split}}
\newcommand{\es}{\end{split}}

\newcommand{\bom}{\mbox{\boldmath $\Om$}}

\newcommand{\bpi}{\mbox{\boldmath $\Pi$}}

\newcommand{\ba}{\mathbf{a}}

\newcommand{\sZ}{\mathsf{Z}}
\newcommand{\sM}{\mathsf{M}}
\newcommand{\sN}{\mathsf{N}}
\newcommand{\sfb}{\mathsf{b}}

\newcommand{\al}{\alpha}
\newcommand{\De}{\Delta}
\newcommand{\de}{\delta}
\newcommand{\ga}{\gamma}

\newcommand{\ka}{\kappa}
\newcommand{\La}{\Lambda}
\newcommand{\la}{\lambda}
\newcommand{\Om}{\Omega}

\newcommand{\Th}{\Theta}
\newcommand{\ep}{\epsilon}
\newcommand{\si}{\sigma}

\newcommand{\vep}{\varepsilon}

\newcommand{\cC}{{\cal C}}
\newcommand{\cD}{{\cal D}}

\newcommand{\cO}{{\cal O}}
\newcommand{\cI}{{\cal I}}
\newcommand{\cH}{{\cal H}}
\newcommand{\cS}{{\cal S}}

\newcommand{\cR}{{\cal R}}

\newcommand{\cK}{{\cal K}}

\newcommand{\cN}{{\cal N}}
\newcommand{\cW}{{\cal W}}

\newcommand{\cL}{{\cal L}}
\newcommand{\cP}{{\cal P}}

\newcommand{\cJ}{{\cal J}}
\newcommand{\cZ}{{\cal Z}}

\newcommand{\bbR}{{\mathbb{R}}}
\newcommand{\bbZ}{{\mathbb{Z}}}

\newcommand{\bbT}{{\mathbb{T}}}

\begin{document}

\title{The Renormalization Group According to Balaban\\ III. Convergence}
\author{ 
J. Dimock
\thanks{dimock@buffalo.edu}\\
Dept. of Mathematics \\
SUNY at Buffalo \\
Buffalo, NY 14260 }
\maketitle

\begin{abstract}
This is   an expository account of Balaban's approach to the renormalization group.  The method is 
illustrated with a treatment of  the the ultraviolet problem  for     the scalar $\phi^4$ 
model  on  toroidal lattice  in dimension  $d=3$.   In this third     paper we   demonstrate convergence of the expansion   and complete the 
proof of  a stability bound.
\end{abstract}

\section{Introduction}

   We   recall  the  general  setup  from part I   \cite{Dim11} and  part II   \cite{Dim12}.
We  are  studying the  $\phi^4$  field theory   on   a toroidal    lattice  of the form  
 \begin{equation}
\bbT_{\sM}^{-\sN}  =   ( L^{-\sN}\bbZ  / L^{\sM} \bbZ )^3  
\end{equation}
The  theory  is  scaled up to the unit lattice   $\bbT^0_{\sM  + \sN }$
and there   the  partition  function  has the form  
\begin{equation}
Z_{\sM,  \sN}  =  \int   \rho^{\sN}_0(\Phi)  d \Phi
\end{equation}
where  for   fields  $\Phi:   \bbT^0_{\sM  + \sN } \to  \bbR$
we have the density  
\begin{equation}  \label{den0}
 \rho^{\sN}_0(\Phi)  
=\exp \left( - S^{\sN}_0(\Phi)  - V^{\sN}_0(\Phi)  \right) 
\end{equation}
with 
\begin{equation}
\begin{split}
S^{\sN}_0(\Phi)  = &  \frac12  \|  \pa  \Phi \|^2  +  \frac12    \bar  \mu^{\sN}_0  \|  \Phi  \|^2 \\
V^{\sN}_0(\Phi)  = & \vep^{\sN}_0     \Vol( \bbT^0_{\sM+\sN} )   +  \frac12   \mu^{\sN}_0      \|  \Phi  \|^2     + \frac14   \la^{\sN}_0  \sum_x  \Phi^4 (x)  \\
\end{split}
\end{equation} 
and very small  positive  coupling  constants   $\la^{\sN}_0 = L^{-\sN}  \la,  \mu^{\sN}_0  =  L^{-2\sN} \mu$,   etc.
The  superscript  $\sN$ is generally omitted  so we have  $\la_0,  \mu_0,  etc.  $.

Our goal  is  to show  that with  intelligent   choices of the counter terms   $ \vep^{\sN}_0 ,   \mu^{\sN}_0$    the partition function  $Z_{\sM,  \sN}  $  satisfies  stability bounds   which are uniform in the ultraviolet 
cutoff  $\sN$   and with  bulk   dependence on the volume parameter  $\sM$.    The  method  is  
the renormalization group  method  of  Balaban (\cite{Bal82a}  - \cite{Bal98c}).   In fact   our primary goal 
is  not the stability bounds,  which are interesting  but  not new,   but rather   the illustration of Balaban's method.

We   repeatedly block  average   starting with  $\rho_0$   given  by  (\ref{den0}).     Given  $\rho_k( \Phi_k)$
we  define   for  $\Phi_{k+1}:  \bbT^1_{\sM + \sN -k}  \to  \bbR$  and block averaging operator  $Q$
  \footnote{  $\cN_{a, \Om}  =  ( 2\pi/  a)^{|\Om|/2} $  where  $|\Om|$ is the number of elements in  $\Om$.}
  \begin{equation} 
\tilde  \rho_{k+1}(  \Phi_{k+1}   ) = 
  \cN_{ aL,   \bbT^1_{\sM+ \sN -k}}^{-1}     \int     \exp \left(
- \frac  12  aL    |\Phi_{k+1}- Q \Phi_{k}|^2   \right)  \rho_k(  \Phi_k)
   d\Phi_k  
\end{equation}
Next   we    scale by  
\begin{equation}
  \rho_{k+1}(  \Phi_{k+1}   )     =     \tilde  \rho_{k+1}(  \Phi_{k+1,L}   ) L^{-|\bbT^{1}_{\sM + \sN -k}|/2}
\end{equation}
Then for any  $k$ the partition function can be expressed as  
\begin{equation}
\sZ_{\sM,\sN} =
\int   \rho^N_k( \Phi_k)d  \Phi_k
\end{equation}

We  quote the main result on these densities from part II.      
It  says that    after  $k$  steps  the 
density can be represented in the form
\begin{equation}     \label{representation1}
\begin{split}
  \rho_{k}(\Phi_k)  =     &  Z_{k}   \sum_{\bpi} \int d \Phi_{k,\bom^c}\  d W_{k,\bpi}\ 
  {K}_{k, \bpi}\  \cC_{k, \bpi}   \\
&   \chi_k ( \La_k)      \exp \Big( -S^+_k(\La_k ) +  E_k(\La_k ) +
 R_{k, \bpi}(\La_k)  +  B_{k,\bpi}( \La_k)         \Big) \\
\end{split}
\end{equation}
where
 \begin{equation}  \label{orca}
\begin{split}
d \Phi_{k,\bom^c}  =  & \prod_{j=0}^{k-1}  
\exp \left(  - \frac{1}{2 } aL^{-(k-j-1)}|\Phi_{j+1}- Q \Phi_{j}|^2_{\Om_{j+1}^c}  \right)
 d\Phi^{(k-j)}_{j, \Om_{j+1}^c}  
\\
d W_{k,\bpi}   =  &  \prod_{j=0}^{k-1}      (2 \pi)^{-|  [ \Om_{j+1}  - \La_{j+1}]^{(j)}|/2}
  \exp  \Big(   - \frac{1}{2 } L^{-(k-j)} | W_j |^2_{\Om_{j+1}- \La_{j+1}}  \Big) 
    dW^{(k-j)}_{j,\Om_{j+1}- \La_{j+1}}
\\ 
  K_{k,\bpi}  =  &
  \prod_{j=0}^{k}  
      \exp\left( c_j|\Om_j^{c,(j-1)}|-S^{+,u}_{j,L^{-(k-j)}}(\La_{j-1}  -  \La_{j}     )+  \Big( \tilde   B_{j,L^{-(k-j)}}\Big) _{\bpi_j}(\La_{j-1}, \La_{j}) \right)     \\
 \cC_{k, \bpi}  =   &      \prod_{j=0}^k  \Big(  \cC_{j, L^{-(k-j)}}\Big)_{ \La_{j-1}, \Om_j, \La_j }  \\
\end{split}
\end{equation}
Here   
\be  \bpi  =  ( \La_0,  \Om_1,  \La_1,  \dots  ,   \Om_k,   \La_k)   \ee
is a  decreasing   sequence of small field regions in $\tk$,  with  $\Om_j, \La_j$   a union of  $L^{-(k-j)}M$ cubes.  
With    $\de \Om_j  =  \Om_j - \Om_{j+1}$ 
our basic variables   are  
\be   \label{nuts}
 \Phi_{k,\bom}   = (  \Phi_{1, \de  \Om_1},    \Phi_{2, \de  \Om_2},  \dots,   \Phi_{k-1, \de \Om_{k-1}}, \Phi_{k, \Om_k})
\ee
  where   $\Phi_{j, \de \Om_{j+1}}:   (\de  \Om_{j})^{(j)}  \to \bbR$. 
\footnote{ If   $X \subset   \tk$   then  $X^{(j)}  \subset   \bbT^{-(k-j)}_{\sM + \sN -k}$ are   the centers of  $L^j$ cubes  in   $X$.}
There  are  also   variables     $ (  \Phi_{0,   \Om^c_1},    \Phi_{1, \Om_1^c}, \Phi_{2, \Om_2^c},   \dots,   \Phi_{k, \Om^c_{k}})$ which play a lesser role.
In      $d \Phi_{k,\bom^c}$   the measure is  
\be
  \label{old97}
d \Phi^{(k-j)}_{j, \Om_{j+1}^c}  = [ L^{-(k-j)/2}]^{|(\Om_{j+1}^c)^{(j)}|}   \prod_{ x \in   [ \Om^c_{j+1}]^{(j)} }  d \Phi_j(x)
\ee
Besides our basic  variables    there  are         auxiliary variables      
\begin{equation}
W_{k, \bpi}  =  (W_{0, \Om_1-  \La_1},  \dots, W_{k-1,  \Om_k -  \La_k}) 
\end{equation}
  with  $W_{j, \Om_{j+1}  - \La_{j+1}}    :    [ \Om_{j+1}  - \La_{j+1}]^{(j)} \to \bbR$.   In  $d W_{k,\bpi}  $  the   measure   $  dW^{(k-j)}_{j,\Om_{j+1}- \La_{j+1}}
$  is defined as in   (\ref{old97}).
We  employ the convention that   $ \La_{-1}, \Om_0$  are  the full torus  $ \bbT^{-k}_{\sM + \sN -k}$.
\bigskip

The    precise statement of the result is the following:

\begin{thm}  \label{maintheorem}    Let   $0< \la < e^{-1}$ and  $0< \bar \mu  \leq  1$.  Let  $\la_k=  L^{-(\sN-k)} \la$   and  $\bar \mu_k  =
L^{-2(\sN -k)}  \bar \mu$ be running coupling constants.    
Let   $L$  be sufficiently large,  let  $M$ be sufficiently large (depending on $L$), and   let  $\la_k$  be sufficiently small  (depending on $L,M$). 
   Let    $\vep_k,    \mu_k$   be the  dynamical  coupling constants  selected   in part I. 
Then  the  representation    (\ref{representation1}),(\ref{orca})    holds with the following properties:
\begin{enumerate} 
\item   
$Z_k$  is the global normalization factor  of   part  I.   It  satisfies  $Z_0 =1$  and 
 \be
Z_{k+1}  =   Z_k \ \cN^{-1}_{a ,  \bbT^1_{\sM + \sN-k}}   (2 \pi)^{| \bbT^0_{\sM +\sN-k}|/2}( \det  C_k  )^{1/2}
\ee

\item  With  $p_k = ( - \log \la_k)^p$   and  $\al_k  =  \max \{   \la_k^{\frac14},  \bar \mu_k^{\frac12}  \}$
the    characteristic function    $ \chi_k ( \La_k) $   enforce   bounds  on   $\La_k$ stronger than:
 \be
     |\Phi_k|  \leq   2 p_k \al_k^{-1}   \hs   |\pa \Phi_k|  \leq 3p_k  
 \ee

\item    The  characteristic functions  
   $\cC_{k,   \La_{k-1},  \Om_k,  \La_k}( \Phi_{k-1}, W_{k-1}, \Phi_k ) $   enforce   bounds    on  $\La_{k-1} - \Om_k$  stronger than 
 \be   \label{mexicanhat} | \Phi_{k-1}|  \leq  2 p_{k-1} \al_{k-1}^{-1}L^{\frac12}   \hs   |\pa   \Phi_{k-1}|  \leq  3 p_{k-1} L^{\frac32} 
 \ee
 and enforce    bounds on    
  $\Om_k - \La_k$  stronger than 
    \begin{equation}   \label{sombrero}
      |\Phi_k|  \leq   \  3    p_{k-1}\al_{k-1}^{-1}L^{\frac12}   
        \hs \    |\pa \Phi_k|  \leq   \  4    p_{k-1}L^{\frac32}
        \hs       |W_{k-1}|  \leq   C_w p_{k-1}  L^{\frac12}    
 \end{equation} 
   for some constant $C_w$. 
  In the  expression   (\ref{orca}) this is scaled down   by  $L^{-(k-j)}$.

\item    The  bare      action   is  $S^{+}_k(\La_k ) =    S^{+}_k(\La_k, \Phi_{k},  \phi_{k, \bom(\La_k^*)} )$
where    $ \phi_{k, \bom(\La^*_k)}$  is a field approximately localized in $\La_k^*$,  an  enlargement of $\La_k$,  and
for     $ \phi: \tk  \to \bbR $ 
\begin{equation}  \label{v1}
\begin{split}
S^{+}_k(\La_k, \Phi_{k}, \phi ) =&   S^*_{k}(\La_k,  \Phi_{k}, \phi )
 +   V_k( \La_k,  \phi)    \\
S^*_k( \La_k,  \Phi_k, \phi)    
 =&\frac{a_k}{2}\| \Phi_k  -  Q_{k}   \phi  \|^2_{\La_k} 
+   \frac12 \| \pa  \phi  \|^2_{*,\La_{k}}   +  \frac 12  \bar \mu_k   \|  \phi\|^2_{\La_k} \\
 V_k( \La_k,  \phi)  = & \vep_k  \Vol (\La_k) +
\frac12  \mu_k  \|  \phi \|^2_{\La_k}  + \frac14  \la_k  \int_{\La_k}   \phi^4  \\ 
\end{split}
\end{equation} 

\item  $  E_k(\La_k)   =   E_k(\La_k, \phi_{k,\bom(\La^*_k)}   ) $     are  the main corrections to the bare action.  For  $\phi:  \tk  \to \bbR$ it  has   the local expansion 
 \begin{equation}
  E_k(\La_k, \phi)  =  \sum_{X \subset \La_k}  E_k(X,  \phi)    
\end{equation} 
where    $X$ is a connected union of $M$-cubes.      $ E_k(X, \phi   )$ depends on $\phi$ in $X$,   is analytic in a certain complex domain   $\phi \in  \cR_k(X, \ep  )$,   
and satisfies     there for  $\beta <  \frac14- 10 \ep$
\begin{equation}  \label{basicE}
|    E_k(X )|  \leq   \la_k^{\beta}   e^{ - \ka d_M(X)  }       
\end{equation} 
where  $Md_M(X)$ is the length of the shortest tree joining the $M$-cubes in  $X$.

\item        $R_{k, \bpi}(\La_k) =  R_{k, \bpi}(\La_k,\Phi_{k}) $  is a tiny remainder and   has   the    local   expansion       
\begin{equation}   R_{k, \bpi}(\La_k,\Phi_{k})  =  \sum_{ X \subset \La_k}   R_{k, \bpi}(X,\Phi_{k})   
\end{equation} 
where    $ R_{k, \bpi}(X, \Phi_{k}) $  is analytic  in  a certain complex domain   $\cP_k(X, 2 \de)$,   
  and satisfies   there  for  a fixed integer $n_0 \geq  4$:
\begin{equation}\label{basicR}
|  R_{k, \bpi}(X ) |  \leq    \la_k^{n_0}    e^{ - \ka d_M(X)  } 
\end{equation}

\item            $    B_{k,\bpi} ( \La_k)  
 =  B_{k,\bpi} ( \La_k,  \Phi_{k,\bom},  W_{k,\bpi})  $ is the active boundary term.
It has  the local  expansion   
\begin{equation}  \label{basicB}
   B_{k,\bpi}( \La_k)   =  \sum_{X \in \cD_k (\bmod \Om_k^c), X  \#  \La_k }  B_{k,\bpi}  (X)  
\end{equation}
where  $X \# \La_k$ means  $X$ intersects both $\La_k$ and $\La_k^c$.
The function   $  B_{k,\bpi}  (X, \Phi_{k,\bom},  W_{k,\bpi})  $    is analytic 
in   a  certain complex domain    $\cP_{k, \bom}$    and   
\be    \label{secretary}
 |W_j|  \leq     B_w \   p_j L^{\frac12(k-j)}    \  \textrm{ on } \   \Om_{j+1} -     \La_{j+1} 
 \ee
and  it    satisfies there  
\begin{equation}
| B_{k,\bpi}  (X) |  \leq  B_0  \la_k^{\beta}  e^{- \ka  d_M(X,  \bmod  \Om_k^c)}
\end{equation}
for some constant  $B_0$  depending   on $L,M$.

\item       $\tilde   B_{k,\bpi}(\La_{k-1}, \La_k)  =\tilde   B_{k,\bpi}(\La_{k-1}, \La_k, \Phi_{k, \bom},  W_{k,\bpi})  $  is  the  inactive   boundary term.    It   depends
on  the   variables  only   in  $\Om_1- \La_k$,   is analytic  in $\cP_{k, \bom}$   and   (\ref{secretary}) and   satisfies there  
   \begin{equation}   \label{someday}
| \tilde   B_{k, \bpi}(\La_{k-1}, \La_k) |  \leq   B_0 \Big |  \La^{(k)}_{k-1} - \La^{(k)}_k  \Big|  =  B_0 \Vol( \La_{k-1} - \La_k )
\end{equation}
 Also it is additive in the connected components of  $\La_k^c$.

 \item  With  $\de \La_{k-1}  =  \La_{k-1} - \La_k$,  the  unrenormalized    action   is  
 $ S^{+,u}_k\Big(\de \La_{k-1},  \Phi_{k,\bom}, \phi_{k, \bom(\La_{k-1}, \Om_k, \La_k)}  \Big)$
where    $ \phi_{k,  \bom(\La_{k-1}, \Om_k, \La_k)  } $  is  a field approximately localized in  $\La_{k-1}^* \cap (\La^c_k)^*$  and 
\begin{equation}  \label{v2}
\begin{split}
S^{+,u}_k(\de \La_{k-1} ,  \Phi_{k,\bom}, \phi) =&   S^*_k(\de \La_{k-1}  ,  \Phi_{k, \bom}, \phi )
 +   V^u_k( \de \La_{k-1} ,  \phi)   \\
 V^u_k( \La,  \phi)  =  &     L^3\vep_{k-1}  \Vol (\La) +
 \frac12   L^2     \mu_{k-1}  \| \phi^2\|_{\La}  + \frac 14 \la_{k}  \int_{\La}   \phi^4  \\ 
\end{split}
\end{equation} 
\bigskip
\end{enumerate}
\end{thm}

\re  The complex domains  $\cR_k(X, \ep  )$, $\cP_k(X, 2 \de)$,  $\cP_{k, \bom}$  are defined in section 3.2 of part II.  
\bigskip

\noindent
\textbf{Convention:}    Throughout the paper  $\one$  stands for a generic constant independent of all parameters,   $C$  stands
for  a generic constant possibly depending on $L$.

\newpage

\section{The last  step}

For the rest of the paper  we   take  $\bar \mu  = 1$  and  fulfill    the condition by  $\la_k$   be sufficiently small by 
requiring that  $\la$   be sufficiently  small.    Then  we   can  run the iteration all the way  to   $k = {\sN}$  and we
have  an expression  back on the original  torus   $\bbT^{-\sN}_{\sM}$.    (In the terminology of paper I we are taking
$K = \sN$ and  $\De  =0$  ).   At  the end our running coupling constants   are   $\bar   \mu_{\sN} =  \bar \mu = 1$  and  $\la_{\sN} = \la$.
The   dynamical coupling constants  are  fixed to satisfy   $\vep_{\sN}=0,   \mu_{\sN} =0$  by  the choice  of  initial 
conditions   $\vep_{0}= \vep^{\sN}_{0},    \mu_{0}  =  \mu_0^{\sN}$.

 The  partition  function  is  now     given  by  $\sZ_{\sM,\sN} =  
\int   \rho_{\sN}( \Phi_{\sN})    d \Phi_{{\sN}} $  and substituting the expression  (\ref{representation1})   for  $ \rho_{\sN}( \Phi_{\sN}) $ it is  
\begin{equation}  \label{representation2}
\begin{split}
\sZ_{\sM,\sN}
 =     &  \  Z_{\sN}  \sum_{\bpi} \int   d \Phi_{{\sN}}     d \Phi_{{\sN}, \bom^c} \  d W_{{\sN},\bpi}\ 
 K_{{\sN}, \bpi} \    \cC_{{\sN}, \bpi}    \\
&   \chi_{\sN} ( \La_{\sN})      \exp \Big( -S_{\sN}(\La_{\sN} ) +  E^+_{\sN}(\La_{\sN} ) +
 R_{{\sN}, \bpi}(\La_{\sN})  +  B_{{\sN},\bpi}( \La_{\sN})         \Big) \\
\end{split}
\end{equation}
Here    we  have  transferred the potential  from  $S^+_{\sN}(\La_{\sN})$ 
to    $ E^+_{\sN}(\La_{\sN} ) $.  

The mass    $\bar  \mu_{\sN}  =  \bar  \mu =1$  will  enable us to control the final integral  over   $\Phi_{\sN}$.
Before estimating this expression we    manipulate it into a form which exhibits the local structure.    These  manipulation are  similar to the general step in the iteration  and they   will occupy the reminder of this section.   The  final estimates come in the next section.

\subsection{minimizers}
We   analyze  further the integral  over $\Phi_{{\sN}}$.
Split   the    integral  on  
\footnote{Let   $r_k  =  ( - \log  \la_k)^r$.   Recall that  for   a union of $M$ cubes $X$ in $\tk$,  $X^*$  is an enlargement   by  $[r_k]$
layers  of  $M$ cubes,   and  $X^{\nat}$ is   a shrinkage of  $X$ by  $[r_k]$ layers of  $M$ cubes.}
\begin{equation}
\Om_{{\sN}+1}   \equiv   \La^{5 \nat}_{\sN}
\end{equation}
With  this definition there is   no new  large field region   ($P_{{\sN}+1} =  \emptyset$)   and  $\Om_{{\sN}+1}$ is a union of  $M$ blocks
(rather than  $LM$ blocks as in the general step).  
We  split  $\Phi_{{\sN},\Om_{\sN}}   =  ( \Phi_{ {\sN},  \Om^c_{{\sN}+1}   }, \Phi_{{\sN},   \Om_{{\sN}+1}}  )$  and analyze  the  integral over   
$\Phi_{{\sN},   \Om_{{\sN}+1}} $  in more detail.
For this    idea would   be to  take the main term in the action 
  $ S^{*}_{\sN}(\La_{\sN},  \Phi_{\sN},  \phi_{{\sN} ,  \bom(\La^*_{\sN})}    )  $  and expand around the minimum  in   $\Phi_{{\sN}, \Om_{{\sN}+1}}$.
\bigskip 

Instead     we  use   the minimizer for   a  related    action     better    suited to    $   \phi_{{\sN} ,  \bom(\La^*_{\sN})}$.
Let 
\be     \label{sinking}
\tilde   \Phi_{{\sN}, \bom(\La^*_{\sN})}   =  \tilde  Q^T_{\bbT^0,  \bom(\La^*_{\sN})}\Phi_{{\sN}} 
 = \Big ([Q_{\sN}^T \Phi_{\sN}]_{\Om_1(\La^*_{\sN})^c},      Q^T_{\bbT^0,  \bom(\La^*_{\sN})}\Phi_{{\sN}} \Big )
\ee   
then  
 $  \phi_{{\sN} ,  \bom(\La^*_{\sN})}=    \phi_{{\sN} ,  \bom(\La^*_{\sN})}( \tilde   \Phi_{{\sN}, \bom(\La^*_{\sN})} )  $  is defined to be   the minimizer in $\phi$   on $\Om_1(\La_{\sN}^*)$  for     
\be      \label{awkward}
 \frac12   \| \ba^{\frac12} ( \tilde  \Phi_{{\sN}, \bom(\La^*_{\sN})}  -  Q_{{\sN},  \bom(\La^*_{\sN})}   \phi )\|^2_{ \Om_1(\La^*_{\sN})} 
  +   \frac12 \| \pa   \phi \|^2  
 + \frac12 \bar   \mu_{\sN}  \|    \phi \|^2
\ee
with   $\phi  =   Q_{\sN}^T \Phi_{\sN}$  on $ \Om_1(\La^*_{\sN})^c$.   

With  $\Om_{{\sN}+1}$  we  introduce       $   \bom'  =  ( \bom(\La_{\sN}^*) ,  \Om_{{\sN}+1})$  and  
$ \de \bom'  =  ( \de \Om'_1,   \dots,    \de  \Om'_{{\sN}}) $.    Then  $\Om_1(\La^*_{\sN})  =  \Om'_1$  and  we further split  (\ref{sinking}) as 
\be     
 \tilde   \Phi_{{\sN}, \bom(\La^*_{\sN})}  
 = \Big ([Q_{\sN}^T \Phi_{\sN}]_{\Om'^{c}_1},  \    \tilde  \Phi_{{\sN}, \de \bom'}, \  \Phi_{{\sN}, \Om_{{\sN}+1}}  \Big )
\ee
where     
\be
\tilde  \Phi_{{\sN}, \de \bom'}  =   \Big(   [Q^T_{{\sN}-1} \Phi_{\sN}]_{\de \Om'_1},  \dots, [Q^T_{1} \Phi_{\sN}]_{\de \Om'_{{\sN}-1}},   \Phi_{{\sN},\de \Om'_{\sN}}  \big)  
\ee
We    ask for the minimizer of   (\ref{awkward})  in   $ \Phi_{{\sN}, \Om_{{\sN}+1}}$ and     $\phi_{\Om'_1}$ 
This is discussed  in a general context in   appendix  \ref{A}.  
the minimum comes  at  $\Phi_{{\sN},   \Om_{{\sN}+1}}  =\Psi_{{\sN}, \Om_{{\sN}+1} }(\de   \bom')$
and  at   $\phi_{\Om'_1}  =     \phi_{{\sN} ,  \de    \bom'}   $  where   
 \be   \label{sombulant}
\begin{split}
\phi_{{\sN} ,  \de    \bom'}  = &   
  \phi_{{\sN} ,  \de    \bom'} \Big(   [ \tilde   \Phi_{{\sN}, \bom(\La^*_{\sN})}  ]_{\Om^c_{{\sN}+1} }  \Big)   \\   \equiv    &     G_{{\sN},  \de   \bom'} 
\Big(     Q^T_{{\sN}, \de  \bom'} \ba \tilde  \Phi_{{\sN}, \de \bom'}   
+  [\De]_{\Om'_1,  \Om'^c_1} [Q_{\sN}^T \Phi_{\sN}]_{\Om'_1}   \Big)   \\
\Psi_{{\sN}, \Om_{{\sN}+1} }(\de   \bom') \equiv   & [   Q_{\sN}      \phi_{{\sN} , \de \bom'}]_{\Om_{{\sN}+1}}  \\
   \end{split}
 \ee
Here    
\be  \label{santa}
  G_{{\sN}, \de  \bom'}    =  \Big [ - \De   +  \bar \mu_{\sN}   +   Q^T_{{\sN},  \de   \bom'}  \ba  \tilde  Q_{{\sN}, \de  \bom'}\Big ]_{\Om'_1}^{-1}
\ee
With  our  choice of  ${\sN}$,  the mass term  $\bar \mu_{\sN}$  in    $\tilde   G_{{\sN},\de \bom'}$  is now substantial.   As  a consequence
the kernel  of  $\tilde   G_{{\sN},\de  \bom'}$   has  exponential decay as we  will see,   and so   $  \phi_{{\sN} ,  \de  \bom'}  $
is  approximately  localized  in $\Om^c_{{\sN}+1}$.
We  also note   the identity  from  appendix   \ref{A} 
\be  \label{snoring}
   \phi_{{\sN} ,  \de    \bom'}  =    \phi_{{\sN} ,  \bom(\La^*_{\sN})} \Big(   [ \tilde   \Phi_{{\sN}, \bom(\La^*_{\sN})}  ]_{\Om^c_{{\sN}+1} }    ,   \Psi_{{\sN}, \Om_{{\sN}+1}} (\de  \bom') \Big)
\ee

 Returning to the original problem  with   $ S^{*}_{\sN}(\La_{\sN},  \Phi_{\sN},  \phi_{{\sN} ,  \bom(\La^*_{\sN})}    )  $   we   expand around the   minimizer    
 by  making  the  translation 
 \be   \label{translate}
  \Phi_{{\sN},   \Om_{{\sN}+1}}   =   \Psi_{{\sN}, \Om_{{\sN}+1}} (\de \bom') +Z 
   \ee   
 and change the integral to an integral over $Z:  \Om^{({\sN})}_{{\sN}+1} \to \bbR$.
Then    by (\ref{snoring})  and  defining   $ \cZ_{\sN}=   \phi_{{\sN}, \bom( \La_{\sN}^* )}(0, Z) $
\begin{equation}
\begin{split}
\phi_{{\sN}, \bom( \La_{\sN}^* )}= &  \phi_{{\sN}, \bom( \La_{\sN}^* )}\Big(    [ \tilde   \Phi_{{\sN}, \bom(\La^*_{\sN})}  ]_{\Om^c_{{\sN}+1} }  ,   \Phi_{{\sN}, \Om_{{\sN}+1}}\Big) \\
 =   &
\phi_{{\sN}, \bom( \La_{\sN}^* )}\Big(   [ \tilde   \Phi_{{\sN}, \bom(\La^*_{\sN})}  ]_{\Om^c_{{\sN}+1} } ,  \Psi_{{\sN}, \Om_{{\sN}+1}} (\de   \bom') \Big)
+  \phi_{{\sN}, \bom( \La_{\sN}^* )}(0, Z) \\
\equiv  &   \phi_{{\sN} ,  \de  \bom'}  +  \cZ_{\sN}  \\
\end{split}
\end{equation}

\begin{lem}  \label{sundry}
\begin{equation}   
\begin{split}
& S^*_{{\sN}}\Big(\La_{\sN},  ( \Phi_{{\sN}, \de   \Om_{\sN}  },  \Psi_{{\sN}, \Om_{{\sN}+1}}(\bom') +Z) ,    \phi_{{\sN} ,  \de    \bom'}  +  \cZ_{\sN} \Big )\\
= & S^*_{{\sN}}(\La_{\sN}-  \Om_{{\sN}+1},   \Phi_{{\sN}, \de   \Om_{\sN}  },  \phi_{{\sN} ,  \de    \bom'}  )  +  \tilde  S_{\sN}( \Om_{{\sN}+1},   \phi_{{\sN} ,  \de    \bom'} )
  +    \frac12 \Big<Z,  \left[ \De_{\bom( \La_{\sN}^* )}   \right]_{\Om_{{\sN}+1}}  Z  \Big>
+    R^{(1)}_{\bpi, \Om_{{\sN}+1}}
    \\
   \end{split}
\end{equation}
where
  $\tilde  S_{\sN}(   \Om,   \phi )   = \frac12 \| \pa   \phi \|^2_{*,  \Om}   + \frac12 \bar   \mu_{\sN}  \|   \phi \|^2_{ \Om}  $
and     $ R^{(1)}_{\bpi, \Om_{{\sN}+1}}$       is a tiny term.
\end{lem}
\bigskip

\re    A ''tiny term'' is  $\cO(\la^{n_0})$   for  our standard integer  $n_0 \geq 4$.   We  are more precise about this when we discuss localization.
\bigskip

\pr With $ \Psi_{\sN}= \Psi_{{\sN}, \Om_{{\sN}+1}} (\de \bom') $  we  expand  in $Z$ and find  as in  lemma 2.4 in part II:
\begin{equation}
\begin{split}
  & S^*_{{\sN}}\Big(\La_{\sN},  ( \Phi_{{\sN}, \de   \Om_{\sN}  },  \Psi_{\sN}) +(0, Z) ,    \phi_{{\sN} ,  \de    \bom'}  +  \cZ_{\sN} \Big ) 
=     S^*_{{\sN}+1}(\La_{\sN}, ( \Phi_{{\sN}, \de   \Om_{\sN}  },  \Psi_{\sN}) ,    \phi_{{\sN} ,  \de    \bom'}  \Big )   \\
+   &  a_k
< Z,   ( \Psi_{{\sN} }  -  Q_{{\sN}}     \phi_{{\sN} ,  \de    \bom'} ) >_{\La_{\sN}} \  +  \sfb_{\La_{\sN}} (\pa   \phi_{{\sN} ,  \de    \bom'}  ,\cZ_{\sN})  \    +S^*_{{\sN}}(\La_{\sN},    (0,Z) ,    \cZ_{\sN}  )     \\
\end{split}   
\end{equation}
Here we have made a cancellation in the linear  terms  using again  (\ref{snoring}). 
The  term    $ \sfb_{\La_{\sN}} (\pa   \phi_{{\sN} ,  \de    \bom'}  ,\cZ_{\sN}) $  is a boundary term  localized  on $\pa  \La_{\sN}$.  It  is tiny  since
 $\cZ_{\sN}$  is  tiny on  $\pa \La_{\sN}$,  and  so  contributes to    $ R^{(1)}_{\bpi, \Om_{{\sN}+1}}$.  The  second   term on the right side of this equation vanishes 
 by the definition of    $\Psi_{\sN}$.
   As in     lemma 2.5 in   part II,  the last      term    can  be  written  
\begin{equation}
\begin{split}  
&S^*_{{\sN}}(\La_{\sN},    (0,Z) ,    \cZ_{\sN}  )\
= \frac12 \Big<Z,  \left[ \De_{\bom( \La_{\sN}^* )}   \right]_{\Om_{{\sN}+1}}  Z  \Big>  +      \dots  
\\
\end{split}
\end{equation}
where the omitted terms  are    tiny and contribute    to   $ R^{(1)}_{\bpi, \Om_{{\sN}+1}}$.
Finally   using  the definition of $\Psi_{\sN}$  again   we have
\be   
   S^*_{{\sN}}\Big(\La_{\sN},  ( \Phi_{{\sN}, \de   \Om_{\sN}  },  \Psi_{\sN}) ,     \phi_{{\sN} , \de  \bom'} \Big )  = 
S^*_{{\sN}}(\La_{\sN}-  \Om_{{\sN}+1},   \Phi_{{\sN}, \de   \Om_{\sN}  },  \phi_{{\sN} ,  \de    \bom'}  )  +  \tilde  S_{\sN}( \Om_{{\sN}+1},   \phi_{{\sN} ,  \de    \bom'} )    
\ee
to complete the proof.
\bigskip

Actually  we  use  a modification  of    (\ref{translate}) as in part II.  The propagator   $ G_{{\sN}, \de\bom'}$
has  a random walk expansion,  explained in more detail  later.   Hence one can introduce weakening parameters
$s =  \{ s_{\square} \}$ for   multiscale cubes $\square$,   and define a weakened version  $ G_{{\sN}, \de  \bom'}(s)$.  This leads  to a weakend field
$    \phi_{{\sN} , \de \bom'} (s)$,   hence a field $     \phi_{{\sN} , \de \bom'} (\square)$   localized in   $\square^*$.   Approximating 
$ \phi_{{\sN} , \de  \bom'} $   by    $\phi_{{\sN} , \de  \bom'}(\square)$  on  $\square$ gives a  localized field   $\phi^{\loc}_{{\sN} , \de  \bom'} $ 
and hence a localized    minimizer    $  \Psi^{\loc}_{{\sN}, \Om_{{\sN}+1}}(\de  \bom')$
  The actual  translation  is then
 \be   \label{translate2}
  \Phi_{{\sN},   \Om_{{\sN}+1}}   =   \Psi^{\loc}_{{\sN}, \Om_{{\sN}+1}}(\de   \bom')  +Z 
   \ee   
This  is done for benefit of the characteristic functions.   But in $ S^*_{{\sN}}(\La_{\sN}-  \Om_{{\sN}+1})  +  \tilde  S_{\sN}( \Om_{{\sN}+1} ) $ and  in  $E^+_{\sN}(\La)$ we immediately  undo it and return   
from  $ \Psi^{\loc}_{{\sN}, \Om_{{\sN}+1}}(\de \bom')$  to  $ \Psi_{{\sN}, \Om_{{\sN}+1}}(\de  \bom')$   at the cost of some more tiny terms  
$ R^{(2)}_{\bpi, \Om_{{\sN}+1}}$  , $ R^{(3)}_{\bpi, \Om_{{\sN}+1}}$.  
We  define   $  R^{(\leq 3)}_{ \bpi,  \Om_{{\sN}+1}}     = R_{{\sN}, \bpi  }  (\La_{\sN})     +     R_{ \bpi,  \Om_{{\sN}+1}}^{(1)} + \dots   +  R_{ \bpi,  \Om_{{\sN}+1}}^{(3)}  $.  

Now  (\ref{representation2}) can be written
\begin{equation}    \label{representation3}
\begin{split}
 \sZ_{\sM,\sN} =     & Z_{\sN} \sum_{\bpi} \int    d  \tilde  \Phi_{{\sN}, \bom^c}\  d W_{{\sN},\bpi}\ 
  K_{{\sN}, \bpi} \    \cC_{{\sN}, \bpi} \    \exp \Big(  - S^*_{{\sN}}(\La_{\sN}-  \Om_{{\sN}+1})  -  \tilde  S_{\sN}( \Om_{{\sN}+1} )   \Big )     \\
&   \int   d  Z \ \chi_{{\sN}}(  \La_{\sN} ) 
 \exp  \Big(
-       \frac12 \Big<Z,  \left[ \De_{\bom( \La_{\sN}^* )}   \right]_{\Om_{{\sN}+1}}  Z  \Big>   +  E^+_{\sN}(\La_{\sN} ) +
 R^{(\leq 3)}_{ \bpi,  \Om_{{\sN}+1}}  + B_{{\sN}, \bpi}(\La_{\sN})    \Big) \\
\end{split}
\end{equation}
Here     
\be
 d  \tilde  \Phi_{{\sN}, \bom^c}  =   d \Phi_{{\sN},  \Om^c_{{\sN}+1}} d \Phi_{{\sN}, \bom^c} 
 \ee
  and     $ S^*_{{\sN}}(\La_{\sN}-  \Om_{{\sN}+1}),\  \tilde  S_{\sN}( \Om_{{\sN}+1} )$
are as  in lemma \ref{sundry}, and  
\be
\begin{split}
 E^+_{\sN}(\La_{\sN}) =  &  E^+_{\sN}(\La_{\sN},    \phi_{{\sN} ,  \de    \bom'}  +  \cZ_{\sN} )  \\
 \chi_{\sN} ( \La_{\sN})  =  &  \chi_{\sN} (\La_{\sN},  \Phi_{{\sN}, \de \Om_{\sN}},      \Psi^{\loc}_{{\sN}, \Om_{{\sN}+1}}(\de   \bom')  +Z )  \\
 \end{split}
 \ee
and $ R_{{\sN}, \bpi  }  (\La_{\sN})   $  and     $ B_{{\sN}, \bpi}(\La_{\sN}) $ also    have their arguments shifted by  (\ref{translate2}).

\subsection{fluctuation integral}

The integral over  $Z$ is   a Gaussian  integral with covariance  
\begin{equation}
\tilde   C_{{\sN},   \bom'}  =[\De_{{\sN},  \bom(\La_{\sN}^*)} ]^{-1}_{\Om_{{\sN}+1}}
\end{equation}
There is now no  term   $aL^{-2}Q^TQ$ in the operator we are inverting, as was the case in earlier steps.  To obtain an ultra local measure we       want to    change      variables  by   $Z =  \tilde C^{1/2}_{{\sN},   \bom'  }W$   where
$W :  \Om^{({\sN})}_{{\sN}+1}  \to  \bbR$.  For this we need to control the operator   $\tilde  C^{1/2}_{{\sN}, \tilde  \bom  }$.
It  has the representation (Appendix  C in part  II with  $a=0$ ):
\begin{equation}
\begin{split}
\tilde C^{1/2}_{{\sN}, \bom'  }  \label{z1}
  =&     \frac{1}{\pi}  \int_0^{\infty}  \frac{dr}{\sqrt{r}} \tilde  C_{{\sN}, \bom', r}    \hs
\tilde C_{{\sN}, \bom', r}  =    \Big[   \De_{{\sN},  \bom(\La^*_{\sN})}  +r \Big]^{-1}_{\Om_{{\sN}+1}}  \\
\end{split}
\end{equation}
And   $\tilde C_{{\sN}, \bom', r} $   has the representation (Appendix  C in part  II with  $a=0$ )
\begin{equation}  \label{z2}
\tilde   C_{{\sN}, \bom', r} =  \Big[  \frac{1}{a_{\sN}+r}  + \left( \frac{a_{\sN}}{a_{\sN}+r}  \right)^2 Q_{\sN}  G_{{\sN}, \de  \bom',  r} Q_{\sN}^T\Big]_{\Om_{{\sN}+1}}
\end{equation}
where 
\begin{equation}  \label{z3}
\begin{split}
 G_{{\sN}, \de  \bom', r}  = &  \Big[ -\De  + \bar \mu_{\sN}  +  Q_{{\sN}, \de \bom'}^T  \ba  Q_{{\sN},  \de \bom' }     +
  \frac{a_{\sN}r}{a_{\sN}+r}  \left[ Q_{\sN}^T  Q_{\sN} \right]_{\Om_{{\sN}+1}}  \Big]_{\Om'_1}^{-1} \\
\end{split}
\end{equation}
The Green's function    $     G_{{\sN},  \de    \bom', r}  $   has  a random walk expansion   and  hence there is a weakened
form    $ G_{{\sN},  \de   \bom', r}(s)  $.    This leads  to  weakened  forms  $\tilde C_{{\sN}, \bom', r} (s)$,      $\tilde C^{1/2}_{{\sN}, \bom'  }(s)$,
  therefore to    $\tilde C_{{\sN}, \bom', r} (\square^*),   \tilde C^{1/2}_{{\sN},   \bom'  }(\square^*)   $   and   so to     $(\tilde C^{1/2}_{{\sN},   \bom'  })^{loc}$.
The  actual  change of  variables  is     then
\be  \label{sondra}
Z =  \Big(\tilde C^{1/2}_{{\sN},   \bom'  }\Big)^{loc}W_{\sN}
\ee
  This localization  is for the benefit  of  
the characteristic functions.   In   $E_k(\La_{\sN})$   we  immediately change back  to  $\tilde  C^{1/2}_{{\sN},   \bom'  }W_{\sN}$.
The new measure in $W_{\sN}$   is replaced  by   the Gaussian measure  $ d\mu_{\Om_{{\sN}+1}} ( W_{\sN}) $  with identity covariance.  
Furthermore
the induced determinant  $  \det  (  (\tilde  C^{1/2}_{{\sN},   \bom'  })^{loc})  $  is  changed back to      $  \det  (  \tilde  C^{1/2}_{{\sN},   \bom'  })  $, and then        to  a global   determinant  $ \det  (\De_{\sN}^{-\frac12}   )$,
the special   case  in which   all small field regions  are the whole torus.   The operator  $\De_{\sN}$  has the representation  $ \De_{\sN}  =   a_{\sN} - a_{\sN}^2 Q_{\sN}G_{\sN}Q_{\sN}^T$  and  $\De_{\sN}^{-\frac12} $  has a representation which is a special case of  (\ref{z1}) - (\ref{z3}).
All  these  replacements  are  
at the cost of  further  tiny  terms    $ R^{(4)}_{\bpi, \Om_{{\sN}+1}}, \dots,    R^{(7)}_{\bpi, \Om_{{\sN}+1}}$
and   an overall volume   factor  $ \exp(  \tilde c_{\sN}  |\Om^{c,({\sN})}_{{\sN}+1}|)$.  See part II for more details.

We  also note that   we  can identify  $Z_{\sN}   \det  (\De_{\sN}^{-\frac12}   ) $   as   the bare normalization factor  $\sZ_{\sM, \sN}(0) $
defined   with $\la_0 =0$ and no counter terms   (i.e. $ V_0 = 0$ ).  
 This is so since if we  run  the  global renormalization
group   as in section 2.2 in   part  I  with   we  find 
\be   
\begin{split}
\sZ_{\sM, \sN}(0)  = &   \int  \rho_{\sN}(\Phi_{\sN})  d \Phi_{\sN}  =   Z_{\sN}  \int   \exp\Big(  - S_{\sN}(\Phi_{\sN}, \phi_{\sN})  \Big)d \Phi_{\sN}\\
= &   Z_{\sN}   \int   \exp\Big(  -  \frac 12  <\Phi_{\sN}, \De_{\sN}  \Phi_{\sN}>  \Big)d \Phi_{\sN}  =Z_{\sN} (\det   \De_{\sN}^{-\frac12} )\\
\end{split}
\ee

With these  changes we find  
\begin{equation}    \label{representation4}
\begin{split}
 \sZ_{\sM,\sN} =     & \sZ_{\sM, \sN}(0)   \sum_{\bpi}   \int      d  \tilde  \Phi_{{\sN}, \bom^c} \  d W_{{\sN},\bpi}\ 
  K_{{\sN}, \bpi} \    \cC_{{\sN}, \bpi} \    \exp \Big(  -  S^*_{{\sN}}(\La_{\sN}-  \Om_{{\sN}+1})  - \tilde  S_{\sN}( \Om_{{\sN}+1} )   \Big )     \\
&  \exp \Big(  \tilde c_{\sN}  |\Om^{c,({\sN})}_{{\sN}+1}|  \Big) 
\int   d\mu_{\Om_{{\sN}+1}} ( W_{\sN})   \ \chi_{{\sN}}(  \La_{\sN} ) 
 \exp  \Big(   E^+_{\sN}(\La_{\sN} ) +
 R^{(\leq 7)}_{ \bpi,  \Om_{{\sN}+1}}  + B_{{\sN}, \bpi}(\La_{\sN})    \Big) \\
\end{split}
\end{equation}
  The field  
 $\cZ_{\sN}  =  a_{\sN} G_{k \bom(\La_{\sN}^*)}  Q_{\sN}^T Z$    has  become  
\be 
\tilde     \cW_{{\sN},  \bom'}  =     a_{\sN}  G_{{\sN},   \bom(\La_{\sN}^*)}  Q_{\sN}^T \ (  \tilde    C^{1/2}_{{\sN},  \bom'}   W_{\sN} )
\ee
and we have  
\be
\begin{split}
 E^+_{\sN}(\La_{\sN}) =  &  E^+_{\sN}(\La_{\sN},      \phi_{{\sN} , \de    \bom'}  +  \tilde    \cW_{{\sN},  \bom'}  )  \\
 \chi_{\sN} ( \La_{\sN})  =  &  \chi_{\sN} \Big(\La_{\sN},  \Phi_{{\sN}, \de \Om_{\sN}},      \Psi^{\loc}_{{\sN}, \Om_{{\sN}+1}}(\de \bom')  + \Big(\tilde  C^{\frac12}_{{\sN},   \bom'  }\Big)^{loc}W_{\sN}\Big )  \\
 \end{split}
 \ee
 with   the change of variables   (\ref{sondra})  also made where appropriate in     $ R^{(\leq 7)}_{ \bpi,  \Om_{{\sN}+1}}$ and  $B_{{\sN}, \bpi}(\La_{\sN}) $.

\subsection{estimates}

We  collect some estimates we need.

\begin{lem}     \label{green1}
The Green's function $   G_{{\sN},  \de \bom'}$   has  a random walk expansion based on multi-scale cubes     for  $\de \bom'$, 
  convergent in $L^2$ and $L^{\infty}$ norms for  $M$ sufficiently large.
 There  are constants  $ C,  \ga_0$   (depending on $L$)  such  that for  $L^{-({\sN}-j)}$ cubes       $ \De_y \subset     \de \Om_j $
and   $L^{-({\sN}-j')}$ cubes   $  \De_{y'}  \subset    \de \Om_{j'} $:
  \begin{equation}  \label{fundamental}
  \begin{split}
 |1_{\De_{y}}  G_{{\sN},  \de \bom'}1_{\De_{y'}} f|\  \leq   \  &   C L^{-2({\sN}-j')}
   e^{ - \frac12 \ga_0  d_{\bom'}(y,y') }   \|f\|_{\infty}  
    \\ 
 L^{-({\sN}-j)} | 1_{\De_{y}} \pa   G_{{\sN},  \de \bom'}1_{\De_{y'}} f|  \    \leq    &  \    C L^{-2({\sN}-j')}
   e^{ - \frac12 \ga_0  d_{\bom'}(y,y') }   \|f\|_{\infty}  \\
   L^{-(1+ \al)({\sN}-j)} | 1_{\De_{y}} \de_{\al}\pa   G_{{\sN},  \de \bom'} 1_{\De_{y'}} f|  \    \leq    &  \   C L^{-2({\sN}-j')}
  e^{ -  \frac12 \ga_0  d_{\bom'}(y,y') }   \|f\|_{\infty}  \\
\end{split}  
\end{equation} 
\end{lem}
\bigskip

\pr   This follows the proof of Theorem  2.2  in   part II,  to which we refer for details.  
In that  theorem, specialized to the case at hand,      the proof is based  on   local inverses   for  multi-scale cubes  $\square$ given by 
\be   
 G_{\sN, \bom(\La_{\sN}^*)} (\square )  = \Big [ - \De   + \bar \mu_{\sN}  +   Q_{k, \bom(\La_{\sN}^*)}^T\ba   Q_{k, \bom(\La_{\sN}^*)}\Big ]_{\tilde \square}^{-1}   
\ee
Now      with      $\bom'  =  (  \bom(\La_{\sN}^*),  \Om_{k+1}  )$     we   modify    this to    $G_{\sN, \de  \bom'}(\square)$  defined  in   (\ref{santa}).   The only difference is that there are no   averaging  operators  in  $\Om_{\sN +1}$.  Then  if  $\tilde \square  \subset   \Om_{\sN+1}$   we have  since  $\bar \mu_{\sN} =1$
\be
G_{\sN, \de  \bom'}(\square) =   \Big [ - \De   + I \Big]_{\tilde \square}^{-1} 
\ee  
This satisfies the same bounds   as      $G_{\sN, \bom(\La_{\sN}^*)  } (\square )$.   Also   the operator  $\cH_{\sN}$ in lemma 31 in part I,  now 
with no averaging  operators  and  $\bar \mu_{\sN} =1$,  satisfies the same  bounds.   Thus the proof goes through as before.
\bigskip
      
\re   The random walk expansion still has $M$ cubes in  $\Om_{\sN+1}$,  unlike the general step where it had  $LM$ cubes.

\begin{lem}    
The Green's function $   G_{{\sN}, \de   \bom', r}$   has  a random walk expansion   convergent in  the   $L^2$ norm for  $M$ sufficiently large.
   It yields the bounds for   all  $r  \geq  0$
 \begin{equation}  \label{green2}
 \|1_{\De_{y}}   G_{{\sN},\de \bom',r} 1_{\De_{y'}} f\|_2\  \leq   \     C L^{-2({\sN}-j')}
   e^{ -  \ga    d_{\bom'}(y,y') }   \|f\|_2 
\end{equation} 
\end{lem}
\bigskip

\pr   This follows the proof of lemma  3.5     in   part II,  to which we refer for details.  Again the absence of averaging operator  in $\Om_{\sN +1}$
is compensated by   $\bar \mu_{\sN} =1$.   
\bigskip

\begin{lem}   \label{green3}     {  \  }
\begin{equation}
\begin{split}            \label{around}
 \Big| \tilde   C^{1/2}_{{\sN},  \bom'}f\Big|, \ \   \Big |( \tilde C^{1/2}_{{\sN},  \bom'})^{\loc} f\Big|    \leq &\     C \|f\|_{\infty}   \\  
 \Big | \de   \tilde C^{1/2}_{{\sN},  \bom'} f  \Big |   =  \Big| \Big( \tilde   C^{1/2}_{{\sN},  \bom'}-
 ( \tilde C^{1/2}_{{\sN},  \bom'})^{\loc}\Big ) f  \Big|                   \leq &\     C    \|f\|_{\infty}   e^{-r_{{\sN}}}  \\  
  \Big|  \tilde  C^{-1/2}_{{\sN},  \bom'}f\Big|, \ \   \Big |   \Big[ (  \tilde C^{1/2}_{{\sN},  \bom'})^{\loc} \Big]^{-1}f   \Big|    \leq  & \     C \|f\|_{\infty}  \\
  \end{split}
\end{equation}
\end{lem}
\bigskip

\bigskip
\pr  $\tilde   C^{1/2}_{{\sN},  \bom'}$   is expressed in terms  of   $   G_{{\sN},\de \bom',r}  $  in   (\ref{z1})-(\ref{z3}).   The results then follow 
 from the previous lemma,  as in  lemma 3.6 in part II.

\subsection{new   characteristic functions}

Since   $  \phi_{{\sN},   \de   \bom'}$  is already tiny inside  $\Om_{{\sN}+1}$  we  do not introduce any new conditions on
this field.   Thus  $Q_k = \emptyset$.  We   still  need   a small field  expansion  to  remove  the non-locality  in   the  characteristic function.
We  therefore introduce
\begin{equation}  \label{skunk}
\begin{split}
1=  &    \sum_{R_{{\sN}+1}  \subset   \Om_{{\sN}+1} }   
   \zeta^w_{{\sN}+1}(   R_{{\sN}+1} )    \chi^w_{{\sN}+1}(  \Om_{{\sN}+1}-  R_{{\sN}+1} )\\
 \end{split}
 \end{equation}
where   $   \chi^w_{{\sN}+1}(   \La_{{\sN}+1})$    enforces   $|W_{\sN}|  \leq   p_{0,{\sN}}$  everywhere in  $\La_{{\sN}+1}$.
The new small field region is  
\be
  \La_{{\sN}+1}  =   \Om_{{\sN}+1}^{5 \nat}   -   R_{{\sN}+1}^ { 5 *} \ \   \textbf{   or    }   \ \     \La^c_{{\sN}+1}  =  ( \Om^c_{{\sN}+1})^{5 *}    \cup     R_{{\sN}+1}^ { 5 *}
\ee
Then  (\ref{skunk})  is rewritten as   
\be
\begin{split}
1  =  &  \sum_{\La_{{\sN}+1}   \subset  \Om_{{\sN}+1}^{5 \nat} } \tilde   \cC_{{\sN}+1} ( \Om_{{\sN}+1},  \La_{{\sN}+1})\chi^w_{\sN}(  \La_{{\sN}+1} )  \\
 \tilde   \cC_{{\sN}+1} ( \Om_{{\sN}+1},  \La_{{\sN}+1}) 
  =  &  \sum_{R_{{\sN}+1}  \subset   \Om_{{\sN}+1} :   \La_{\sN+1}  =   \Om_{{\sN}+1}^{5 \nat}   -   R_{{\sN}+1}^ { 5 *} }
    \zeta^w_{{\sN}+1}(   R_{{\sN}+1} )    \chi^w_{{\sN}+1}( ( \Om_{{\sN}+1}-  R_{{\sN}+1}) - \La_{\sN+1} )   \\
\end{split}
\ee
This  is    inserted  under the integral signs in  (\ref{representation4})  and then  the sum is taken  outside the integrals.

The characteristic  functions  are  now  
\be      \label{other} 
 \cC_{{\sN}, \bpi}\  \chi_{{\sN}}(  \La_{\sN} ) \tilde  \cC_{{\sN}+1} ( \Om_{{\sN}+1},  \La_{{\sN}+1})\ \chi^w_{\sN}(  \La_{{\sN}+1} )
   =    \tilde \cC_{{\sN}+1, \bpi^+}  \   \chi_{\sN}(  \La^{**}_{{\sN}+1} )
    \chi^w_{\sN}(  \La_{{\sN}+1} )\ee
where    with   $\bpi^+  =   (\bpi,  \Om_{{\sN}+1},  \La_{{\sN}+1})$
\be    \tilde \cC_{{\sN}+1, \bpi^+} =   \cC_{{\sN}, \bpi}\  \chi_{{\sN}}(  \La_{\sN}  - \La_{{\sN}+1}^{**} ) 
   \tilde   \cC_{{\sN}+1} ( \Om_{{\sN}+1},  \La_{{\sN}+1})
\ee

\begin{lem}   \label{lemon}
 $ \tilde \cC_{{\sN}+1, \bpi^+}$  is   independent  of  $W_{\sN}$ in  $\La_{{\sN}+1}$  and  
on   the support $ \tilde \cC_{{\sN}+1, \bpi^+}  \ \chi^w_{\sN}(  \La_{{\sN}+1} )$   
\be     \chi_{{\sN}}(  \La^{**}_{{\sN}+1} )    =  1  \ee
\end{lem}
\bigskip

\pr  We  must show that  $  \Psi^{\loc}_{{\sN}, \Om_{{\sN}+1}}(\de  \bom')  + (\tilde  C^{1/2}_{{\sN},   \bom'  })^{\loc}W_{\sN}$  is 
in the space  $ \cS_{\sN}(\square)$   for  any  $M$-cube $\square   \in   \La^{**}_{{\sN}+1}$.  We  show separately that     $  \Psi^{\loc}_{{\sN}, \Om_{{\sN}+1}}(\de  \bom')$    and   $ (\tilde  C^{1/2}_{{\sN},   \bom'  })^{\loc}W_{\sN}$  are
in     $\frac12   \cS_{\sN}(\square)$.

The functions   $ \tilde \cC_{{\sN}+1, \bpi^+}  \ \chi^w_{\sN}(  \La_{{\sN}+1} )$  force that
  $|W_{\sN}|  \leq   p_{0,\sN}$  on $\La^{4*}_{\sN+1}$.  Then  by   (\ref{around})  we have  on   $\La^{3*}_{\sN+1}$  
\be  
 | (\tilde  C^{1/2}_{{\sN},   \bom'  })^{\loc}W_{\sN}  |  \leq  C p_{0,{\sN}}    =   \Big(   C \frac{p_{0,{\sN}} }{p_{\sN} }  \Big) p_{\sN}  
\ee
and the derivative satisfies the same bound since we are on a unit lattice.    By  lemma  3.1  in  part II it follows that  for  $\square  \subset 
   \La^{**}_{\sN+1}$ that   
$  (\tilde  C^{1/2}_{{\sN},   \bom'  })^{\loc}W_{\sN} $  is  in  $ ( C p_{{\sN},0}/p_{\sN} ) \cS_{\sN}(\square)$.  
But   for  $\la$  sufficiently small and  $p_0 < p$
\be     C \frac{p_{0,{\sN}} }{p_{\sN} }  =  C    ( - \log \la  )^{p_0 -p}   \leq  \frac 12
\ee
Therefore   $  (\tilde  C^{1/2}_{{\sN},   \bom'  })^{\loc}W_{\sN}   \in   \frac12   \cS_{\sN}(\square)$.

For the second point    note that  $\phi_{\sN, \de \bom'}$  depends on  $\Phi_{\sN}$   on  $\de  \Om_{\sN}  =\Om_{\sN} - \Om_{\sN+1}$.
But  $\cC_{\sN,  \bpi}$  gives a bound on  $\Om_{\sN} - \La_{\sN}$  and   $ \chi_{{\sN}}(  \La_{\sN}  - \La_{{\sN}+1}^{**} ) $  gives a 
bound  on   $\La_{\sN} - \Om_{\sN +1}$.   These  imply   $|\Phi_{\sN}|  \leq  C p_{\sN} \al_{\sN}^{-1}  \leq    C p_{\sN} \la_{\sN}^{-\frac14}$.
Since  $\La^{3*}_{{\sN}+1} $   is  at least   $2[r_{\sN}]$  layers of $M$ blocks  away  from  $\Om^c_{{\sN}+1}$,  the estimates
on $ G_{\sN,  \de \bom'}$    give that   on  $\La^{3*}_{{\sN}+1}$  we have $   |  \phi_{\sN, \de \bom'}|   \leq   e^{- r_{\sN}} $.
Then  $  \Psi_{\sN, \Om_{\sN+1}}(\de  \bom') =  [Q_{\sN}     \phi_{\sN,  \de   \bom'}]_{\Om_{\sN +1}} $  satisfies a similar bound  as does
 $ \Psi^{\loc}_{\sN, \Om_{\sN+1}}(\de  \bom')$. 
It follows easily   that   $ \Psi^{\loc}_{{\sN}, \Om_{{\sN}+1}}(  \de \bom')  \in   \frac12   \cS_{\sN}(\square)$  for  $\square  \subset   \La^{**}_{{\sN}+1}$  .
This completes the proof.
\bigskip

We  use this result in   (\ref{representation4}).
 We also    split the measure   $ d \mu_{\Om_{\sN +1}}(W_{\sN}) $   on  $\La_{\sN +1}$  and     identify
 \begin{equation}
 d W_{{\sN}+1,  \bpi^+}  =      d W_{{\sN},  \bpi} \    d \mu_{\Om_{\sN +1}- \La_{\sN+1}}(W_{\sN}) 
 \end{equation}
 and the
 ultralocal  probability measure
  \begin{equation}
d\mu^*_{\La_{{\sN}+1}}(W_{\sN})  \equiv   (\cN^w_{ {\sN},  \La_{{\sN}+1}})^{-1}  \chi^w_{k}(  \La_{{\sN}+1}) 
 d \mu_{\La_{\sN+1}}(W_{\sN}) 
\end{equation}     
The normalizing factor  $\cN^w_{ {\sN},  \La_{{\sN}+1}}$  has the form    $ \exp(-\vep_{\sN}^{(0)}\Vol(\La_{{\sN}+1}) ) $.    
We  also define $ \de    E_{\sN}^+$  by  
\be    E_{\sN}^+ (X, \phi + \cW)  =     E_{\sN}^+ (X, \phi)  +  \de    E_{\sN}^+ (X, \phi, \cW)
\ee
Then     (\ref{representation4})  becomes
\begin{equation}    \label{representation5}
\begin{split}
 \sZ_{\sM,\sN} =     &  \sZ_{\sM,\sN}(0)    \sum_{\bpi^+}   \int     d  \tilde \Phi_{{\sN}, \bom^c}\   d W_{{\sN}+1,  \bpi^+} \ 
 K_{{\sN}, \bpi} \    \tilde \cC_{{\sN}+1, \bpi^+} \       \\
& \  \exp \Big( \tilde c_{\sN}  |\Om^{c,({\sN})}_{{\sN}+1}| -  S^*_{{\sN}}(\La_{\sN}-  \Om_{{\sN}+1})  - \tilde  S_{\sN}( \Om_{{\sN}+1})   \Big )\
\tilde   \Xi_{{\sN},  \bpi^+}  
 \\
\end{split}
\end{equation}
 where  we  have  isolated  the     fluctuation integral 
 \begin{equation}
 \begin{split}
\tilde    \Xi_{{\sN},  \bpi^+}  
= &  
  \exp\Big(-\vep_{\sN}^{(0)}\Vol(\La_{{\sN}+1})  +  E^+_{\sN}(\La_{\sN}) \Big)  \\
  & \int  \    d\mu^*_{\La_{{\sN}+1}}(W_{\sN})    \exp \Big(  \de  E^+_{\sN}(\La_{\sN} )  +  R^{(\leq 7)}_{ \bpi, \Om_{\sN +1}}  +    B_{{\sN}, \bpi} (    \La_{\sN})     \Big)  \\
  \end{split}
\end{equation}

 \subsection{localization}
 
 We next localize the expressions   in the fluctuation integral.

  \begin{lem}    \label{first}   For  complex   $|\Phi_{{\sN}, \de \Om_{\sN}}|  \leq   \la_{\sN}^{-\frac14 - \de}$    and     $  |W_{\sN}| \leq  B_wp_{\sN}$:
 \begin{equation}   \label{city}
 \begin{split}
& \de  E^{+}_{\sN} \big(\La_{\sN},    \phi_{{\sN},  \de   \bom'}   + \tilde     \cW_{{\sN},\bom'}  )
 =   \sum_{X \in \cD_{\sN}: X \subset   \La_{{\sN}+1}}   ( \de  E^+_{\sN})^{\loc}( X ) \\
 +& 
 \sum_{X  \in  \cD_{\sN}(\bmod \Om^c_{{\sN}+1}), X \#   \La_{{\sN}+1}}  B^{(E)}_{{\sN}, \bpi^+}(X)  \   +  \  \tilde B_{{\sN}+1, \bpi^+}   \textrm{ terms } \\
\end{split}
\end{equation}
where
\begin{enumerate}
\item     The leading terms     $ (\de  E^+_{\sN})^{\loc}( X) = (\de  E^+_{\sN})^{\loc}( X,   W_{\sN})$   depends on  $ W_{\sN}$ only in $X$, are analytic,   and satisfy
\begin{equation}
 |   ( \de   E^+_{\sN})^{\loc}( X  )|   \leq      \cO(1)  \la^{\beta}  e^{- (\ka- \ka_0-2) d_{M}(X)  }
\end{equation}
\item   The boundary terms   $ B^{(E)}_{{\sN}, \bpi^+}(X)=  B^{(E)}_{{\sN}, \bpi^+}(X,\Phi_{{\sN}, \de \Om_{\sN}},    W_{\sN} )  $  depends on $\Phi_{{\sN}, \de \Om_{\sN}},    W_{\sN}$ only in $X$,  are analytic,       
 and satisfy 
  \begin{equation}         \label{boundary}
|B^{(E)}_{{\sN}, \bpi^+}(X )|   \leq    \cO(1) \la^{\frac14 - 10 \ep}  e^{ -  ( \ka-2\ka_0-3)  d_{M}(X,  \bmod \Om^c_{{\sN}+1})  }
\end{equation}
\item    $\tilde B_{{\sN}+1, \bpi^+}$  terms   are bounded by   $  C|\La^{({\sN})}_{\sN} - \La^{({\sN})}_{{\sN}+1}| =   C\  \Vol(\La_{\sN} - \La_{{\sN}+1})$.
\end{enumerate}
\end{lem}
\bigskip

\pr  We  are studying   $ \de E^{+}_{\sN} \big(\La_{\sN} ) =  \sum_{X \subset  \La_{\sN}} \de E^{+}_{\sN} (X  ) $
where    $ \de E^{+}_{\sN}(X )  =  \de  E^{+}_{\sN}(X,    \phi_{{\sN},  \de   \bom'},  \tilde   \cW_{{\sN},\bom'}  ) $.  Our  assumptions  imply
$ |    \phi_{{\sN},  \de   \bom'}  |   \leq    C   \la_{\sN}^{-\frac14 - \de}$  also for derivatives,    and    $|\tilde  \cW_{k, \bom'}|  \leq    C  p_{\sN}$.
Since   $\de  < \ep$   these bounds  put  us  well inside the domain of analyticity   $\cR_{\sN}$  for  $\la_{\sN}= \la$ sufficiently small.
   So the  basic   bound  $ |  \de   E^+_{\sN}( X  )|   \leq      \la^{\beta}  e^{- \ka d_{M}(X)  }$
 is satisfied.    

To  localize   we  introduce weakened fields    $  \phi_{{\sN},\de   \bom'}(s)$   and    $\tilde \cW_{{\sN},\bom'}(s)$   based on the random walk 
expansions for  $  G_{{\sN},  \de   \bom'}(s)$  and   $\tilde C^{\frac12}_{{\sN}, \bom'}(s)$.   These satisfy the same bounds.  We  proceed as in the proof of lemma 3.15 in
part II with the following modifications.  (1.)  The  random walk is based has  $M$-cubes, not $LM$-cubes, in $\Om_{\sN+1}$,  (2.)  There is no reblocking,
(3.)   The decoupling expansion  can be done   for   $   \phi_{{\sN},\de  \bom'}$   and  $\tilde   \cW_{{\sN},\bom'}$   simultaneously.
The   result  is   an expansion  
\be    \label{soho}
  \de  E^{+}_{\sN} \big(\La_{\sN} )  =  \sum_{X  \in   \cD_{\sN},   X   \cap  \La_{\sN}  \neq  \emptyset}   ( \de  E^{+}_{\sN})' (X, \Phi_{{\sN}, \de \Om_{\sN}},    W_{\sN}  )  
\ee
 where   
 \be    
  |   (  \de  E^+_{\sN})'( X  )|   \leq      \cO(1)  \la^{\beta}  e^{- (\ka- \ka_0-2) d_{M}(X)  }
 \ee

 Now     terms  in  (\ref{soho})     with  $X \subset  \La_{{\sN}+1}$   depend only on   $W_{\sN}$ in $X$  and  are identified as
 the  terms  $ ( \de  E^+_{\sN})^{\loc}( X) $.    If      $ X \#  \La_{\sN}$    we   add on any connected component  of  $\Om^c_{{\sN}+1}$
 which is connected to $X$  to get   $X^+  \in      \cD_{\sN}(\bmod \ \Om^c_{{\sN}+1})$.    Terms  in   (\ref{soho})  with  $X \# \La_{\sN}$
 are partially summed  by  the $X^+$ they determine and this  gives the boundary terms $ B^{(E)}_{{\sN}, \bpi^+}(X)$  which satisfy the bound   (\ref{boundary}).   Finally
 terms  in  (\ref{soho})   with  $X  \subset  \La^c_{{\sN}+1}$   are the    $\tilde B_{{\sN}+1, \bpi^+}$ terms.   This completes the proof. 
 \bigskip

  \begin{lem}    \label{second}  
  For  complex   $|\Phi_{{\sN}, \de \Om_{\sN}}|  \leq   \la_{\sN}^{-\frac14 - \de}$   and     $  |W_{\sN}| \leq  B_wp_{\sN}$: 
    \begin{equation}
  R^{(\leq   7)}_{ \bpi, \Om_{{\sN}+1}}   = \sum_{X \in \cD_{{\sN}}: X  \subset  \La_{{\sN}+1} }    R^{\loc}_{{\sN},\bpi^+}( X) 
  +    \sum_{  X \in \cD_{{\sN}}( \bmod  \Om^c_{{\sN}+1}), X  \# \La_{{\sN}+1}}     B^{(R)}_{{\sN}, \bpi^+}( X)  
  + \ \  \tilde B_{{\sN}+1, \bpi^+}   \textrm{  terms  }  
 \end{equation}
 Here   $  R^{\loc}_{{\sN},\bpi^+}( X,  W_{\sN}) $  and $  B^{(R)}_{{\sN}, \bpi^+}( X,   \Phi_{{\sN},\de  \Om_{\sN}},  W_{{\sN}})  $ 
 are  strictly localized,  analytic  in the fields,  and  satisfy    
 \begin{equation} \label{lulu0}
\begin{split}
|   R^{\loc}_{{\sN},\bpi^+}( X) |  \leq  &   \cO(1)  \la^{n_0}   e^{-(\ka - \ka_0-2) d_{M}(X)}  \\
| B^{(R)}_{{\sN}, \bpi^+}( X) | \leq  &      \cO(1) \la^{n_0}   e^{-(\ka- 2 \ka_0 - 3) d_{M} (X,  \bmod  \Om^c_{{\sN}+1})}  \\
 \end{split}
\end{equation}
  \end{lem}
  \bigskip

\pr  The function   $  R^{(\leq  7)}_{ \bpi^+, \Om_{k+1}} $  has  many parts.  Consider the  original term 
$R^{(0)}_{\bpi,  \Om_{\sN+1}}  \equiv   R_{k,\bpi}(\La_{\sN})$.    After the     change of  variables  this     has the form     
\begin{equation}   
R_{\sN, \bpi}(\La_{\sN})  =   \sum_{X \in \cD_{\sN},  X  \subset  \La_{\sN}}  
   R_{\sN, \bpi  }(X, \Phi_{\sN, \de \Om_{\sN}}, \Psi^{\loc}_{\sN,\Om_{\sN+1}}(\de \bom')  +   (\tilde  C_{\sN,\bom'}^{1/2})^{\loc}W_{\sN})  \\ 
\end{equation}
In  addition to   $|\Phi_{\sN, \de \Om_{\sN}}|  \leq   \la_{\sN}^{-\frac14 - \de}$  we  have  $|\Psi^{\loc}_{k,\Om_{k+1}}(\de  \bom')|
 \leq  C  \la_{\sN}^{-\frac14 - \de}$  and     $|(\tilde   C_{{\sN},\bom'}^{1/2})^{\loc}W_{\sN}|  \leq  C  p_{\sN} \leq  C\la_{\sN}^{-\frac14 - \de }$.  
Then     for  $\square  \subset \La_{\sN}$  and local fields 
\be  \label{stupor}
 \phi_{\sN, \bom(\square)}  =\phi_{\sN, \bom(\square)}\Big(\Phi_{{\sN}, \de \Om_{\sN}}, \Psi^{\loc}_{{\sN},\Om_{{\sN}+1}}( \de  \bom')  +   (\tilde   C_{{\sN},\bom'}^{1/2})^{\loc}W_{\sN}\Big)  
\ee
 we have the estimates     $|\phi_{\sN, \bom(\square)}|  \leq    C \la_{\sN}^{-\frac14 - \de }  $ 
and     $|\pa   \phi_{\sN, \bom(\square)}|  \leq    C \la_{\sN}^{-\frac14 - \de }  $.
  These are the estimates give that  $(\Phi_{{\sN}, \de \Om_{\sN}}, \Psi^{\loc}_{{\sN},\Om_{{\sN}+1}}( \de  \bom')  +   (\tilde   C_{{\sN},\bom'}^{1/2})^{\loc}W_{\sN})$  is in  $C \cP_{\sN}(\square, \de)$   and hence  in  $C \cP_{\sN}(\La_{\sN}, \de)$.        But for  $\la_{\sN} = \la$ sufficiently small   $ C \la_{\sN}^{-\frac14 - \de } 
    \leq    \la_{\sN}^{-\frac14 - 2\de } $  so this field is in    $ \cP_{\sN}(\La_{\sN}, 2\de)$.  Thus  we are in the analyticity domain for 
      $ R_{\sN,\bpi}(X)$  and   
 can use   the bound    $|   R_{{\sN},\bpi}( X) |  \leq      \la^{n_0}   e^{-\ka  d_{M}(X)} $.

 To  localize   we  introduce weakened fields    $\Psi^{\loc}_{\sN,\Om_{\sN+1}}(\de   \bom',s)$   and    $(\tilde C_{k,\bom'}^{1/2})^{\loc}(s)W_k$.    These satisfy the same bounds.  We  proceed   with a decoupling expansion   as in the proof of lemma 3.16 in
part II,  except that  the  random walk  has no    $LM$-cubes  and there is   no reblocking.  As in the previous lemma
the   result  is   an expansion  
\be   
  R_{{\sN}, \bpi}(\La_{\sN} )  =  \sum_{X  \in   \cD_k,   X   \cap  \La_{\sN}  \neq  \emptyset}    (R_{{\sN}, \bpi})' (X, \Phi_{{\sN}, \de \Om_{\sN}},    W_{\sN}  )  
\ee
 where   $ (   R_{{\sN}, \bpi})'( X  )$  is strictly localized and analytic and  satisfies 
 \be    
  |   (   R_{{\sN}, \bpi})'( X  )|   \leq      \cO(1)  \la^{n_0}  e^{- (\ka- \ka_0-2) d_{M}(X)  }
 \ee
 Now    divide the terms  by  $X \subset  \La_{{\sN}+1}$,  $ X  \# \La_{{\sN}+1}$,  and  $X  \subset  \La^c_{{\sN}+1}$
 and   get a contribution for each of the three types of terms.

 The other contributions to  $  R^{(\leq  7)}_{ \bpi^+, \Om_{k+1}} $ are treated similarly,  see  lemma  3.16  in part II.  This 
 completes the proof.
 \bigskip

 For the  next  result  we  recall that the analyticity domain for  $B_{\sN, \bpi}$  is    $|W_j|  \leq  B_w p_jL^{\frac12(\sN-j)}$ on $\Om_{j+1} - \La_{j+1}$
 for  $j=1, \dots,   \sN-1$  and  $\Phi_{k, \bom}$ in   
 \be    
    \cP_{{\sN} ,\bom}   = \bigcap_{j=1}^{{\sN}-1}\Big[  \cP'_j(\de  \Om_j, \de) \Big]_{L^{-({\sN}-j)}}  \cap  \cP_{\sN}( \Om_{\sN} - \Om_{\sN}^{2\nat},  \de)  \cap   \cP_{\sN}(\Om_{\sN}^{ 2  \nat}, 2 \de)  
    \ee
We   modify  it  to a complex   domain for  
\be 
\Phi_{\sN, \de \bom^+}  \equiv   ( \Phi_{1, \de \Om_1}, \dots,   \Phi_{\sN, \de \Om_{\sN}})
\ee  
which is  
 \be    
 \begin{split}
 \tilde     \cP_{{\sN} ,\bom^+}   = &\bigcap_{j=1}^{{\sN-1}}\Big[  \cP'_j(\de  \Om_j, \de) \Big]_{L^{-({\sN}-j)}}
  \    \cap  \  \cP_{\sN}( \Om_{\sN} - \Om_{\sN}^{2\nat},  \de)  \\
&  \cap \   \cP_{\sN}(\Om_{\sN}^{ 2  \nat}-  \Om^*_{\sN +1}, 2 \de)  
 \   \cap   \   \Big   \{    |\Phi_{\sN,  \de   \Om_{\sN}} |  \leq  \la_{\sN}^{-\frac14 -   \de}  \Big\} \\
\end{split}    
 \ee
This is contained in the domain $   |\Phi_{\sN,  \de   \Om_{\sN}} |  \leq  \la_{\sN}^{-\frac14 -   \de} $   used in  lemma  \ref{first} and lemma \ref{second},   but still  is large enough to contain the
domain  specified by the characteristic functions.

 \begin{lem}    \label{third}   For    $  \Phi_{{\sN},  \de  \bom^+}   \in   \tilde     \cP_{{\sN} ,\bom} $  and   $|W_j|  \leq  B_w p_jL^{\frac12(\sN-j)}$
   \begin{equation}
 \begin{split}
 B_{{\sN}, \bpi}(\La_{\sN})  
  =  &  \sum_{       X  \in \cD_{{\sN}}(\bmod  \Om^c_{{\sN}+1}), X \#   \La_{{\sN}+1}     }   
    B^{(B)}_{{\sN}, \bpi^+}(X) 
   +   \tilde  B_{{\sN}+1, \bpi^+}  \textrm{  terms }
\end{split}
\end{equation}
Here      $ B^{(B)}_{{\sN}, \bpi^+}(X,    \Phi_{{\sN},  \de    \bom^+}, W_{{\sN}+1,\bpi^+}, W_{{\sN}, \La_{{\sN}+1}})   $  is    strictly local in the fields,  analytic,   
and satisfies
 \begin{equation}
|  B^{(B)}_{{\sN}, \bpi^+}(X) |  \leq   
  \la^{n_0}   e^{ -( \ka  - \ka_0  -3)   d_{M}(X, \bmod  \Om^c_{{\sN}+1}) }
\end{equation}
 \end{lem}
\bigskip

\pr   We  are studying 
\begin{equation}  \label{spitoon}
 B_{{\sN}, \bpi}(\La_{\sN})    =
 \sum_{     X \in \cD_{\sN}(\bmod  \Om^c_{{\sN}}), X \#   \La_{\sN}   }           B_{{\sN}, \bpi}  \Big(X,   \Phi_{{\sN},  \de  \bom^+},    \Psi^{\loc}_{{\sN}, \Om_{{\sN}+1}}(\de  \bom') 
 +   (\tilde   C_{{\sN}, \bom'}^{1/2})^{\loc}W_{{\sN}} , W_{{\sN},\bpi}\Big)  
\end{equation}
We   claim that under our assumptions   the field    $ ( \Phi_{{\sN},  \de  \bom^+},    \Psi^{\loc}_{{\sN}, \Om_{{\sN}+1}}(\de  \bom') 
 +   (\tilde   C_{{\sN}, \bom'}^{1/2})^{\loc}W_{{\sN}} )$  is  in   the  domain  $\cP_{ \sN, \bom}$.   This is a statement about the fields
 $\phi_{\sN, \bom}(\square)$ defined in (\ref{stupor}).
  If  $\square \subset    (\Om^{**}_{\sN +1})^c$   the statement    is inherited from the
 definition  of  $ \tilde     \cP_{{\sN} ,\bom^+} $.     On the other hand  if  $\square    \subset    \Om^{**}_{\sN +1}$ then,   as  in   the proof of the previous lemma,   the bounds
 $ |\Phi_{\sN,  \de   \Om_{\sN}} |  \leq  \la_{\sN}^{-\frac14 -   \de}  $   and  $|W_{\sN}| \leq   p_{\sN}$   imply that the field is in    $ \cP_{\sN}(\square  , 2\de)$ and hence in  
$ \cP_{\sN}( \Om^{**}_{\sN +1}, 2 \de)$  as  required.  Hence the claim  is verified  (twice in $\Om^{**}_{\sN +1}  -  \Om^{*}_{\sN +1}$). 
 Thus  we  are in   the analyticity domain  for  $B_{\sN, \bpi}$    and    have the bound
 \be | B_{{\sN},\bpi}  (X) |  \leq  B_0  \la^{\beta}  e^{- \ka  d_M(X,  \bmod  \Om_{\sN}^c)}  \ee

Terms with  $X  \subset  \Om^c_{{\sN}+1}$  are already localized and qualify   as   $\tilde  B_{{\sN}+1, \bpi^+}$  terms.
The remaining terms have   $X \cap  \Om_{{\sN}+1}  \neq  \emptyset$.
For these we localize  by  introducing     $\Psi^{\loc}_{{\sN},\Om_{{\sN}+1}}(\bom',s) $   and   $ ( C_{{\sN},\bom'}^{1/2})^{\loc}(s)W_{\sN}$   and  making
a decoupling expansion.    This follows the  proof of lemma  3.17  in part II,  except  that  the random walk has no  $LM$-cubes 
and   there is no reblocking.  After   adding on appropriate    connected components    of  $\Om^c_{{\sN}+1}$  the   result is   
\be    \label{soho7}
  B_{{\sN}, \bpi} (\La_{\sN} )  =  \sum_{X \# \La_{\sN},    X   \cap  \Om_{{\sN}+1}  \neq  \emptyset}   ( B_{{\sN}, \bpi} )' (X,   \Phi_{{\sN},  \de    \bom^+}, W_{{\sN}+1,\bpi^+}, W_{{\sN}, \La_{{\sN}+1}} )
   +   \tilde  B_{{\sN}+1, \bpi^+}  \textrm{  terms }  
\ee
 where   now the sum is over   $X  \in   \cD_{\sN}( \bmod \ \Om^c_{{\sN}+1})$  and  where  
 \be    
  |   ( B_{{\sN}, \bpi} )'( X  )|   \leq      \cO(1)  B_0  \la^{\beta}e^{- (\ka- \ka_0-2) d_{M}(X,  \bmod  \Om^c_{{\sN}+1})  }
 \ee
Terms with  $X  \subset   \La^c_{{\sN}+1}$  are  $ \tilde  B_{{\sN}+1, \bpi^+}  \textrm{  terms }$.   Terms with   $X \# \La_{{\sN}+1}$
are   the terms  $ B^{(B)}_{{\sN}, \bpi^+}(X)$.   The  stronger  bound  with  $ \la^{n_0} $  is obtained from the separation of 
$\La_{{\sN}+1}$  and  $\La^c_{\sN}$.   This completes the proof.
\bigskip

Now  all the  active     boundary  terms  can be combined into a single boundary term   
\be     B_{{\sN}, \bpi^+}^{\loc}(X)  =   B^{(E)}_{{\sN}, \bpi^+}(X)  +   B^{(R)}_{{\sN}, \bpi^+}(X)  +    B^{(B)}_{{\sN}, \bpi^+}(X) 
\ee
analytic  in   $\tilde \cP_{\sN, \bom^+}  \times    \{   |W_j|  \leq  B_w p_jL^{\frac12(\sN-j)} \}$  and satisfying the various  stated bounds.
All the  inactive   boundary  terms    $ \tilde  B_{{\sN}+1, \bpi^+}  \textrm{  terms }  $   are   
combined into a single  term    $ \tilde  B_{{\sN}+1, \bpi^+} (\La_{\sN}, \La_{{\sN}+1}) $ 
analytic  in   the same domain      satisfying    there 
 \be  \label{ninety}
    |  \tilde B_{{\sN}+1, \bpi^+}(\La_{\sN}, \La_{{\sN}+1} ) |  \leq    C|\La^{({\sN})}_{\sN} - \La^{({\sN})}_{{\sN}+1}|
\ee

 The fluctuation integral is then 
 \begin{equation}  \label{stone}
 \begin{split}
\tilde   \Xi_{{\sN},  \bpi^+}  
=   &
 \exp\Big(-\vep_{\sN}^{(0)}\Vol(\La_{{\sN}+1})  +  E^+_{\sN}(\La_{\sN})  +    \tilde  B_{{\sN}+1, \bpi^+} (\La_{\sN}, \La_{{\sN}+1})  \Big)   \\ 
  & \int  \    d\mu^*_{\La_{{\sN}+1}}(W_{\sN})  \exp \Big( ( \de  E^+_{\sN})^{\loc}(\La_{\sN+1} )  
+ R^{\loc}_{{\sN}, \bpi^+}  (    \La_{\sN+1})    +    B^{\loc}_{{\sN}, \bpi^+} (    \La_{\sN+1})     \Big) 
\end{split}
\end{equation}

\subsection{cluster expansion}
A cluster expansion is  now  carried out as  in section 3.14 in part II,   and   in the resulting  localization expansion we
identify  leading,   tiny,  and boundary terms.    We   find      
\begin{equation}  \label{aplomb}
\begin{split}
& \int  \    d\mu^*_{\La_{{\sN}+1}}(W_{\sN})  \exp \Big( ( \de  E^+_{\sN})^{\loc}(\La_{\sN+1} )  
+ R^{\loc}_{{\sN}, \bpi^+}  (    \La_{\sN+1})    +    B^{\loc}_{{\sN}, \bpi^+} (    \La_{\sN+1})     \Big) \\
& =   
 \exp  \Big(       E^\#_{{\sN}}(\La_{{\sN}+1}) +     R^\#_{{\sN}, \bpi^+}(\La_{{\sN}+1} ) +    B^\#_{{\sN},  \bpi^+}(  \La_{{\sN}+1}  )     \Big)\\
\end{split}
\end{equation}
Here    $ E^\#_{\sN}(\La_{\sN+1} ) ,   R^\#_{{\sN}, \bpi^+}(\La_{\sN+1} )    $  have local expansions  like $ E^\#_{\sN}(\La_{\sN +1} )  =  \sum_{X  \subset  \La_{\sN+1}}  E^\#_{\sN}( X)$    with  $X  \in  \cD_{{\sN}}$.   These  are 
are independent of all fields and satisfy  
\be 
\begin{split}  \label{spirit}
   |E^\#_{\sN}( X)|  \leq   &\  \one  \la^{\beta}  e^{-(\ka - 6\ka_0 -6)d_M(X) }  \\
     |R^\#_{{\sN}, \bpi^+}( X)|  \leq   &\one   \la^{n_0}  e^{-(\ka - 6\ka_0 -6)d_M(X)}   \\
\end{split}
\ee
Also  $ B^{\#}_{{\sN}, \bpi^+}(\La_{{\sN}+1})=  \sum_{X \#  \La_{{\sN}+1}}  B^{\#}_{{\sN}, \bpi^+}(X) $  with   $X  \in  \cD_{{\sN}}(\bmod\  \Om^c_{{\sN}+1})$. 
Here     $   B^\#_{{\sN}, \bpi^+}(X   )   $   is a function  of  $( \Phi_{{\sN}, \de    \bom^+}, W_{{\sN}+1,\bpi^+})$,     
is  analytic  in    $\tilde \cP_{\sN, \bom^+}  \times    \{   |W_j|  \leq  B_w p_jL^{\frac12(\sN-j)} \}$,
and satisfies there 
\be
    |  B^\#_{{\sN}, \bpi^+}(X)  |   \leq    \cO(1)  B_0  \la^{\beta}e^{- (\ka- 6\ka_0-6) d_{M}(X,  \bmod  \Om^c_{{\sN}+1})  }  
\ee

Now  insert      (\ref{aplomb})  back  into  (\ref{stone}),  and then   (\ref{stone})  back into    (\ref{representation5}),   and obtain  
 \begin{equation}    \label{representation6}
\begin{split}
 \sZ_{\sM,\sN} =     &Z_{\sM, \sN}(0) \sum_{\bpi^+}   \int     d   \tilde  \Phi_{{\sN}, \bom^c}\   d W_{{\sN}+1,  \bpi^+} \ 
    K_{{\sN}+1, \bpi} \    \tilde \cC_{{\sN}+1, \bpi^+} \       \\
&   \exp  \Big(   \tilde c_{\sN}  |\Om^{c,({\sN})}_{{\sN}+1}|   -  \vep_{\sN}^0 \Vol(  \La_{{\sN}+1} )   -  S^*_{{\sN}}(\La_{\sN}-  \Om_{{\sN}+1})  - \tilde  S_{\sN}( \Om_{{\sN}+1})   +  E^+_{\sN} (\La_{\sN})   \\      +&  E^\#_{{\sN}}(\La_{{\sN}+1}) +     R^\#_{{\sN}, \bpi^+}(\La_{{\sN}+1} ) +    B^\#_{{\sN},  \bpi^+}(  \La_{{\sN}+1}  )   + \tilde B_{{\sN}+1, \bpi^+} (\La_{\sN}, \La_{{\sN}+1})  \Big)\\
\end{split}
\end{equation}
\bigskip

\subsection{final  localization}

We  would  like to write the action in the final  small field region   $\La^c_{{\sN}+1}$  as  a sum  over pieces concentrated in the various connected components.   However  there is  still some  dependence  on the field  $\phi_{{\sN}, \de \bom'}$   which
is  defined all  over  $\La_{{\sN}+1}$  (but is tiny there)   and penetrates into  $\La^c_{{\sN}+1}$.     This gives weak coupling 
between  the connected components  of  $\La^c_{{\sN}+1}$.   We  have  to exhibit this  coupling in a form we can use.

The  terms we have to consider are 
\be    S^*_{{\sN}}(\La_{\sN}-  \Om_{{\sN}+1})  + \tilde  S_{\sN}( \Om_{{\sN}+1}) =
 S^*_{{\sN}}(\La_{\sN}-  \Om_{{\sN}+1})  + \tilde  S_{\sN}( \Om_{{\sN}+1}- \Om^{\nat}_{{\sN}+1})  +   \tilde  S_{\sN}( \Om^{\nat}_{{\sN}+1}) 
\ee
and     
\be    E^+_{\sN} (\La_{\sN})   =   E^+_{\sN} (\La_{\sN}- \Om^{\nat}_{{\sN}+1}) +   E^+_{\sN} (\Om^{\nat}_{{\sN}+1})   
\ee
We  first deal with   the   last two  terms  which contain  $\La_{\sN +1}$.  

\begin{lem}     For  $|\Phi_{{\sN}, \de \Om_{\sN}} |  \leq  \la^{-\frac14 - \de}$: 
\be   \label{undone}
 \begin{split}
   E^+_{\sN} (\Om^{\nat}_{{\sN}+1}, \phi_{{\sN}, \de \bom'}) -   E^+_{\sN} (\Om^{\nat}_{{\sN}+1}, 0)      = 
    &   \sum_{Y  \cap \Om^{\nat}_{{\sN}+1} \neq  \emptyset,  Y  \cap  \de \Om_{\sN} \neq  \emptyset  }  R^{*,(1)}_{{\sN}, \bpi^+}(Y,\Phi_{{\sN}, \de \Om_{\sN}})  \\
  \tilde  S_{\sN}( \Om^{\nat}_{{\sN}+1},  \phi_{\sN,  \de  \bom'}) 
  = &   \sum_{Y \cap  \Om^{\nat}_{{\sN}+1} \neq  \emptyset, Y  \cap  \de \Om_{\sN} \neq  \emptyset  }  R^{*,(2)}_{{\sN}, \bpi^+}(Y,\Phi_{{\sN}, \de \Om_{\sN}})  \\
\end{split}
\ee
where  the terms on the right  are strictly localized and  satisfy
\be  
  | R^{*,(i)}_{{\sN}, \bpi^+}(Y)| \leq   \la^{n_0}  e^{ - (\ka- \ka_0 -2)  d_M(Y) }  \hs   i=1,2  
   \ee
\end{lem}
\bigskip

\pr  This follows the proof  of lemma 3.22  in part II. Consider   $E^+_{\sN} (\Om^{\nat}_{{\sN}+1})$.  
Since   $\Om^{\nat}_{{\sN}+1}$  and  $\Om_{\sN+1}$   are  separated   
by   $[r_{\sN}] $ layers   of  $M$-cubes we have        $| \phi_{{\sN},  \de   \bom'}|  \leq  e^{-r_{\sN}/2}$ on $\Om^{\nat}_{{\sN}+1}$. 
(See a similar argument in lemma \ref{lemon}).
Then we can  write   for  $X  \subset \Om^{\nat}_{{\sN}+1}$ 
\be  
    E^+_{\sN} (X,    \phi_{{\sN}, \de   \bom'})  - E^+_{\sN} (X,0)  
=  \frac {1}{2\pi i}  \int_{|t|  =  e^{r_k/2}}   
\frac{dt}{t(t-1)}   E_{\sN}(X,  t    \phi_{{\sN}, \de  \bom'})
 \ee 
 and  obtain the bound     
 \be   | E^+_{\sN} (X, \phi_{{\sN},\de \bom'})  - E^+_{\sN} (X,0) | \leq   e^{-r_{\sN/2}}  e^{-\ka d_M(X) }
 \ee 

  Next  replace   $ \phi_{{\sN},  \de   \bom'}$  by the weakened  version  $ \phi_{{\sN},\de    \bom'}(s)$,  and   
 then     $ E^+_{\sN} (X,  \phi_{{\sN}, \de   \bom'}(s)) - E^+_{\sN} (X,0) $ has the same bound.  
 Then     $  E^+_{\sN} (X,    \phi_{{\sN}, \de   \bom'})  - E^+_{\sN} (X,0)   =  \sum_{Y \supset X} E^+_{\sN}(X,Y)$ where $Y$ is connected and   
    \be    \label{snoopy}
 E^+_{\sN}(X,Y,  \Phi_{\sN, \de \Om_{\sN}})   =    \int   ds_{Y-X}  \frac  { \pa   }{ \pa s_{Y-X} }\Big[  E^+_{\sN}(X, \phi_{{\sN},  \de   \bom'}(s)) -  E^+_{\sN}(X,0)  \Big]_{s_{Y^c} =0, s_X=1}  
\ee
which is strictly localized  in  $Y$.
In  the random walk expansion for   
 $ \phi_{{\sN}, \de   \bom'}(s)$  here   only paths  which   
 start in   $X$ and finish in $\de \Om_{\sN}$  contribute.     In  addition  
the condition  $s_{Y^c}  =0$ imposes that  only  paths  in  $Y$ contribute.  Thus paths must  intersect  $Y \cap   \de  \Om_{\sN}$.   If  
$Y \cap  \de  \Om_{\sN}  = \emptyset$  then  $ \phi_{{\sN},  \de   \bom'}(s)=0 $  and  so  $ E^+_{\sN}(X,Y)  =0$.   Thus we can assume  $Y \cap  \de  \Om_{\sN} \neq  \emptyset$.

  If  we  now  define   
\be  
   R^{*,(1)}_{{\sN},  \bpi^+}(Y)    =  \sum_{X \subset  Y  \cap  \Om^{\nat}_{\sN +1}}    E^+_{\sN} (X,Y) 
    \ee
then we have  (\ref{undone}). 
Using Cauchy bounds    we  obtain   the stated bound   on $ R^{*,(1)}_{{\sN},  \bpi^+}(Y)$   by the usual analysis.    The  analysis of
 $ \tilde  S_{\sN}( \Om^{\nat}_{{\sN}+1})$  is similar.    This completes the proof.
\bigskip

Recall that  the field  $\phi_{\sN, \de \bom '}$  is  defined  in   (\ref{sombulant})   in terms of  $\bom' =  ( \bom(\La_{\sN}^*),  \Om_{\sN+1} )$
and  $G_{\sN,  \de \bom'}$  defined in (\ref{santa}).   We  modify  this  to a more local field  by introducing  
\be
  \bom''   \equiv  \bom(\La_{\sN},  \Om_{\sN+1}, \Om^{ \nat}_{\sN+1})  \equiv   \bom'  \cap  \bom( (\Om^{2 \nat}_{\sN+1})^c )
  \ee
and  the Green's   function    
\be    \label{santa2}
  G_{\sN,   \bom''}    =  \Big [ - \De   +  \bar \mu_{\sN}   +   Q^T_{\sN,     \bom''}  \ba  \tilde  Q_{\sN, \de  \bom''}  \Big ]_{\Om''_1}^{-1}
\ee
The  field     $\phi_{\sN, \bom ''}$  is defined  just  as $\phi_{\sN, \de \bom '}$  in    (\ref{sombulant})   but   with    $G_{\sN,  \de \bom'}$ 
replaced  by    $ G_{{\sN},   \bom''}  $   (and  still with vanishing field in  $\Om_{\sN +1}$).   
 The field   $\phi_{\sN, \de \bom ''}$   is localized in a region slightly larger than
   $ \La^*_{\sN} \cap \Om^{2 \nat}_{\sN+1}=\La^*_{\sN}   \cap   (  \Om^c_{\sN+1} )^{2*}$   and hence outside of  $\La_{\sN +1}$.

\begin{lem}  \label{sour}     For  $|\Phi_{{\sN}, \de \Om_{\sN}} |  \leq  \la^{-\frac14 - \de}$ 
\be 
\begin{split}  \label{sweet}
 E^+_{\sN}( \La_{\sN}  - \Om_{\sN+1}^{\nat},      \phi_{{\sN},  \de  \bom'}  )
  = & E^+_{\sN}(\La_{\sN}    - \Om_{\sN+1}^{\nat},     \phi_{\sN,   \bom''})
   \\
&+ \sum_{Y \cap  (\La_{\sN}  - \Om_{\sN+1}^{\nat} )  \neq  \emptyset} R^{*, (3)}_{{\sN}, \bpi^+}(Y, \Phi_{{\sN},  \de \Om_{{\sN}+1}}) 
 \\
   S^{*}_{{\sN}}(\La_{\sN} - \Om_{{\sN}+1},  \Phi_{{\sN}, \de \Om_{\sN}},    \phi_{{\sN},  \de   \bom'}  )
  = &   S^{*}_{{\sN}}(\La_{\sN}  - \Om_{{\sN}+1},  \Phi_{\sN, \de \Om_{\sN}},   \phi_{\sN,   \bom''})
   \\
&+ \sum_{Y \cap  ( \La_{\sN} - \Om_{{\sN}+1} )  \neq  \emptyset }  R^{*, (4)}_{{\sN}, \bpi^+}(Y,  \Phi_{{\sN},  \de \Om_{{\sN}+1}} )  
  \\
 \tilde  S_{{\sN}}(\Om_{{\sN}+1}  - \Om^{\nat}_{{\sN}+1},      \phi_{{\sN},  \de  \bom'}  )  
 = & \tilde  S_{{\sN}}(\Om_{{\sN}+1}  - \Om^{\nat}_{\sN+1}, \phi_{\sN,   \bom''}    )
  \\
&+ \sum_{Y \cap (\Om_{{\sN}+1}  - \Om^{\nat}_{{\sN}+1} )  \neq  \emptyset } 
 R^{*, (5)}_{{\sN}, \bpi^+}(Y,  \Phi_{{\sN},  \de \Om_{{\sN}+1}} )    \\
\end{split}
\ee
where the terms on the right are strictly localized     and  satisfy 
\be   \label{wonder}
 | R^{*, (i)}_{{\sN},  \bpi^+}(Y)  |
\leq \one   \la^{n_0} e^{- (\ka- \ka_0 -2)  d_M(Y)}     \hs   i=3,4,5
\ee
\end{lem}
\bigskip

\re   The leading terms on the right  are  bounded but   not   small.    
We  do not need them for large field bounds,   but have made a point to localize them in a way that preserves positivity so they can be estimated
in  the exponential.
 \bigskip

\pr      Consider  $ E_{\sN}(\La_{\sN}  - \Om_{\sN+1}^{\nat}  )$.     First  note that  on     $\La_{\sN}  - \Om_{\sN+1}^{\nat}$
we  have   $|\phi_{\sN, \de \bom'}|,   |\phi_{\sN,  \bom''}|  \leq  C \la_{\sN}^{-\frac14 - \de}$.  
 Therefore with either field  $|E_{\sN}(X) | \leq    \la^{\beta}e^{ - \ka d_M(X)}  $   and  since  $V_{\sN}(\square, \phi)  =   \la_{\sN} \int \phi^4$
 we have     $|V_{\sN}(\square)| \leq   C M^3\la_{\sN}^{ - 4 \de}$.
 Therefore      $E^+_{\sN}(X) =  E_{\sN}(X) - V_{\sN}(X)$  satisfies   for  $X \subset   \La_{\sN}  - \Om_{\sN+1}^{\nat}$
 \be   
 |E^+_{\sN}(X)  | \leq     C M^3\la_{\sN}^{ - 4 \de}e^{ - \ka d_M(X)}
 \ee 
But  also we  have  on    $ \La_{\sN}  - \Om_{\sN+1}^{\nat}$
\be
  |  \phi_{{\sN}, \de  \bom'}    -   \phi_{{\sN},   \bom''}|  \leq     e^{-r_{\sN}/2}  
\ee
 This holds  since  $G_{{\sN}, \de \bom'}$  and    $G_{{\sN} , \bom''} $
have  random walk expansions differing only in  $ \Om_{\sN +1}^{2 \nat}$  which is at least  $[r_{\sN}]$ layers of   $M$-cubes  away  from   $\La_{\sN}  - \Om_{\sN+1}^{\nat}$.    Hence we can  write    
\be
  E^+_{\sN}( X,      \phi_{{\sN},  \de  \bom'}  )  - E^+_{\sN}(X,      \phi_{{\sN},  \de  \bom''}  )
 =    \frac {1}{2\pi i}  \int_{|t|  =  e^{r_{\sN}}/2}    \frac{dt}{t(t-1)}  
  E^+_{\sN}\Big(X,    \phi_{{\sN}, \de  \bom''}+ t ( \phi_{{\sN}, \de  \bom'}    -   \phi_{{\sN},   \bom''} )\Big)  
\ee
and have the estimate
\be | E^+_{\sN}( X,      \phi_{{\sN},  \de  \bom'}  )  - E^+_{\sN}(X,      \phi_{{\sN},    \bom''}  )|
\leq  e^{-r_{\sN}/2}    C M^3\la_{\sN}^{- 4 \de}e^{ - \ka d_M(X)}\
\leq  \la_{\sN}^{n_0+1}  e^{- \ka  d_M(X)}   
\ee
The  localization  now proceeds  more or less   as in lemma  3.22  in  part II  and gives the representation   (\ref{sweet}) with the bound  (\ref{wonder}).   The  terms $ S^{*}_{{\sN}}, \tilde  S_{{\sN}}$ are
treated  similarly.

\subsection{summary}

We  rearrange all these terms  and insert them into  (\ref{representation6}).  
First we  write    
\be 
  E^+_{\sN} (\Om^{\nat}_{{\sN}+1}, 0)       =    E^+_{\sN} (\Om^{\nat}_{{\sN}+1}- \La_{\sN+1}, 0)     +   E^+_{\sN} (\La_{{\sN}+1}, 0)    +
\sum_{X \subset  \Om^{\nat}_{{\sN}+1},  X  \#  \La_{\sN+1} }     E^+_{\sN} (X, 0)  
\ee
The  first  term  is absorbed into  $ \tilde B_{{\sN}+1, \bpi^+} (\La_{\sN}, \La_{{\sN}+1})$.   We  also write
 \be
     E^+_{\sN}(\La_{\sN}    - \Om_{\sN+1}^{\nat},     \phi_{\sN,   \bom''})
 =       E_{\sN}(\La_{\sN}    - \Om_{\sN+1}^{\nat},     \phi_{\sN,   \bom''})  -   V_{\sN}(\La_{\sN}    - \Om_{\sN+1}^{\nat},     \phi_{\sN,   \bom''})
\ee
and the first term  here is absorbed into    $ \tilde B_{{\sN}+1, \bpi^+} (\La_{\sN}, \La_{{\sN}+1})$.

We  collect  field independent terms in  $\La_{\sN +1}$  by defining
\be   
    E^*_{{\sN}}(\La_{{\sN}+1}) = -  \vep_{\sN}^0 \Vol(  \La_{{\sN}+1} )   +  E^+_{\sN} (\La_{{\sN}+1}, 0) 
       + E^\#_{{\sN}}(\La_{{\sN}+1})   \ee
Each of these is expressed as a sum over polymers  $X  \subset  \La_{\sN +1}$  and we have
\be     \label{lumbar2}
| E^*_{{\sN}}(X)  |  \leq    \one \la^{\beta} e^{-(\ka - 6 \ka_0 - 6)d_M(X)}
\ee
We  collect  boundary terms  by defining   
\be 
  B^*_{{\sN}, \bpi^+}(\La_{{\sN}+1})  =       B^\#_{{\sN},  \bpi^+}(  \La_{{\sN}+1}  )  
  +   \sum_{i=1}^5  \sum_{X \#  \La_{\sN+1}} R^{*, (i)}_{{\sN}, \bpi^+}(X)
+ \sum_{X \subset  \Om^{\nat}_{{\sN}+1},  X  \#  \La_{\sN+1} }     E^+_{\sN} (X, 0)  
\ee
Each of these can be expressed as    sum over polymers  $X \in   \cD_{\sN}( \bmod \ \Om^c_{\sN+1} )  $.  In the case of   $ R^{*, (i)}_{{\sN}, \bpi^+}$
this involves adjoining connected components of  $\Om^c_{\sN +1}$  as in part  H of lemma 3.15 in part II.   Hence   $ B^*_{{\sN}, \bpi^+}(\La_{{\sN}+1})$  is a sum of such $X$,  and on the domain  of   $  B^\#_{{\sN},  \bpi^+}$   
and satisfies   
\be 
 | B^*_{{\sN}, \bpi^+} (X ) |       \leq   \one  B_0 \la^{\beta} e^{- (\ka- 6\ka_0-6) d_{M}(X,  \bmod  \Om^c_{{\sN}+1})  } 
\ee

It is convenient to make a further adjustment here.
Each    $X \in  \cD_{\sN}  ( \bmod  \  \Om^c_{{\sN}+1})$    determines a  $Y \in  \cD_{\sN}  ( \bmod  \  \La^c_{{\sN}+1})$
by adjoining connected components of  $\La^c_{{\sN}+1}$.   Define a new  $ B^*_{{\sN}, \bpi^+} (Y )$  by  summing over all terms  $X$
yielding the same  $Y$.   Then we  have    
\be
  B^*_{{\sN}, \bpi^+}(\La_{{\sN}+1})=  \sum_{Y \in  \cD_{\sN}(\bmod    \La^c_{\sN+1}),  Y \#  \La_{\sN+1}}  B^*_{\sN, \bpi^+}(Y) 
\ee 
where now 
\be  \label{slinky}
 |B^*_{{\sN}, \bpi^+}(Y) |   \leq
  \one  B_0 \la^{\beta} e^{- (\ka- 7\ka_0-7) d_{M}(Y,  \bmod  \La^c_{\sN+1})  } 
 \ee
See appendix  \ref{B} for details on   this  step.

The  remaining  terms  coming from   $ R^{*, (i)}_{\sN+1, \bpi^+}(X)$  have    $X \subset  \La_{\sN+1}^c $    and  $X \cap  \La_{\sN}  \neq  \emptyset$
and can be    absorbed into   $\tilde B_{\sN+1, \bpi^+} (\La_{\sN}, \La_{\sN+1})$.  With  all these additions it still satisfies the bound   
 \be
 | \tilde B_{{\sN}+1, \bpi^+} (\La_{\sN}, \La_{{\sN}+1}) |  \leq   C   |   \La^{(\sN)}_{\sN} -   \La^{(\sN +1)}_{\sN}| 
\ee
Finally     we  define  (with   $S^*_{{\sN}},  \tilde  S_{\sN},  V_{\sN}$   are evaluated at  $\phi_{\sN,  \bom''}$)
\be   
\begin{split}
\label{springtime}
  \tilde  K_{{\sN}+1, \bpi^+} = K_{{\sN}+1, \bpi}  
  \exp  \Big(   &     \tilde c_{\sN}  |\Om^{c,({\sN})}_{{\sN}+1}|         
 -  S^*_{{\sN}}(\La_{\sN}-  \Om_{{\sN}+1})  - \tilde  S_{\sN}( \Om_{{\sN}+1}-  \Om^{\nat}_{{\sN}+1}) \\   &
  - V_{\sN}(\La_{\sN}- \Om^{\nat}_{\sN +1})    +  \tilde B_{{\sN}+1, \bpi^+} (\La_{\sN}, \La_{{\sN}+1})    \Big) \\
 \end{split}
 \ee
Altogether then   (\ref{representation6}) has become
 \begin{equation}    \label{representation7}
\begin{split}
 \sZ_{\sM,\sN} = &\    \sZ_{\sM,\sN}(0)     \sum_{\bpi^+}   \int     d  \tilde    \Phi_{{\sN}, \bom^c}\   d W_{{\sN}+1,  \bpi^+} \ 
  \tilde   K_{{\sN}+1, \bpi^+} \    \tilde \cC_{{\sN}+1, \bpi^+} \       \\
&   \exp  \Big(     E^*_{{\sN}}(\La_{{\sN}+1}) +      R^\#_{{\sN}, \bpi^+}(\La_{{\sN}+1} ) +   B^*_{{\sN},  \bpi^+}(  \La_{{\sN}+1}  ) 
 \Big)\\
\end{split}
\end{equation}

\section{large field   bounds}

\subsection{rearrangement}

There are some further rearrangements  before commencing with   the final bounds.
We   start by making   some  Mayer expansions in (\ref{representation7}).
Since    $ E^*_{{\sN}+1}(\La_{{\sN}+1}) =  \sum_{X   \subset  \La_{{\sN}+1}}    E^*_{{\sN}+1}(X)$   and  the same for the
tiny terms  we have  
\begin{equation}
\begin{split}
 \exp \Big (      E^*_{{\sN}+1}(\La_{{\sN}+1})  +  R^\#_{\sN, \bpi^+}(\La_{{\sN}+1})  \Big)
 =  & \prod_{X \subset   \La_{{\sN}+1}    }  
\Big(\cI_{\bpi^+}(X)+1   \Big) 
 =     \sum_{  \{X_{\al}  \}  }   \prod_{\al}   \cI _{\bpi^+} (X_{ \al} )\\
 \end{split}
\end{equation}
where for  $X   \in \cD_{\sN}$
\begin{equation}
\cI_{\bpi^+}(X)   =   \exp \Big  (    E^*_{{\sN}}(X)     +  R^\#_{\sN, \bpi^+}(X)   \Big)   -1    
\end{equation}
and the   sum  is  now  over  collections of distinct       $\{ X_{\alpha}  \}$    in    $\La_{{\sN}+1}$.
We  also have that    $ B_{{\sN}+1,  \bpi^+}(  \La_{{\sN}+1})  =   \sum_{Y  \# \La_{{\sN}+1}}  B_{{\sN}+1,  \bpi^+}(Y)$  and then
 \begin{equation}
\begin{split}
 \exp \Big(   B_{{\sN}+1,  \bpi^+}(  \La_{{\sN}+1})    \Big)
 =  & \prod_{Y \#   \La_{{\sN}+1}    }  \Big(\cJ_{\bpi^+}(Y)+1   \Big) 
 =     \sum_{  \{Y_{\si}  \}  }   \prod_{\si}   \cJ_{\bpi^+} (Y_{ \si} )\\
 \end{split}
\end{equation}
where for   $Y  \in  \cD_{\sN}( \bmod \  \La^c_{\sN+1}) $
\begin{equation}
\cJ_{\bpi^+}(Y)  =  \exp   \Big(   B_{{\sN}+1,  \bpi^+}( Y)  \Big) -1
\end{equation}   
and the   sum  is  now  over  collections of distinct       $\{ Y_{\si}  \}$   which cross      $\La_{{\sN}+1}$.

Classify the terms in the sum over  $\Pi^+=(\La_0,  \Om_1,  \La_1, \dots,   \Om_{{\sN}+1},  \La_{{\sN}+1}) $   by  
$\Theta  \equiv   \La^c_{\sN+1}$,  a union of $M$ cubes.   We  write   $\sum_{\bpi^+} ( \cdots)   = \sum_{\Theta} \sum_{\bpi^+:  \La^c_{\sN+1}= \Theta}$.   The sum  over $\Theta $  is written as a sum over its connected components  $\{  \Theta_{\ga}  \}$.
The  sums  over   $ \{X_{\al}  \}  , \{Y_{\si}  \}$  only depend on   $\Theta  =   \cup_{\ga}  \Theta_{\ga}$   and so  can come  outside the sum over
$\bpi^+$.
Then we  have   
 \be   \label{sum}
  \sZ_{\sM,\sN} =      \sZ_{\sM,\sN}(0)     \sum_{\{ \Theta_{\ga}  \}}   \sum_{  \{Y_{\si}  \}  }          \sum_{  \{X_{\al}  \}  }  
    \sum_{\bpi^+:  \La^c_{{\sN}+1} =\Theta}      \cL_{\bpi^+}\Big ( \{ X_{ \al} \} ,  \{ Y_{\si} \} \Big )  
    \ee
    where
\be
   \cL_{\bpi^+}\Big (\{ X_{ \al} \} ,  \{ Y_{\si} \} \Big )    \equiv      \int      d  \tilde \Phi_{{\sN}, \bom^c}\   d W_{{\sN}+1,  \bpi^+} \ 
\tilde   K_{{\sN}+1, \bpi^+}  \  \tilde      \cC_{{\sN}+1, \bpi^+} 
       \prod_{\si}   \cJ_{\bpi^+} (Y_{ \si} ) \prod_{\al}   \cI _{\bpi^+} (X_{ \al} ) 
\ee
The sum in  (\ref{sum})  is over  disjoint   $ \Theta_{\ga} $,     over distinct    $\{X_{\al}  \}$  satisfying    $X_{\al}  \subset  \Theta^c$
  and   over   distinct   $ \{Y_{\si}\} $  
satisfying      $Y_{\si}  \#   \Theta$  and  $Y_{\si}  \in  \cD_{\sN}( \bmod \  \Th) $.

\begin{figure}[t] 
\begin{picture}(250,250)(-120,-20)
\linethickness{2pt}
\put(-20,20){\framebox(100,60)}
\put(40,40){\framebox(20,20)}
\put(40,140){\framebox(20,20)}
\put(110,140){\framebox(80,20)}
\put(20,100){\framebox(180,80)}
\put(70,70){\framebox(100,50)}
\put(188, 105){\text{$Y_1$} }
\put(157, 75){\text{$X_1$} }
\put(68,25){\text{$Y_2$} }
\put(48,43){\text{$\Theta_3$} }
\put(48,143){\text{$\Theta_2$} }
\put(178,143){\text{$\Theta_1$} }
\end{picture}
\caption{A possible connected component of $U$  \label{stinger} }
\end{figure}
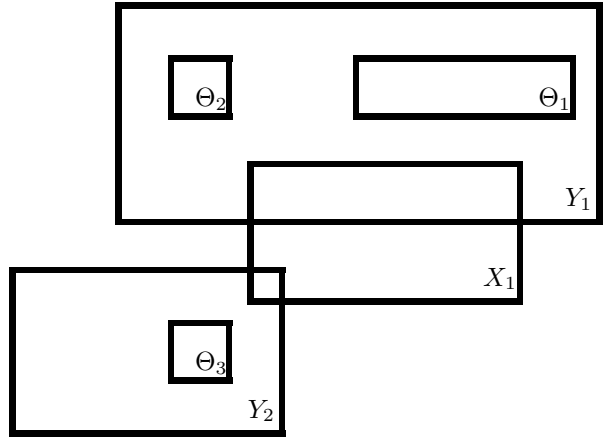

Let   $U$  be the union of       $\{ \Theta_{\ga}  \},  \{Y_{\si}  \},     \{X_{\al}  \}$  and let  $\{  U _{\ell} \}$  be the connected components of  $U$,
  where now   we say  $X,Y$  are connected
if they have  a cube $\square$ in common. See figure  \ref{stinger}.   We  write the sum as  
\be  \label{s1}
   \sum_{\{ \Theta_{\ga}  \},  \{Y_{\si}  \},     \{X_{\al}  \}   }   =  \sum_{U}    \sum_{\{ \Theta_{\ga}  \},  \{Y_{\si}  \},     \{X_{\al}  \} \to  U}
   =  \sum_{\{U_{\ell}\} }   \prod_{\ell}      \sum_{\{ \Theta_{\ga}  \},  \{Y_{\si}  \},     \{X_{\al}  \} \to  U_{\ell}}\\
\ee
   The statement  that   $\tilde B_{k, \bpi}(\La_{k-1},  \La_k^c)   $   
is  additive in the connected  components of  $\La_k^c$   means   that   
\be 
  \tilde B_{k, \bpi}(\La_{k-1},  \La_k^c)  = \sum_{\ga}  \tilde B_{k, \bpi  \cap  \Theta_{\ga}}(\La_{k-1} \cap  \Theta_{\ga},  \La_k^c  \cap  \Theta_{\ga}) 
\ee
Thanks to the change of fields  in  lemma \ref{sour}  a similar decomposition  holds for   the other  contributions to  $  \tilde   K_{{\sN}+1, \bpi^+}$.
Thus   
 $ \tilde   K_{{\sN}+1, \bpi^+}  =  \prod_{\ga}  \tilde   K_{{\sN}+1, \bpi^+ \cap  \Theta_{\ga}  }$  and hence  
\be
  \tilde   K_{{\sN}+1, \bpi^+}  =  \prod_{\ell}  \tilde   K_{{\sN}+1, \bpi^+ \cap U_{\ell} }
 \ee
 The same holds for  $\tilde      \cC_{{\sN}+1, \bpi^+} $   and thus
 \be  \label{s2}
 \begin{split}  
  \sum_{\bpi^+:  \La^c_{{\sN}+1} =\Theta}   \cL_{\bpi^+} \Big (\{ X_{ \al} \} ,  \{ Y_{\si} \} \Big )    
   =  & \sum_{\bpi^+:  \La^c_{{\sN}+1} =\Theta}   \prod_{\ell}   \cL_{\bpi^+\cap  U_{\ell}}\Big (   \{X_{\al}  \}  \cap  U_{\ell} ,   \{ Y_{\si} \} \cap  U_{\ell}\Big)   \\
 =   &  \prod_{\ell}\ \  \sum_{\bpi^+:  \La^c_{{\sN}+1} =\Theta \cap  U_{\ell}}    \cL_{\bpi^+\cap  U_{\ell}}\Big (   \{X_{\al}  \}  \cap  U_{\ell} ,   \{ Y_{\si} \} \cap  U_{\ell}\Big)   \\
  \end{split} 
 \ee
  Combining      (\ref{s1})  and (\ref{s2})   yields  
\begin{equation}  \label{stingray}
  \sZ_{\sM,\sN} =      \sZ_{\sM,\sN} (0)   \sum_{\{U_{\ell}  \}}     \prod_{\ell}  \cK(   U_{\ell}  )
 \end{equation}
 where the sum is over  disjoint connected $\{U_{\ell}  \}$ and
 where  for connected $U$
 \begin{equation}
 \cK(U) =      \sum_{\{ \Theta_{\ga}  \}, \{Y_{\si}  \}, \{X_{\al}  \}  \to  U  }  \ \ 
    \sum_{\bpi^+:   \La^c_{{\sN}+1} =\Theta}       \cL_{\bpi^+} \Big( \{ X_{ \al} \} ,  \{ Y_{\si} \}  \Big)  
\end{equation}
$\cK(U)$    is  invariant  under  $M$-lattice symmetries.
Our  goal  is to get a good  bound on $\cK(U)$    so we  can exponentiate the expansion (\ref{stingray}).

\subsection{first   bounds}

The characteristic functions put us in the analyticity domains for the various functions.   
Thus we can use the bound       (\ref{lumbar2})  on    $  E^*_{{\sN}+1}(X) $  and   the stronger  bound    (\ref{spirit})  on  $R^\#_{\sN,\bpi^+}(X)$
  and conclude that  
    \begin{equation} 
|\cI_{\bpi^+}(X)| \leq   \cO(1)   \la^{\beta}   e^{- (\ka- 6 \ka_0-6) d_M  (X)  }
\end{equation}
The  estimate   (\ref{slinky})    on   $ B_{{\sN}+1, \bpi^+}(Y)$   gives         
\begin{equation} 
|\cJ_{\bpi^+}(Y)| \leq    \one  B_0 \la^{\beta} e^{- (\ka- 7\ka_0-7) d_{M}(Y,  \bmod  \La^c_{\sN+1})  } 
\end{equation}
Also  recall that  
 \begin{equation}
\begin{split}
 K_{{\sN},\bpi}  =  &
  \prod_{j=0}^{{\sN}}  
      \exp\left( c_j|(\Om^c_j)^{(j-1)}|-S^{+,u}_{j,L^{-({\sN}-j)}}(\La_{j-1}  -  \La_{j}     )+ ( \tilde   B_{j,L^{-({\sN}-j)}}) _{\bpi_j}(\La_{j-1}, \La_{j}) \right)     \\
\end{split}
\end{equation}
By  (\ref{someday}) we have  $| \tilde   B_{\sN,  \bpi}(\La_{\sN-1}, \La_{\sN}) |  \leq     B_0 |    \La^{(\sN)}_{\sN-1} -    \La^{(\sN)}_{\sN}| $.    
The right side    is scale invariant,   so the scaled characteristic functions   imply  that         
\be
| (\tilde   B_{j,L^{-({\sN}-j)}}) _{\bpi_j}(\La_{j-1}, \La_{j})|      \leq  B_0 |     \La^{(j)}_{j-1} -    \La^{(j)}_{j}|
\ee
The function   $ \tilde  K_{\sN+1,  \bpi^+}$  has   $ K_{{\sN},\bpi} $  and a similar  factor for the last step,   see  (\ref{springtime}).
Here  we  use     $| \tilde B_{{\sN}+1, \bpi^+} (\La_{\sN}, \La_{{\sN}+1})|  \leq    B_0   |  \La^{({\sN})}_{{\sN}}-    \La^{({\sN})}_{{\sN}+1}   | $.
We   also use the positivity  (  $V_{\sN}(\square,  \phi) =  \la  \int_{\square} \phi^4$)  to estimate  
\be
\Big| \exp\Big (  
 -  S^*_{{\sN}}(\La_{\sN}-  \Om_{{\sN}+1})  - \tilde  S_{\sN}( \Om_{{\sN}+1}-  \Om^{\nat}_{{\sN}+1})   - V_{\sN}(\La_{\sN}- \Om^{\nat}_{\sN +1})   \Big) \Big|  \leq   1
\ee
 Now    with  $\ka'  =  \ka - 7 \ka_0 -7$  
\begin{equation}  \label{under}
\cK(  U)   \leq        \sum_{   \{ \Theta_{\ga} \},  \{Y_{\si} \} ,  \{X_{\al}  \} \to  U}  \  \cK'(\Theta)  \ \   
 \prod_{\al} \one   \la^{\beta}   e^{- \ka' d_M  (X_{\al})  }   \prod_{\si}  \la^{\beta}  \one B_0  e^{- \ka' d_M  (Y_{\si},    \bmod   \Theta)  } \\
\end{equation}
where     
\begin{equation}
\begin{split}
\cK'(\Theta)  
 =  &  \sum_{\bpi^+: \La^c_{{\sN}+1}= \Theta}  
  \int    d \tilde  \Phi_{{\sN}, \bom^c}\  d W_{\sN+1,\bpi^+}\  \tilde    \cC_{{\sN}+1, \bpi^+}\\
  &    
       \prod_{j=0}^{{\sN}}        \exp\left(  c_j|(\Om^c_j)^{(j-1)}|-S^{+,u}_{j,L^{-({\sN}-j)}}(\La_{j-1} -  \La_{j}     )
       +  B_0  |    \La^{(j)}_{j-1} -   \La^{(j)}_{j}|    \right)  \\
 &
      \exp\left(      \tilde c_{{\sN}+1}  |(\Om^c_{{\sN}+1})^{({\sN})}|         
+  B_0   |  \La^{({\sN})}_{{\sN}}-    \La^{({\sN})}_{{\sN}+1}  |    \right) \\
 \end{split}
\end{equation}

Next we take a closer look  at    the characteristic functions  which we  can write as
\be   \label{soup}
 \tilde      \cC_{{\sN+1}, \bpi^+}  =       \prod_{j=-1}^{\sN-1}  \Big(  \cC_{j+1, L^{-({\sN}-j-1)}}\Big)_{ \La_{j}, \Om_{j+1}, \La_{j+1} } \ \ 
\  \chi_{{\sN}}(  \La_{\sN}  - \La_{{\sN}+1}^{**} ) 
   \tilde   \cC_{{\sN}+1} ( \Om_{{\sN}+1},  \La_{{\sN}+1})
   \ee
Recall  that   $ \cC_{j+1, \La_{j}, \Om_{j+1}, \La_{j+1}} =  ( \cC^0_{j+1,L^{-1}})_{ \La_{j}, \Om_{j+1}, \La_{j+1}}$   where 
if  $\La_{j}$ is a union of  $M$ cubes and  $\Om_{j+1},\La_{j+1}$ are  unions  of  $LM$ cubes 
\begin{equation}
 \cC^0_{j+1, \La_{j}, \Om_{j+1}, \La_{j+1}} =   \cC_{j}^q(\La_{j}, \Om_{j+1})  \chi_{j} ( \La_{j}-  \La^{**}_{j+1})  \chi^q_{j}(   \Om_{j+1}- \La^{**}_{j+1})  
   \cC'_{j+1}(\Om_{j+1},\La_{j+1})    
\end{equation}
and 
\begin{equation}   \label{lonely}
\begin{split}
&\cC^q_{j}(\La_{j},   \Om_{j+1} )  =   \sum_{P_{j+1}  \subset  \bar \La_{j}:  \Om_{j+1}  =   (\bar  \La_{j})^{5 \nat  } -    P^{ 5*}_{j+1} }  
 \zeta^q_{j}( P_{j+1} )  \chi^q_{j}((\bar  \La_{j}-P_{j+1})- \Om_{j+1} )\\
&  \cC'_{j+1}(\Om_{j+1},\La_{j+1})    
=     \sum_{     Q_{j+1},R_{j+1} \to \La_{j+1}    }
 \zeta^0_{j+1}( Q_{j+1} ) 
 \zeta^w_{j}(  R_{j+1}  )  \\
 &  \hspace{1in}  \chi^0_{j+1}(   \Om^{\nat}_{j+1} - (Q_{j+1} \cup \La_{j+1}) )
    \chi^w_{j}(  \Om_{j+1} - (R_{j+1} \cup  \La_{j+1})  ) \\
    \end{split}
 \end{equation}
where the sum is over unions  of     $LM$ cubes  $P_{j+1},Q_{j+1}, R_{j+1}$.
Hence  if   $\Om_{j+1}, \La_{j+1}$ are unions of  $L^{-({\sN}-j-1)}M$ cubes   as in  (\ref{soup}),    then  
   $ ( \cC_{j+1,L^{-({\sN}-j-1)}})_{ \La_{j}, \Om_{j+1}, \La_{j+1}} =  ( \cC^0_{j+1,L^{-({\sN}-j)}})_{ \La_{j}, \Om_{j+1}, \La_{j+1}}$
depends on   
\be
\begin{split}
&(\cC^q_{j, L^{-({\sN}-j)  }})(\La_{j},   \Om_{j+1} )
 =  
 \sum_{P_{j+1}  \subset \bar \La_{j}:  \Om_{j+1}  =   (\bar   \La_{j})^{5 \nat  } -    P^{ 5*}_{j+1} }  
 \zeta^q_{j, L^{-({\sN}-j)}   }( P_{j+1} )  \chi^q_{j, L^{-({\sN}-j)  }}((\bar  \La_{j}-P_{j+1})- \Om_{j+1} ) \\
&(\cC'_{j+1, L^{-({\sN}-j)} })(\Om_{j+1},\La_{j+1})    
=     \sum_{     Q_{j+1},R_{j+1} \to \La_{j+1}    }
 \zeta^0_{j+1, L^{-({\sN}-j)} }( Q_{j+1} ) 
 \zeta^w_{j, L^{-({\sN}-j)} }(  R_{j+1}  )\\
 &   \hspace{1in} \chi^0_{j+1,L^{-({\sN}-j)}}(   \Om^{\nat}_{j+1} - (Q_{j+1} \cup \La_{j+1}) )
    \chi^w_{j, L^{-({\sN}-j)} }(  \Om_{j+1} - (R_{j+1} \cup  \La_{j+1})  ) \\
    \end{split}
 \end{equation}
where now the sum is over  unions of $L^{-({\sN}-j-1)}M$ cubes     $P_{j+1},Q_{j+1}, R_{j+1}$.  The last  step is treated similarly.

The     sum  over  $\bpi^+ = (\La_0, \Om_1, \La_1, \dots  \Om_{{\sN}+1}, \La_{{\sN}+1})$  can now     can be written as a sum over  regions 
  $\{ P_j,Q_j, R_j\}_{j=0}^{{\sN}+1}$ with the convention   that      $P_0,R_0,Q_{{\sN}+1},  P_{{\sN}+1} = \emptyset $.       Each of   these is  a union of  $L^{-({\sN}-j)}M$   cubes  in  $\bbT^{-{\sN}}_{\sM}$  (except   $R_{{\sN}+1}$ is still $M$ cubes)   .
  They    determine  
$\Om_j,  \La_j$ recursively   by the following rules
\begin{equation}     \label{understood}
\begin{split}
 \Om^c_{j+1}  = &  (\bar \La_j)^{c,5*} \cup  P^{5*}_{j+1}   \hspace{1.3in}   P_{j+1}  \subset  \bar  \La_{j}  \\
\La^c_{j+1}  =&  (\Om^c_{j+1})^{5*}  \cup  Q^{5*}_{j+1}  \cup  R^{5*}_{j+1}   
  \hspace{.8in}    Q_{j+1}  \subset    \Om^{\nat}_{j+1}, \  R_{j+1}  \subset    \Om_{j+1}  
\\ 
\end{split}
\end{equation}

Now   we  drop  the   characteristic functions  (i.e. estimate them  by  one)  with the following exceptions.    We   retain the large field 
functions  $ \zeta^q_{j  }( P_{j+1} )\  \zeta^0_{j+1 }( Q_{j+1} ) \ \zeta^w_{j }(  R_{j+1}  )$.   We  also  keep  some small field characteristic  
functions in a modified form.     
We  define  for  $\Phi_k$  on      $\bbT^0_{\sN + \sM -k}$   
\be 
  \tilde  \chi_k(X,  \Phi_k)  =  \prod_{x \in X \cap   \bbT^0_{\sN + \sM -k}}  \chi\Big(  |\Phi_k(x)|  \leq   \la_k^{-\frac14- \de}\Big ) 
    \ee
Then   $ \tilde  \chi_{j,  L^{-(\sN-j)}}(\de \Om_j,  \Phi_j)  =   \tilde  \chi_j( L^{\sN-j}\de \Om_j, \Phi_{j,L^{\sN-j}}) $  enforces  $|\Phi_j|  \leq   \la_j^{-\frac14 - \de} L^{\frac12(\sN-j)}$
on  $ \de \Om_j^{(j)}$.    We  know that this is implied by the other characteristic functions,  see  (222)  in part II.  Similarly  we  can  introduce
$  \tilde  \chi_{0,  L^{-k}}(\La_0- \Om_1,  \Phi_0) $.  Everything else is dropped.

   Then    we  have 
\begin{equation}  \label{sally3}
\begin{split}
\cK'(\Theta)  
 \leq &    \sum_{ \{ P_j,Q_j, R_j\}:  \La^c_{{\sN}+1} =  \Theta }     
  \int    d \tilde  \Phi_{{\sN}, \bom^c}\ d W_{{\sN+1},\bpi^+}     \tilde  \chi_{0,  L^{-\sN}}(\La_0- \Om_1)   \prod_{j=1}^{{\sN}}   \tilde  \chi_{j,  L^{-(\sN-j)}}(\de \Om_j) \\ 
   & \prod_{j=-1}^{{\sN}}   \zeta^q_{j, L^{-({\sN}-j)}   }( P_{j+1} )\  \zeta^0_{j+1, L^{-({\sN}-j)} }( Q_{j+1} ) \ \zeta^w_{j, L^{-({\sN}-j)} }(  R_{j+1}  )  \\  
  &    
       \prod_{j=0}^{{\sN}}        \exp\left(  c_j|(\Om^c_j)^{(j-1)}|-S^{+,u}_{j,L^{-({\sN}-j)}}(\La_{j-1}  -  \La_{j}     )
       +  B_0  |  \La^{(j)}_{j-1} -   \La^{(j)}_{j}|    \right)  \\
 &
      \exp\left(      \tilde c_{{\sN}+1}  |(\Om^c_{{\sN}+1})^{({\sN})}|         
  +  B_0   |  \La^{({\sN})}_{{\sN}}-    \La^{({\sN})}_{{\sN}+1}  |    \right) \\
 \end{split}
\end{equation}

\subsection{small factors}

We  continue the estimate on  $\cK'(\Theta)$.
Recall that 
\be
\begin{split}
d \tilde  \Phi_{{\sN}, \bom^c}  = & d \Phi_{{\sN}, \Om^c_{\sN+1}}\  \prod_{j=0}^{{\sN}-1}  
\exp \left(  - \frac{1}{2 } aL^{-({\sN}-j-1)}|\Phi_{j+1}- Q \Phi_{j}|^2_{ \Om^c_{j+1}}  \right)
 d\Phi^{({\sN}-j)}_{j, \Om^c_{j+1}}  \\
\end{split}
\ee
We   split each    exponential  into  two   factors     $\exp \left(  - \frac{1}{4 } aL^{-({\sN}-j-1)}|\Phi_{j+1}- Q \Phi_{j}|^2_{ \Om^c_{j+1}}  \right)$.   The  first  is estimated to give small factors and the second is  integrated over.    Since  $P_{j+1} \subset  \Om^c_{j+1}$ the first is  smaller than  
  $\exp \left(  - \frac{1}{4 } aL^{-({\sN}-j-1)}|\Phi_{j+1}- Q \Phi_{j}|^2_{P_{j+1}}  \right)$.

 \begin{lem}    For    $\Phi_j:  \bbT^{-({\sN}-j)}_{\sM} \to \bbR$   and  $\Phi_{j+1}:  \bbT^{-({\sN}-j-1)}_{\sM} \to \bbR$
 and      $P_{j+1}$  a union of  $L^{-({\sN}-j-1)} M$   cubes in   $ \bbT^{-{\sN}}_{\sM}$:
 \begin{equation}
\exp \left(  - \frac14 aL^{-({\sN}-j-1)}|\Phi_{j+1}- Q \Phi_{j}|^2_{P_{j+1}}  \right)  \zeta^q_{j, L^{-({\sN}-j)}   }( P_{j+1} )   
 \leq   \exp \Big(  - \frac14  aL p^2_{j}M^{-3}| P^{(j+1)}_{j+1}|   \Big)
\end{equation}
\end{lem} 
 \bigskip

 \pr    It   suffices to prove this back on the lattice  where the term was born.   We  scale up  by   $ L^{{\sN}-j} $ 
 and claim that    for   $\Phi_{j}:  \bbT^{0}_{\sM+\sN -j} \to \bbR$
 and  $\Phi_{j+1}:  \bbT^{1}_{\sM+\sN -j} \to \bbR$  and   $P_{j+1}$  a union of  $LM$-cubes in   $ \bbT^{-j}_{\sM+\sN -j}$:
 \begin{equation}  \label{scaled1}
 \exp \left(- \frac14  aL    |\Phi_{j+1}- Q \Phi_{j}|^2_{P_{j+1}}  \right) \zeta^q_{j}( P_{j+1}, \Phi_{j}, \Phi_{j+1} )  
   \leq   \exp \Big(  - \frac14 aLp^2_jM^{-3}| P^{(j+1)}_{j+1}| \Big)
\end{equation}
 Keep  in mind that    $| P^{(j+1)}_{j+1}| $ is invariant under scaling.
The left side of  (\ref{scaled1}) can be written:     
 \begin{equation}
     \prod_{\square  \subset   P_{j+1}}  \exp \left(-  \frac14   aL   |\Phi_{j+1}- Q \Phi_{j}|^2_{\square}  \right)
       \zeta^q_{j}( \square ) 
 \end{equation}   
 where  the product is over  the  $LM$  cubes.    
 The characteristic function 
 $ \zeta^q_{j+1}(\square )$  enforces that there is at least on   point   in  $\square $  such that
 $|\Phi_{j+1}  - Q \Phi_{j}  |   \geq   p_{j}$.   
  Therefore  
 \begin{equation}
 \exp \left(- \frac14   aL   |\Phi_{j+1}- Q \Phi_{j}|^2_{\square}  \right)  \zeta^q_{j}(\square )
 \leq      \exp \Big(     -  \frac14  aL p_{j}^2     \Big)
 \end{equation}
 The result now follows since  the number  of  $L$-cubes  in  $P_{j+1}$   is   $| P^{(j+1)}_{j+1}|$
 so the number of  $LM$ cubes in   $P_{j+1}$ is    $M^{-3}| P^{(j+1)}_{j+1}|$.
 This completes the proof.
\bigskip

Next consider
\be   d W_{{\sN},\bpi^+}   =    \prod_{j=0}^{{\sN}}      (2 \pi)^{-|  [ \Om_{j+1}  - \La_{j+1}]^{(j)}|/2}
  \exp  \Big(   - \frac{1}{2 } L^{-({\sN}-j)} | W_j |^2_{\Om_{j+1}- \La_{j+1}}  \Big) 
    dW^{({\sN}-j)}_{j,\Om_{j+1}- \La_{j+1}}
\ee
We  break the exponent into two pieces     $\exp(   - \frac14 L^{-(\sN-j)} | W_j |^2_{\Om_{j+1}- \La_{j+1}} )$.
 The  first gives  small factors and the second gives convergence of the integral.  
 Since  $R_{j+1}  \subset   \Om_{j+1}- \La_{j+1}$  the      first is  smaller than    $\exp(   - \frac14 L^{-(\sN-j)} | W_j |^2_{R_{j+1}} )$.

  \begin{lem} 
 \begin{equation}
\exp\Big(   - \frac14 L^{-({\sN}-j)} | W_j |^2_{R_{j+1}} \Big)  \zeta^w_{j, L^{-({\sN}-j)} }(  R_{j+1}  ) 
    \leq   \exp \Big(  -\frac14 p^2_{0,j}M^{-3}| R^{(j+1)}_{j+1}| \Big)
\end{equation}
\end{lem} 
 \bigskip

 \pr  Scaled up to   a unit lattice this says
 \begin{equation}
 \exp \left(- \frac{1 }{4 }   |W_j|^2_{R_{j+1}}  \right)  \zeta^w_{j}( R_{j+1} ) 
    \leq   \exp \Big(  -\frac14 p^2_{0,j}M^{-3}|  R^{(j+1)}_{j+1}| \Big)
\end{equation}
The left side can be written:     
 \begin{equation}
     \prod_{\square  \subset   R_{j+1}}  \exp \left(- \frac14   |W_j|^2_{\square}  \right)
       \zeta^w_{j}( \square ) 
 \end{equation}   
  where  the product is over  the  $LM$  cubes.
 The characteristic function 
 $ \zeta^w_{j}(\square )$  enforces that there is at least on   point   in  $\square $  such that
 $ |W_j  |   \geq   p_{0,j}$.
 Therefore  
 \begin{equation}
 \exp \left(- \frac{1 }{4 }   |W_j|^2_{\square}  \right)  \zeta^w_{j}(\square )
 \leq      \exp \Big(     -    \frac{1}{4} p_{0,j}^2     \Big)
 \end{equation}
 and the result follows  as before.
 \bigskip

   Next   we extract   the small  factors from the action.    We  first note   some bounds on the potential   which are independent  of field 
size.    We   have  in general for    $\la >0$  
\be    \label{seltzer}
\begin{split}
V( \La;   \vep,  \mu,  \la) 
 \equiv   &   \    \vep \  \Vol (\La) + \frac12  \mu  \int_{\La} \phi^2  + \frac14  \la  \int_{\La}   \phi^4  \\ 
\geq     &     -  |\vep|  \Vol (\La) -  \frac12 | \mu|    \int_{\La} \phi^2  + \frac14  \la  \int_{\La}   \phi^4  \\ 
\geq    &-   \Big(| \vep|  + \frac 14  \mu^2  \la^{-1}\Big )   \Vol(\La)    \\
 \end{split}
 \ee
 The last step follows since   $ -  \frac12 | \mu|  x^2  + \frac14  \la x^4  $  has the minimum value  $-   \frac14  \mu^2 \la^{-1}$.
 
  Note  also that  
 \be    \prod_{j=-1}^{{\sN}-1}   \  \zeta^0_{j+1, L^{-({\sN}-j)} }( Q_{j+1} ) 
 =   \prod_{j=0}^{{\sN}}  \  \zeta^0_{j, L^{-({\sN}+1-j)} }( Q_{j} ) 
 = \prod_{j=0}^{{\sN}}  \  \zeta_{j, L^{-({\sN}-j)} }( Q_{j} ) 
 \ee
 Thus  we   can use the following estimate:

  \begin{lem}  \label{citizen}  Assume the small field bounds in  (\ref{sally3}).  Then  there   is a  constant  $c_2$  (depending on  $L$) such that  for    $j=0, \dots  {\sN}$  and $\de \La_{j-1} = \La_{j-1} - \La_j$:
 \begin{equation}  \label{snicker}
  \exp  \left(    -      S^{+,u}_{j,L^{-({\sN}-j)}}(\La_{j-1} - \La_j)  
  \right)     \zeta_{j,  L^{-(\sN-j)}  }( Q_j) 
  \leq   \exp \Big(  C \la_j^{ \beta}   |\de  \La^{(j)}_{j-1} |  -c_2 p^2_jM^{-3}|Q^{(j)}_{j}|  \Big)
\end{equation}
\end{lem} 
 \bigskip

\re    This  estimate   is  more involved
  because the action     $ S^{+,u}_{j}(\La_{j-1}  - \La_{j})$   is a  function  of  
  $\phi_{j, \bom(\La_{j-1}, \Om_{j}, \La_{j} ) }$,   but   $\zeta_j(Q_j)$    expressed  in term  of a different field,  namely    
    $ \phi_{j,  \bom }(\square)$.   We  need to make  a connection  and we do it via the fundamental fields.  
  For this  the lemma  \ref{sylvan1} in the appendix will  be important.  
\bigskip

\pr 
(A.)  The bound scales  up  to   
 \begin{equation}  \label{snicker1}
  \exp  \left(    -      S^{+,u}_{j}(\La_{j-1} - \La_j)  
  \right)     \zeta_j ( Q_{j}) 
  \leq   \exp \Big(  C \la_j^{ \beta}   |\de  \La^{(j)}_{j-1} |  -  c_2 p^2_jM^{-3}|Q^{(j)}_{j}|  \Big)
\end{equation}
which is what we  prove.
Now we are on   $\bbT^{-j}_{\sM + \sN -j}  $   and  $\La_{j-1}$ is a union of  $L^{-1}M$ cubes   and  $\La_j, Q_j$ are unions of 
$M$ cubes.

 Split  the  quartic  term in  the      potential in half and write
 \be  
  S^{+,u}_j( \de  \La_{j-1}   )  =    S^*_j( \de  \La_{j-1}   ) +  V^u_j( \de \La_{j-1})
 =   \hat   S_j( \de  \La_{j-1}   )     +  V_j (\de \La_{j-1},  L^3 \vep_{j-1},L^2 \mu_{j-1},  \frac 12 \la_j)
 \ee    
 where  
 \be  
  \hat   S_j( X)    \equiv     S^*_j(X)  +   \frac18 \la_j  \int_{X}  \phi^4_{j, \bom(\La_{j-1}, \Om_{j}, \La_{j} ) } 
 \ee
 is now non-negative.   Then  since  $\vep_{j-1}    \leq   \one \la_{j}^{\beta}$
   and  $\mu_{j-1} \leq   \one   \la_{j}^{\frac12  + \beta} $ we have  by
 (\ref{seltzer})  
 \be 
 \begin{split}
   \exp \Big(  -  S^{+,u}_j( \de  \La_{j-1}   ) \Big ) 
    \leq &  \exp \Big (-  \hat   S_j( \de  \La_{j-1}   )  \Big )  \exp  \Big (  C \la_j^{ \beta}   \Vol(\de  \La_{j-1} )  \Big)  \\
     \leq &  \exp \Big (-  \hat   S_j( Q^*_j  )  \Big )  \exp  \Big (  C \la_j^{ \beta}  |\de  \La^{(j)}_{j-1} |  \Big)  \\
  \end{split}  
 \ee
The  second inequality  follows since  $Q^{5*}_j \subset     \La^c_{j} $   and  $Q_j^{5*}  \subset  \Om_j^{5*}  \subset  \bar \La_{j-1}  $
imply   $ Q^{5*}_j \subset    \bar   \La_{j-1} - \La_j$  and hence   $Q_j^* \subset  \La_{j-1} - \La_j$.
The  bound (\ref{snicker1})   is now reduced to    
\begin{equation}  \label{snicker2} 
  \exp  \left(    -    \hat  S_{j}(Q_{j}^{*})  
  \right)     \zeta_{j}( Q_{j}) 
  \leq   \exp \Big(  -c_2 p^2_jM^{-3}|Q^{(j)}_{j}|  \Big)
  \end{equation}

\noindent(B.)
 Let    $R_1 = 2R+1$  where  $R$  is   the parameter   which enters the definition  of   $\phi_{j, \bom(\square)}$,   see
 section 3.1.5 in part II.  
  \begin{equation}
\begin{split}
\hat S_{j}(Q_{j}^{*})   =  &   \sum_{\square'  \subset  Q_{j}^{*}}\hat S_{j}(\square')   
=        \sum_{\square'  \subset   Q_j^{*}}    \sum_{\square:   \square^{\sim R_1} \supset  \square'}  
(2 R_1 +1)^{-3}       \hat S_{j}(\square')  \\
 \geq   &    \sum_{\square  \subset   Q_j}  
   \sum_{\square'  \subset  \square^{ \sim  R_1} }  (2 R_1 +1)^{-3}       \hat S_{j}(\square')  
  =     \sum_{\square  \subset   Q_j}      (2 R_1 +1)^{-3}           \hat S_{j}(\square^{  \sim   R_1})  \\
\end{split}
\end{equation}
Here  $\square,  \square'$  are  $M$ cubes,  and    we use that 
   $\square  \subset  Q_j  $  and  $\square'  \subset  \square^{\sim R_1}$ 
 imply   $\square'  \subset   Q^{*}_j $  and    $  \square^{\sim R_1} \supset  \square'$,   so we  are summing over a smaller set  in the
 fourth expression as opposed to the third expression.

 Now (\ref{snicker2})  follows if we can   show  for  $ \square  \subset   Q_j$  with  $R_2= 2R_1 +1$
 \begin{equation}  \label{snicker3}
  \exp  \Big(   -  R_2^{-3}       \hat S_{j}(\square^{\sim R_1})  \Big)    \zeta_{j}(\square) 
\leq      \exp (  -  c_2      p^2_{j}  )     
\end{equation}
\bigskip

\noindent (C.) 
There  is a    constant   $c_1$ (small, depending on $L$) such that if   $    | \pa   \Phi_{j}  |  \leq  c_1 p_{j}$ 
   and $  |  \Phi_{j}   | \leq c_1 \al_{j}^{-1}   p_{j}$  on    $\square^{\sim R_1}$  then on
      $\tilde \square$
 \begin{equation}  \label{notable}
\begin{split}
 | \Phi_j  -  Q_j    \phi_{j, \bom_j(    \square)})|  \leq &  p_{j} \\
 |\pa   \phi_{j, \bom_j(    \square)}|  \leq &  p_{j}   \\   
 |  \phi_{j, \bom_j(   \square)})|  \leq &  \al_{j}^{-1}  p_{j}  \\
\end{split}
\end{equation} 
This follows  by a slight variation of    lemma  3.1 in part II,  and needs   $\square$  well inside $ \Om_{j}$  which we have since  $Q_j \subset  \Om_j^{\nat}$.
 This implies that     $ \chi_{j}(\square) =1$   and hence     $ \zeta_{j}(\square) =  1 - \chi_j(\square) =0$    and so the inequality (\ref{snicker3}) holds. 
 Thus  we  can restrict attention to fields such     that  either    $    | \pa   \Phi_{j}  |  \geq   c_1  p_{j}$
  or  $  |  \Phi_{j}   |  \geq   c_1  \al_{j}^{-1} p_{j}$
hold  somewhere   in $\square^{\sim R_1}$.   
\bigskip

\noindent (D.)
 If  $    | \pa   \Phi_{j}  |  \geq  c_1   p_{j}$  for some bond in  $\square^{\sim R_1}$,  then   by   lemma  \ref{sylvan1} in appendix  \ref{C} 
 there is a constant  $c'_0= \one$   so with 
 $\phi =   \phi_{j,  \bom(\La_{j-1}, \Om_{j}, \La_{j} ) } $
\begin{equation}  \label{lowerbound}
\begin{split}
\hat S_j(\square^{\sim R_1})  \geq    & 
\  \frac{a_j}{2}  \|  \Phi_{j}- Q_{j}   \phi   \|^2_{ \square^{\sim R_1}}  
 +  \frac12 \| \pa   \phi     \|^2_{*, \square^{\sim R_1}} 
 + \frac12  \bar   \mu_j  \|   \phi  \|^2_{ \square^{\sim R_1}} \\
 \geq  & \   c'_0    \Big(    \|  \pa \Phi_{j} \|_{ \square^{\sim R_1}} ^2   +  \bar \mu_j  \| \Phi_{j}\|_{ \square^{\sim R_1}} ^2      \Big)
   \geq    c'_0 c_1^2 p_{j}^2  \\
 \end{split}
 \end{equation}
This is sufficient to prove  (\ref{snicker3})  if   $c_2  \leq    c'_0 c_1^2  R_2^{-3} $.
\bigskip

\noindent(E.)   On the other hand suppose        $    | \pa   \Phi_{j}  |  \leq c_1  p_{j}$
everywhere  in $\square^{\sim  R_1}$, and that      $  |  \Phi_{j}   |  \geq  c_1   \al_{j}^{-1} p_{j}$ at some point  in $\square^{\sim R_1}$.    
If  also    $\al_j \equiv  \max \{  \bar \mu_j^{\frac12}, \la_j^{\frac14}  \} =  \bar \mu_j^{\frac12}$  then    $  |  \Phi_{j}   |  \geq   c_1  \bar \mu  _{j}^{-\frac12} p_{j}$ at some point.  Again using    (\ref{lowerbound})  we  have the sufficient bound   
\be  \hat  S_j(\square^{\sim R_1})   \geq    c'_0  \bar \mu_j  \| \Phi_{j}\|_{ \square^{\sim R_1}} ^2    \geq  c'_0c_1^2 p_j^2
\ee
\bigskip

 \noindent (F.)  We  are now reduced to the case   $    | \pa   \Phi_{j}  |  \leq c_1  p_{j}$
everywhere  in $\square^{\sim R_1}$,   $  |  \Phi_{j}   |  \geq  c_1   \al_j^{-1} p_{j}$ at some point   in $\square^{\sim R_1}$,     and    $\al_j =   \la_j^{\frac14}$. 
We  want to show that the field  
\be       \label{piggy}
  \phi_{j,  \bom( \La_{j-1},  \Om_j,  \La_j)}  =         \phi_{j,  \bom( \La_{j-1},  \Om_j,  \La_j)}   \Big(   \tilde  Q^T_{\bbT^{-1},  \bom(  \La^*_{j-1})} \Phi_{j-1, \Om^c_{j}},
\          \tilde  Q^T_{j,\bbT^{0},  \bom(  \La^{c,*}_{j})  }  \Phi_{j, \Om_j}   \Big)
 \ee
 is  large    on the set   $\square^{\sim R_1}$.     The   $\Phi_{j-1}$  term   on  $\Om_j^c$   can be  safely ignored   since   $Q_j  \subset  \Om_j^{\nat}$.
The difference between the field with it and without it   is  $\cO(e^{-r_j})$.   Here  we  need to use small field bounds for  $\Phi_{j-1},  \Phi_j$.

\
Next we use  the identity  for a unit lattice point  $y  \in    \square^{\sim R_1}  $  and  $x$ in a neighborhood of  $\De_y$
\be     \label{samsum}
\begin{split}
\Big[  \phi_{j,  \bom(\La_{j-1}, \Om_{j}, \La_{j} ) }  \Big(0,    \tilde  Q^T_{\bbT^{0},  \bom(  \La^{c,*}_{j})  }  \Phi_{j, \Om_j} \Big) \Big](x)   
 = & \Big[  \phi_{j,  \bom(\La_{j-1}, \Om_{j}, \La_{j} ) }
  \Big(0,    \tilde  Q^T_{\bbT^{0},  \bom(  \La^{c,*}_{j})  }(\Phi_{j} -  \Phi_{j}(y)  ) \Big) \Big](x)   \\
   +  & \Phi_{j}(y)    - \Big[  \bar \mu_j   G_{j, \bom(\La_{j-1}, \Om_{j}, \La_{j} )}  \cdot  1 \Big](x)   \Phi_{j}(y)    \\
 \end{split}
\ee
See the proof of lemma 3.1 in part II   for a similar identity.
The first term  is bounded by   $Cc_1p_j$  by the bound on $\pa \Phi_j$. 
The last term  is bounded by $  \bar \mu_j  C |\Phi_j(y)| $   and since     $  \bar   \mu^{\frac12}_j  \leq  \al_j  =  \la_j^{\frac14}$
 and  $|\Phi_j(y)|  \leq   \la_j^{\frac14-\de}$  this is bounded by    $ \la_j^{\frac 12} C   \la_j^{- \frac 14 - \de} p_j \leq p_j$.   Here again we are using small field bounds. 
Thus we  have   for  $x$ in a neighborhood of  $\De_y$:
\begin{equation}
| \phi_{j,  \bom(\La_{j-1}, \Om_{j}, \La_{j} ) } (x) -  \Phi_j(y)|   \leq   C p_j
 \end{equation} 

Now     if  $  |  \Phi_{j}(y)   |  \geq   c_1  \la_{j}^{-\frac14}  p_{j}$  at  some point  $y$   in $\square^{\sim R_1}$,  then  for  $\la_j \leq  \la$  sufficiently small,   the last inequality implies that  
  $  |  \phi_{j,  \bom(\La_{j-1}, \Om_{j}, \La_{j} ) }   |\geq   \frac12  c_1   \la_{j}^{-\frac14} p_{j}$    at  all points  in  some unit square   $x \in  \De_y $.   Then we  get the small factor from the potential in  $\hat S_j(\square^{\sim  R_1}) $  : 
  \begin{equation}
 \hat  S_j(\square^{\sim R_1})   \geq    \frac18 \la_j  \int_{\square^{\sim R_1}}   \phi_{j,  \bom(\La_{j-1}, \Om_{j}, \La_{j} ) }  ^4
  \geq \frac18     \la_j  \int_{\De_y}    \phi_{j,  \bom(\La_{j-1}, \Om_{j}, \La_{j} ) }  ^4
  \geq    \frac{1}{128} c_1^4  p_j^4
  \end{equation}
  This is sufficient  for  (\ref{snicker3})  if   $c_2  \leq     \frac{1}{128} c_1^4  R_2^{-3}$.   Thus   (\ref{snicker3})  is established.
  \bigskip
  
  \noindent  (G.)  A  remark on the case
  $j=1$.  In this case it is not   $\phi_{1,  \bom( \La_{0},  \Om_1,  \La_1)} $ we  are considering,   but  a modification 
  $\phi_{1,  \bom(  \Om_1,  \La_1)} $   where  $\bom( \Om_1,  \La_1) =  \Om_1  \cap   \bom(  \La^{c,*}_{1}) $.  
       The only    $\Phi_0$ dependence just comes from a  term  $ \De_{\Om_1, \Om^c_1}  \Phi_0$,  and    hence from  $\Phi_0$ near  $\pa \Om_1$.
   Here   we do   have a small field bound    and   so   the   above argument goes  through.     \bigskip
  
 \noindent (H.)   $j=0$   is a special  case.    In this case it suffices to show  on   $\bbT^0_{\sM +\sN}  $ that for an  $M$-cube 
 $\square \subset  Q_0$
 \be 
   \exp \Big(   - \frac 12\| \pa \Phi_0 \|^2_{\square}  - \frac   12\bar \mu_0  \| \Phi_0 \|^2_{\square}   
     - \frac 14  \la_0  \int_{\square}  \Phi_0^4  \Big)  \zeta_0 (\square) \ 
 \leq    \    \exp (  -  c_2      p^2_0  )     
 \ee
 where   $\zeta_0 (\square) $ enforces that  either   $| \pa  \Phi_0 |  \geq   p_0$  or  $|\Phi_0|   \geq   \al_0^{-1  } p_0$  at some point
 in   $\square$.      Splitting into the two  cases   $\al_0  =   \bar \mu_0^{\frac12}  $   and    $\al_0  =  \la_0^{\frac 14}$   this follows
 directly.    (Actually     $\al_0  =  \la_0^{\frac 14}$   will  always hold  for $\sN$ sufficiently large.)
  This completes the proof.
 \bigskip 
 
\res  
\begin{enumerate}
\item The last  lemma works  as well with  $\exp ( - \frac12  S^{+,u}_j(\La_{j-1}- \La_j ))$  rather than 
 $\exp ( -  S^{+,u}_j(\La_{j-1}- \La_j ))$.  Then we  would have an extra factor     $\exp ( - \frac12  S^{+,u}_j(\La_{j-1}- \La_j ))$ 
 to use  for the convergence of the integrals.   We  do  not need to do this since we still have  small field characteristic  functions
 to enforce the convergence.      However the small field characteristic functions  are  not available  for  $j=0$,  so in this case 
 we  do  make the split  and have  a factor   $\exp ( - \frac12  S^{+,u}_0( \La^c_0 ) )=\exp ( - \frac12  S^{+}_0( \La^c_0 ) )$ left over.
 \item
 We collect  the small factors generated by the previous three lemmas.    With a further shift of indices and taking account that  
 $P_0,  R_0,  P_{{\sN}+1}, Q_{{\sN}+1}  = \emptyset$   they  are   
 \be 
  \prod_{j=0}^{\sN+1}   \exp \Big(  - \frac14  aL p^2_{j-1}M^{-3}| P^{(j)}_{j}|   - c_2 p^2_{j}M^{-3}| Q^{(j)}_{j}|   -\frac14 p^2_{0,j-1}M^{-3}| R^{(j)}_{j}| \Big)  
 \ee
    Assuming    $c_2  \leq  \frac14,  \frac14aL$  and using     $p_j  \geq  p_{0,j}$  and  
 that   $p_{0, j-1}   \geq     p_{0,j}$   this is bounded by   
 \be     \label{sangria}
    \prod_{j=0}^{{\sN}+1}     \exp \Big(  -c_2p^2_{0,j}M^{-3}  (    | P^{(j)}_{j}| +   | Q^{(j)}_{j}|+ | R^{(j)}_{j}| )  \Big)  
 \ee
 There  is also the factor   $ \exp (  C \la_j^{ \beta}   |\de  \La^{(j)}_{j-1} |  )  \leq  \exp (  C \la_j^{ \beta}   |(  \La^c_j)^{(j)} |  ) $ from  lemma  \ref{citizen}.
\end{enumerate}

 \subsection{final integrals}
  The  remaining integral  over    $W_{{\sN}, \bpi^+}$  in  (\ref{sally3})   is
 \be    \label{span}
\begin{split}
& \int  \prod_{j=0}^{{\sN}}      (2 \pi)^{-|  [ \Om_{j+1}  - \La_{j+1}]^{(j)}|/2}
  \exp  \Big(   - \frac{1}{4 } L^{-({\sN}-j)} | W_j |^2_{\Om_{j+1}- \La_{j+1}}  \Big) 
    dW^{({\sN}-j)}_{j,\Om_{j+1}- \La_{j+1}}  
   =   \prod_{j=0}^{\sN}   2^{ | \Om_{j+1}^{(j)} - \La^{(j)}_{j+1} |/2  }   \\
   \end{split}
\ee
The  last line follows by   the change of variables  $W_j \to  \sqrt 2  W_j$  which takes us back to a probability measure.
\bigskip

Thus      the   remaining  integrals  in $\cK'(\Theta)$    are      bounded  by  
 \begin{equation}  \label{spiffy}
\begin{split}  
&\int      \prod_{j=0}^{{\sN}}   d\Phi^{({\sN}-j)}_{j, \Om_{j+1}^c}
 \exp \left(  - \frac{1}{4 } aL^{-({\sN}-j)}|\Phi_{j}- Q \Phi_{j-1}|^2_{ \Om^c_{j}}  \right)
  \\
   & \hs  \exp\left( -S^{+}_{0,L^{-{\sN}}}(   \La^c_0     ) \right)  \tilde  \chi_{0,  L^{-N}}(\La_0- \Om_1)   \prod_{j=0}^{{\sN}} 
     \tilde  \chi_{j,  L^{-(N-j)}}(\de \Om_j) 
  \\
 \end{split}
\end{equation}
We  do the integrals   for    $ j={\sN}, {\sN}-1, \dots, 2,  1$ in that order.   In each  case  we   first  scale up  by  $N-j$  
so that  $\Om_j$  is  a union   of  $M$ cubes  and    $\Om_{j+1}$  is a union of $LM$ cubes    in  $\bbT^{-j}_{\sN +\sM -j}$,       and $\Phi_j$ is a function on 
 $( \Om^c_{j+1} )^{(j)} \subset   \bbT^0_{\sN +\sM -j}$.
Split    the integral over   $\Phi_{j, \Om^c_{j+1}}$   into an integral  over   $ \Phi_{j,\de   \Om_{j}^c}$  and an integral over
 $\Phi_{j, \Om_{j}^c}$.     The  first integral is    
 \be  
   \int  d \Phi_{j, \de \Om_j} \   \tilde  \chi_{j}(\de \Om_j,  \Phi_j)   = \Big [ 2 \la_j^{-\frac14 - \de} \Big]^{  |(\de  \Om_j)^{(j)}| }
=  \exp \Big ( \one   (- \log  \la_j )   |( \de \Om_j)^{(j)}| \Big)   
\ee
 The second integral is  
  \be  
  \int     d \Phi_{j, \Om^c_{j}}  \exp \left(  - \frac{1}{4 } a  |\Phi_{j}- Q \Phi_{j-1}|^2_{ \Om^c_{j}} \right) =   \cN\Big( \frac12a,  (\Om^c_{j})^{(j)}\Big)  
 =    \exp  \Big( \one   |(\Om^c_j)^{(j)}|   \Big) 
\ee
These combine to give    a bound  $ \exp  \Big( \one   (- \log  \la_j )    |(\Om^c_{j+1})^{(j)}|   \Big) $

For  $j=0$  we   scale  up  by  $L^{\sN}$   and then split the integral  over   $\Phi_{0, \Om^c_1}$  
  into   integrals over    $\Phi_{0, \La_0 - \Om_1}$   and    $ \Phi_{0,\La_{0}^c}$.
For the first we have as before    
\be  
    \int  d \Phi_{0, \La_0 - \Om_1} \  \tilde \chi_{0}(\La_0- \Om_1,  \Phi_0)    \leq      \exp \Big( \one    (- \log  \la_0 )   |\La^{(0)}_0- \Om^{(0)}_1|   \Big)  
    \leq           \exp \Big( \one    (- \log  \la_0 )   | (\Om^c_1)^{(0)}|   \Big)
\ee
For the second we  have  
\be    
\begin{split}
& \int  d  \Phi_{0,\La_0^c}   \exp   \Big( -\frac12  S^+_{0}( \La_0^c, \Phi_0 ) \Big   ) 
 \leq     \int  d  \Phi_{0,\La_0^c}
\exp    \Big( -\frac14 \bar \mu_0 \|  \Phi_0\|^2_{ \La^c_0}   ) \Big)  \\
   &  \hs  =
 \cN\Big( \frac12  \bar \mu_{0},  | (\La^c_0)^{(0)} | \Big)  \leq     \exp \Big(\one  (- \log \bar \mu _0 )  |(\La^c_0)^{(0)} | \Big) 
  \leq    \exp \Big( \one   (-\log  \la_0 )  |(\La^c_0)^{(0)} |  \Big)         \\
\end{split}
\ee
The last step follows  since
 \be
     -   \log   \bar \mu_0     =      2\sN \log  L       \leq  2  (  - \log \la   +   \sN \log  L)   =  2 ( - \log \la_0)   
\ee 

Putting this together   the integrals (\ref{spiffy}) are bounded by  
\be   
   \exp\left(       \one   (- \log  \la_0 )    |(\La^c_0)^{(0)} | + \sum_{j=0}^{{\sN}}     C  ( - \log \la_{j})  |  ( \Om_{j+1}^c)^{(j)}|       \right)
   \ee
But   
$ |(\Om^c_{j+1})^{(j)}|  =  L^3   |(\Om^c_{j+1})^{(j+1)}|    \leq    L^3   |(\La^c_{j+1})^{(j+1)}|$
and   $- \log\la_{j}  = - \log \la_{j+1} + \log L   \leq   -2 \log\la_{j+1}$.    Hence  the  above  expression can be bounded by 
\be   
   \exp\left(  \sum_{j=0}^{{\sN+1}}     C  ( - \log \la_{j})  |  ( \La_{j}^c)^{(j)}|    \right)
   \ee
A  factor of this   form    also     bounds       the right side  of  (\ref{span}).   It  also bounds contribution from   
 factors  like   $    \exp ( c_j |( \Om^c_j)^{j-1}|) $ and      $  \exp  ( B_0  |  \La_{j-1}^{(j)} -  \La_{j}^{(j)}|_M)$ in  (\ref{sally3}).

Combining these estimates with  (\ref{sangria}) we have finally
  \begin{equation}   \label{randall}
|\cK'(\Theta)| \leq       
  \sum_{ \{ P_j,Q_j, R_j\},:  \La^c_{{\sN}+1} =  \Theta }      \exp\left(  \sum_{j=0}^{{\sN}+1}    C  ( - \log \la_{j})  |  ( \La_j^c)^{(j)}|    - c_2   p^2_{0,j} 
    M^{-3}  \big(     |P^{(j)}_j| +|Q^{(j)}_j|  + |R^{(j)}_j|  \big)  \right)      
\end{equation}
Here    $\Om_j,  \La_j$  are defined from    $ P_j,Q_j, R_j$  by (\ref{understood})  and      $P_0,R_0,Q_{{\sN}+1},  P_{{\sN}+1} = \emptyset $. 
At this point all the fields are gone.

\subsection{convergence}
We  estimate the last  sum.   This analysis is more or less model independent,    we    follow  \cite{Bal82b}.

Let us return to  the general step in the analysis.   We  have   $ \La_k^c$   defined  by sequences
 $\{P_j,Q_j,R_j\}_{j=0}^k$   which are unions  of  $L^{-(k-j)} M $ cubes  in $\tk$.
Let  $\cC_j$  be   the   set of all    $L^{-(k-j)}  M$  cubes   in   $P_j \cup Q_j  \cup R_j$.   The number of elements in this set  is the same  as the number of $M$ cubes  when    $P_j,Q_j,R_j$ are scaled by  $L^{k-j}$  up to   
 $\bbT^{-j}_{\sM +\sN -j}$.    In this  case  $|P_j^{(j)}|$ is  the  number of unit cubes so  
 \begin{equation}  \label{ugh1}
| \cC_j|  =  M^{-3}     \Big(   |P^{(j)}_j| +|Q^{(j)}_j|  + |R^{(j)}_j|   \Big)   
\end{equation}

\begin{lem}
\begin{equation}  \label{ugh2}
\Vol (\La_k^c)  =|(\La_k^c)^{(k)}|  \leq  \cO(1)  (M r_k)^3   \Big(|\cC_0|  +   \dots | \cC_k|  \Big)
\end{equation}
\end{lem}
\bigskip

\pr  
We   first  claim that   $\La_k^c $  can be  covered by  all of the following
\begin{equation}  \label{claim}  
\begin{split} 
|\cC_0| &  \textrm{  cubes of width }  \leq M \Big(  L^{-k}(1+22[r_0])+  22L^{-(k-1)} [r_1] + \dots +   22L^{-1}[ r_{k-1}]      + 22 [r_k] \Big) \\
|\cC_1| &  \textrm{  cubes of width }  \leq M\Big(  L^{-(k-1)}(1+22[r_1]) +  22L^{-(k-2)} [r_2] + \dots +   22L^{-1} [r_{k-1}]      + 22 [r_k] \Big)  \\
&\dots   \\
|\cC_{k-1}| &  \textrm{  cubes of width }   \leq  M \Big( L^{-1}(1+22[r_{k-1}])     + 22 [r_k]  \Big)\\
|\cC_{k}| &  \textrm{  cubes of width }  \leq    M(1 + 22[ r_k])  \
\end{split}
\end{equation}

 The  proof is by induction on  $k$,  just as in the proof  of the main theorem in part II.   First we show the statement is true 
 for  $k=0$.     We  have  $\La_0^c =  Q_0^{5*}$  and  $\cC_0$  is  all  $M$ cubes  $\square$ in  $Q_0$.    Then 
 \be    \La_0^c =  \bigcup_{\square  \subset  \cC_0} \square^{5*}  \ee
 Hence    $\La_0^c$ is covered by   $|\cC_0|$  cubes  of width   $M(1+10[r_0])  \leq  M(1+22[r_0])    $  as  required.

Now     assume it is true for  $k$  and   we   prove it for  $k+1$.  Before  scaling  $\La^c_{k+1}$,  a  union of  $LM$ cubes 
in   $\tk$,  and is generated by  
\be 
 \La^c_{k+1}  =  (\bar  \La^c_k)^{10*}  \cup  P_{k+1}^{10*}  \cup   Q_{k+1}^{5*}  \cup  R_{k+1}^{5*}   
   \ee
 The covering of  $\La^c_k$ is also a covering of the smaller set  $\bar  \La_k^c$.   Each cube in this covering  is enlarged to a 
 cube  which is a union of standard  $LM$ cubes  (adding less than   $2LM$  to  the width)  and then further  enlarged with  by adding $10[r_{k+1}]$
 layers of $LM$ cubes.  The overall enlargement is less than    $22LM[r_{k+1}] $.   Thus we have a covering  
   of   $ (\bar  \La^c_k)^{10*} $   by    
  \begin{equation}  \label{claim2}  
\begin{split} 
|\cC_0| &  \textrm{  cubes of width }  \leq M \Big(  L^{-k}(1+22[r_0])+  22L^{-(k-1)} [r_1] + \dots       + 22 [r_k]   +22L[r_{k+1}]\Big) \\
|\cC_1| &  \textrm{  cubes of width }  \leq M\Big(  L^{-(k-1)}(1+22[r_1]) +  22L^{-(k-2)} [r_2] + \dots      + 22 [r_k]  +22L[r_{k+1}\Big)  \\
&\dots   \\
|\cC_{k-1}| &  \textrm{  cubes of width }   \leq  M \Big( L^{-1}(1+22[r_{k-1}])     + 22 [r_k]   +22L[r_{k+1}]\Big)\\
|\cC_{k}| &  \textrm{  cubes of width }  \leq    M \Big((1 + 22[ r_k])     +22L[r_{k+1} ]\Big)
\end{split}
\end{equation}
 In   addition  if $\cC_{k+1}$ is the $LM$ cubes  in   $  P_{k+1}  \cup   Q_{k+1} \cup  R_{k+1}$   we    can  cover   $  P_{k+1}^{10*}  \cup   Q_{k+1}^{5*}  \cup  R_{k+1}^{5*}$  by  
 \begin{equation}  \label{newnew}
 |\cC_{k+1}|   \textrm{  cubes of width } \leq  LM(1 + 22[r_{k+1}])
\end{equation}
  The actual   $\La_{k+1}^c$  in  $\bbT^{-k-1}_{\sN + \sM -k-1}$   is obtained by scaling    down by   $L^{-1}$.   Thus it  is covered by  the cubes  in 
 (\ref{claim2})  and  (\ref{newnew})  with  widths scaled down by  $L^{-1}$.   This is the claim  for  $k+1$.    Thus  (\ref{claim}) 
 is established.

It  follows that 
\begin{equation}
\begin{split}
\Vol ( \La_k^c)   \leq  & M^3  \Big(  L^{-k}(1+22[r_0])+  22L^{-(k-1)} [r_1] + \dots +   22L^{-1}[ r_{k-1}]      + 22 [r_k] \Big)^3| \cC_0|  \\
+  &  M^3  \Big(  L^{-(k-1)}(1+22[r_1]) +  22L^{-(k-2)} [r_2] + \dots +        22 L^{-1} [r_{k-1}]  +22[r_{k}]\Big)  ^3| \cC_1|  \\
+  & \cdots  \\
+   &    M^3 \Big( L^{-1}(1+22[r_{k-1}])     + 22 [r_k]  \Big)^3| \cC_{k-1}|  \\
+  & M^3    (1 + 22[ r_k]) ^3    | \cC_k|  \\
\end{split}
\end{equation}
However 
\begin{equation}
r_{k-j}   \leq   ( 1 + j \log L  )^r  r_k  
\end{equation}
So  the  first term    is bounded by 
\be
M^3 r_k^3 \Big[   \sum_{j=0}^k  L^{-j} ( 1 + j \log L )   \Big]^3       | \cC_0| 
  \leq   \cO( 1)M^3r_k^3   | \cC_0| 
  \ee 
The  other  sums in the other terms are even smaller  and so  
$\Vol ( \La^c_k) \leq  \cO(1)M^3 r_k^3    \Big(|\cC_0|  +   \dots | \cC_k|  \Big)$.

\begin{lem}
\begin{equation}  \label{sushi}
\cK'(\Theta)   \leq    \la^{n_0}  e^{ - \ka'  | \Theta |_M  }
\end{equation}
\end{lem}
\bigskip

\pr   In   (\ref{randall}) 
we     use the last  result to   estimate  
\be
\begin{split}
& \sum_{j=0}^{{\sN}+1}   ( - \log \la_j)  |  ( \La^c_j )^{(j)}|  
\leq \  \one \sum_{j=0}^{{\sN}+1}  ( - \log \la_j)  ( Mr_j)^3  \sum_{i=0}^j     |\cC_i| \\
 &  =  \one \sum_{i=0}^{{\sN}+1}M^3  |\cC_i|  \sum_{j=i}^{{\sN}+1}  ( - \log \la_j)  r_j^3      
 \leq  \ \one   \sum_{i=0}^{{\sN}+1}   M^3  (- \log \la_i)^{3r +2}     |\cC_i|   \\
\end{split}
\ee
In  the  last  step we  use      $r_j^3 =  (-\log \la_j)^{3r} $   and   $ - \log \la_j  \leq  - \log \la_i$  and  $N-i \leq  (N-i) \log L  - \log \la   =   - \log \la_i$.

Also   in (\ref{randall})   replace     $ M^{-3} \sum_{j=0}^{{\sN}+1}     |P^{(j)}_j| +|Q^{(j)}_j|  + |R^{(j)}_j|   $  by     $ \sum_{j=0}^{{\sN}+1} |\cC_j|$
and then  split it   into  three equal pieces. Then   since  $p_{0,j}  =  (- \log  \la_j)^{p_0 }$  we  have  
 \begin{equation}
\begin{split}
|\cK'(\Theta)| \leq     &  
  \sum_{ \{ P_j,Q_j, R_j\},:  \La^c_{{\sN}+1} =  \Theta}   
     \exp\Big(    \sum_{j=0}^{{\sN}+1} \Big( C  M^3  (- \log \la_i)^{3r +2} - \frac13 c_2  (-\log \la_j)^{2 p_0} \Big)     |\cC_j|     \Big)  \\  
&  \hs  \hs  \exp \Big(     \sum_{j=0}^{{\sN}+1}  - \frac13 c_2      p^2_{0,j}    |\cC_j|  \Big)   
     \exp \Big(     \sum_{j=0}^{{\sN}+1}  - \frac13 c_2      p^2_{0,j}    |\cC_j|  \Big)   
 \\
\end{split}
\end{equation}
We  can  assume  $3r+2   < 2 p_0$.  Then  for  $- \log \la$ and hence $-\log \la_j$ sufficiently large, the first exponential is bounded by one. The second exponential is bounded   using   (\ref{ugh2})  again.   With a new constant  $c_2'$      it is less than
\begin{equation}
\begin{split}
& \exp\Big(  \sum_{j=0}^{{\sN}+1}        - \frac13 c_2       p^2_{0,j}       |\cC_j|     \Big)    \leq 
  \exp\Big(   - \frac13 c_2   p^2_{0,{\sN}}  \sum_{j=0}^{{\sN}+1}      |\cC_j|      \Big) 
 \leq      \exp\Big(  - c_2'    p^2_{0,{\sN}}   ( M r_{\sN} )^{-3}  | (\La^c_{{\sN}+1})^{({\sN})} |    \Big)  \\
& \hs   \hs
     =      \exp\Big(   - c_2'   (-\log \la)^{2p_0- 3r}   |\Theta|_M      \Big)  \leq  
  \la^{n_0}       
      e^{ - \ka  | \Theta |_M  }
      \end{split}
 \end{equation}
In the last  we  use    $|  \Theta |_M \geq  1$  and assume      $\frac12   c_2'   (-\log \la)^{2p_0- 3r}  \geq   \ka$.      

Now  we have    
  \begin{equation}
\begin{split}
|\cK'(\Theta)| \leq     &   \la^{n_0}   e^{ - \ka  | \Theta |_M  }
  \sum_{ \{ P_j,Q_j, R_j\},:  \La^c_{{\sN}+1} =  \Theta}   
      \exp\Big(  - \frac13  c_2     p^2_{0,j}      \sum_{j=0}^{{\sN}+1} | \cC_j|       \Big)   
   \\
\end{split}
\end{equation}
Now  drop all conditions  on  $P_j, Q_j, R_j$ except that they   are  unions  of $L^{-(\sN-j)}M$ cubes   $\square$  contained in $\Theta$.
Each sum is estimated separately.     For  $P_j$ we   use   $  |\cC_j|  \geq  |P_j|_{L^{\sN-j}M}$   and estimate   
\be
\begin{split}
 \sum_{P_j  \subset  \Theta }   \exp \Big(  - \frac19 c_2      p^2_{0,j}    |P_j|_{L^{(N-j)}M}  \Big)   
        =&  \prod_{\square \subset  \Theta}  (  1 +    e^{  -  \frac19  c_2 p^2_{0,j} } )
\leq     \prod_{\square \subset  \Theta }  \left (  1+  \la_j^{n_0} \right)\\  
\leq   &  \prod_{\square \subset    \Theta}  e^{  \la_j^{n_0}   }   
\leq      \exp  \Big (   \la_j^{n_0} | \Theta|_{L^{(N-j)}M} \Big)  \\
\end{split}
\ee
The estimates on $Q_j, R_j$  are the same.   ($R_{N+1}$ has $M$-cubes,  not  $LM$ cubes,  but this only improves things.) 
Our bound becomes
  \begin{equation} \label{sally}
\begin{split}
|\cK'(\Theta)| \leq     &   \la^{n_0}   e^{ - \ka  | \Theta |_M  } 
         \exp\Big(  \sum_{j=0}^{{\sN}+1}  3  \la_j^{n_0}    | \Theta|_{L^{(\sN-j)}M}   \Big)    
 \\
\end{split}
\end{equation}
But  $ | \Theta|_{L^{(\sN-j)}M}    =  L^{3(\sN-j)}   | \Theta|_M $   and  $ \la_j^{n_0}  =  L^{- (N-j)n_0}  \la^{n_0}$.  Since  $n_0 \geq  4$ the sum 
in the exponential is bounded by   
\be
 3  \la^{n_0}    | \Theta |_{M}  \sum_{j=0}^{{\sN}+1}   L^{- (N-j)(n_0-3)}   \leq   | \Theta  |_M
\ee
Since  $\ka -1 >  \ka'$  this yields    the desired bound   $  \la^{n_0}  \exp  (    - \ka'  |\Theta| _M  ) $. 
\bigskip

\re   The sum in (\ref{randall})  factors over the connected components  $\{  \Theta_{\ga} \}$ of  $\Theta$.  The bound (\ref{sushi})
holds separately for each factor.  Using also  $|\Theta_{\ga}|_M  \geq  d_M( \Theta_{\ga})$ we have  
\be
  \cK'(\Theta)   \leq    \prod_{\ga}    \la^{n_0}  e^{ - \ka'  | \Theta_{\ga} |_M  }
\ee

\subsection{the stability bound}

We  return to the estimate on $\cK(U)$   where  $U \in \cD_{\sN}$ is a connected union of $M$ cubes in  $\bbT^{-\sN}_{\sM}$.   Substitute   
the  bound  on $\cK'(\Theta)$   into  the bound    (\ref{under})  on $\cK(U)$  and    find  
 \begin{equation}  \label{under2}
\begin{split}
|\cK(U)|  \leq     &   \sum_{ \{ \Theta_{\ga}  \}, \{Y_{\si}  \}, \{X_{\al}  \}   \to  U      } \prod_{\ga}  \la^{n_0}
  e^{    - \ka'  d_M(\Theta_{\ga})   }  
    \prod_{\al} \one  \la^{\beta}   e^{- \ka' d_M  (X_{\al})  }  
    \prod_{\si}  \one \la^{\beta}   e^{- \ka' d_M  (Y_{\si},    \bmod   \Theta ) }
 \\
\end{split}
\end{equation}

\begin{lem}
\begin{equation}
 \sum_{\ell}  ( d_M(\Theta_{\ell}) +1 )   +  
     \sum_{\al} ( d_M  (X_{\al})   +1)  
   +  \sum_{\si}  ( d_M  (Y_{\si},    \bmod   \Theta ) +1)  \geq   d_M  (U) 
\end{equation}
\end{lem}

\pr  
Let  $\tau_{\si}$  be a minimal tree   intersecting every cube  in  $Y_{\si} \cap  \Theta^c$ 
 of length $\ell(\tau_{\si})  = M d_M(Y_{\si}, \bmod \Theta)$,   let
$\tau_{\gamma}$   be  a  minimal  tree intersecting every  cube  in $\Theta_{\gamma}$  of length      $\ell(\tau_{\ga})  = M d_M(\Theta_{\ga})$,
and  let   $\tau_{\al}$   be  a  minimal  tree intersecting every  cube  in $X_{\al}$  of length      $\ell(\tau_{\al})  = M d_M(X_{\al})$.
  Also  consider the graph consisting of pairs    from   $\{ \Theta_{\ga}  \},  \{X_{\al}  \}, \{Y_{\si}  \}  $
which intersect.   Consider   a subgraph which is a spanning tree.  For every  pair  in the spanning tree  take a cube  $\square$
in the intersection  and  introduce a   line between the two  points in $\square$ which are vertices of the trees.
 This  line has length    $  \leq  M$   and the number of lines  is less than the number of elements  in   
 $ \{ \Theta_{\ga}  \}, \{Y_{\si}  \}, \{X_{\al}  \}  $.
Now   the tree $\tau$ formed from   $\tau_{\si}, \tau_{\ga},  \tau_{\al}$  and the connecting lines has length 
\begin{equation}
\ell(\tau )  \leq   M \left(  \sum_{\ga}  ( d_M(\Theta_{\ga}) +1 )   +  
     \sum_{\al} ( d_M  (X_{\al})   +1)  
   +  \sum_{\si} (  d_M  (Y_{\si},    \bmod   \Theta ) +1)  \right)
   \end{equation}
But  $\tau$  spans  all  cubes   in $U$  (cubes in  $Y_{\si}  \cap  \Theta$  are included because of $\Theta $, not  $Y_{\si}$)   and
so  $\ell(\tau)   \geq   Md_M(U)$  which  gives the result.

\begin{lem}
\begin{equation}  \label{swat}
|\cK(U)|  \leq   \one   \la^{\beta/2}     e^{-  ( \ka'- \ka_0 -  1 ) d_M  (U)  }    
\end{equation}
\end{lem}
\bigskip

\pr   
The previous    result  enables   us to  extract a factor  $  e^{- (  \ka' -  \ka_0) d_M  (U)  }$   from the sum,     
Furthermore  since at least  one of    $\{ \Theta_{\ga}  \},  \{X_{\al}  \}, \{Y_{\si}  \}  $
must be nonempty,    we  can pull out an overall factor  of   $\la^{\beta/2}$.  Now drop    all restrictions  on  $\Theta_{\ga},
X_{\al},  Y_{\si}$   except that they are contained in  $U$  and for   $\Theta =  \cup_{\ga}  \Theta_{\ga}$  that    $Y_{\si} \#  \Theta$
   and $Y_{\si}  \in 
\cD_{\sN}( \bmod \Theta   )$.   Then 
 \begin{equation}  \label{under3}
\begin{split}
&|\cK(U)|  \leq      \la^{\beta/2}     e^{-( \ka'-  \ka_0) d_M  (U)  }    
  \sum_{ \{ \Theta_{\ga}   \}\textrm{ in }  U   } \prod_{\ga} \one  \la^{\beta/2} e^{     -\ka_0    d_M(\Theta_{\ga})     } \\
&
\left( \sum_{  \{   Y_{\si}  \}    \textrm{ in }  U     } 
   \prod_{\si}  \cO(1)  \la^{\beta/2}      e^{- \ka_0 d_M  (Y_{\si}, \bmod  \Theta)  }   \right)
\left( \sum_{   \{X_{\al} \} \textrm{ in }  U   }    \prod_{\al}  \cO(1)   \la^{\beta/2}   e^{- \ka_0 d_M  (X)  }  \right)
 \\
\end{split}
\end{equation}
  Using   $\sum_{X \subset U}  \exp( - \ka_0 d_M(X) ) \leq  \one  |U|_M$ we  have the estimate
\begin{equation}
\begin{split}
 \sum_{   \{X_{\al}   \}   \textrm{ in }  U }    \prod_{\al}   \cO(1)  \la^{\beta/2}   e^{- \ka d_M  (X_{\al})  }  
\leq &  \sum_{N=0}^{\infty}\frac{1}{N!}   \sum_{(X_1, \dots,  X_n)}  \prod_{i=1}^n   \cO(1)   \la^{\beta/2}   e^{- \ka d_M  (X_i)  } \\
\leq    &  \sum_{N=0}^{\infty}\frac{1}{N!}   \Big(  \cO(1)  \la^{\beta/2} |U|_M \Big)^N \\
=   & \exp \Big(  \cO(1)  \la^{\beta/2} |U|_M  \Big)
\end{split}
\end{equation}
Here the  second sum is over sequences of   polymers  $(X_1, \dots,  X_n)$.   
The sum  over  $\{ Y_{\si}  \}$ is estimated similarly  now using
\be
\begin{split}
  \sum_{Y \subset  U: Y \# \Theta,  Y \in \cD_{\sN}(  \bmod     \Theta)}        e^{- \ka_0 d_M  (Y, \bmod  \Theta)  } 
  \leq &  \sum_{\square \subset   U- \Theta}  \ \  \sum_{Y \supset  \square,     Y \in \cD_{\sN}(  \bmod     \Theta)}     e^{- \ka_0 d_M  (Y, \bmod  \Theta)  } \\
   \leq  & \one  |U- \Theta|_M   \leq \one  |U|_M  \\
\end{split}
\ee
Finally  the  sum over  $\{  \Theta_{\ga}  \}$ is  estimated  just as the sum over   $ \{X_{\al}   \} $.
Thus we  have
\begin{equation}   \label{toot}
|\cK(U)|  \leq        \la^{\beta/2}     \exp \Big(   - (\ka'- \ka_0)  d_M  (U)     +  \cO(1)  \la^{\beta/2} |U|_M \Big)
\end{equation}
This is sufficient since   $|U|_M  \leq   \one( d_M(U)  + \one)$ and $\la$ is small.
\bigskip

We  are now ready to prove the main result:

\begin{thm}    \label{major1}  Let  $\bar \mu =1$  and let   $\la$  be sufficiently small.  
Then    there is  a choice  of    counterterms  
     $\vep^{\sN}_0,  \mu^{\sN}_0$
  such  that  
\begin{equation}  \label{soso}
  \sZ_{\sM,\sN}  =   \sZ_{\sM,\sN}(0)   \exp \Big(    \sum_X  \cH(X)    \Big)
\end{equation}
where  the sum is over connected unions of $M$ cubes  $X  \subset  \bbT^{-\sN}_{\sM}$   and  
\begin{equation}  \label{such}
|\cH(X)|  \leq    \cO(1)   \la^{\beta/2}     e^{- \ka_0  d_M  (X)  }    
\end{equation} 
 \end{thm}
\bigskip

\pr  
 Recall that  
\begin{equation}
  \sZ_{\sM,\sN}  =   \sZ_{\sM,\sN}(0)   \sum_{   \{  U_{\ell}   \} }   \prod_{\ell}  \cK( U_{\ell}  )
\end{equation}
Since  $\cK(U)$  satisfies the bound  (\ref{swat}),  and   since    $ \cO(1)   \la^{\beta/2}$ is small,
by   a  standard  theorem   (Appendix   B,  part  I)  we  can exponentiate the sum 
to  the form  (\ref{soso})   with
 \begin{equation}
|\cH(X)|  \leq    \cO(1)   \la^{\beta/2}     e^{- (\ka'-  4\ka_0 -4)   d_M  (X)  }    
\end{equation}
\bigskip
Since   we  can assume   $\ka'-  4\ka_0 -4 =     \ka-  11\ka_0 -11    \geq  \ka_0$   we have the result.

As  a corollary we have the stability bound.  Earlier versions can be found in    \cite{GlJa73},  \cite{FeOs76},  \cite{Bal82a},  \cite{Bal82b},   \cite{BDH95}.

\begin{cor}  (stability)  \label{major2}
 \begin{equation}
\exp \Big(  -   \la^{\eta}  \Vol ( \bbT_{\sM} )  \Big)    \leq     \frac{ \sZ_{\sM,\sN} }{  \sZ_{\sM,\sN}(0)}    \leq   \exp \Big(   \la^{\eta} \Vol ( \bbT_{\sM} )  \Big)      
\end{equation}
for  some     $ \eta >0$  independent  $\sM, \sN$. 
 \end{cor}
\bigskip

\pr   This follows  with $\eta = \beta/2$ since     (with  $  \Vol(  \bbT_{\sM } )  = \Vol (  \bbT^{- \sN}_{\sM } )$ ) 
\begin{equation}
 | \sum_{X   \subset  \bbT_{\sM}^{-\sN}} \cH(X) |  \leq  \   \one   \la^{\beta/2} |  \bbT^{- \sN}_{\sM } |_M  =   \
 \one    M^{-3}   \la^{\beta/2}  \Vol(  \bbT_{\sM } )   \leq  \  \la^{\beta/2}  \Vol(  \bbT_{\sM } ) 
\end{equation}
\bigskip

\re  The analysis  can be adapted to treat correlation functions as  well. See  particularly   \cite{BaOc99}
for  an indication of how this would go.

\appendix 

\section{minimizers}  \label{A}

For  a    sequence of small field regions  $\bom  =  ( \Om_1,    \dots  ,  \Om_k) $   and fields   $\Phi_{k, \bom}=     
 (\Phi_{1 \de \Om_1}, \dots,  \Phi_{k-1, \de \Om_{k-1}},  \Phi_{k, \Om_k})$ in $\tk$  as in (\ref{nuts})   
 we  consider the action
\be  \label{box1}
S(\Om_1,  \Phi_{k, \bom},   \phi_{k, \bom} )
=  
 \frac{1}{2}  \|\ba^{1/2}   (\Phi_{k,\bom}- Q_{k, \bom}  \phi_{k, \bom}   )  \|^2_{\Om_1}  +  \frac12 \Big<  \phi_{k, \bom}  ,   (-\De  +   \bar    \mu_k )  \phi_{k, \bom}  \Big> 
\ee
where   \be
  Q_{k,    \bom}  \phi  =  (  [Q_1 \phi]_{\de  \Om_1}, \dots,  [Q_{k-1} \phi]_{\de  \Om_{k-1}}, [Q_k \phi]_{\Om_k} )  
  \ee
and     $\phi_{k, \bom}$ is the minimizer in  $\phi_{\Om_1}$ of   
\be  \label{box2}
S(\Om_1,  \Phi_{k, \bom},   \phi )
=  
 \frac{1}{2}  \|\ba^{1/2}   (\Phi_{k,\bom}- Q_{k, \bom}  \phi  )  \|^2_{\Om_1}  +  \frac12 \Big<  \phi  ,   (-\De  +   \bar    \mu_k )  \phi  \Big> 
\ee
defined for  $\phi:  \tk  \to  \bbR$.

Given  a  new   region $\Om_{k+1}  \subset  \Om_k$  we  want to find the minimizer   of    $S(\Om_1,  \Phi_{k, \bom},   \phi_{k, \bom}  )$
 in  $\Phi_{k, \Om_{k+1}}$
with  all other variables   fixed.  With       $\bom^+  =  ( \Om_1,    \dots  ,  \Om_{k+1}) $ and  $\de \Om_k =  \Om_k- \Om_{k+1}$  these    are  
\be     \Phi_{k,   \de  \bom^+}  \equiv  (\Phi_{1 \de \Om_1}, \dots,  \Phi_{k, \de \Om_k})
\ee   
This is the  same  as the    minimizer  of   $S(\Om_1,  \Phi_{k, \bom},   \phi )=    S(\Om_1,  \Phi_{k,\de \bom^+}, \Phi_{k, \Om_{k+1}},  \phi )$ in  both   $ \Phi_{k, \Om_{k+1}}$  and    $\phi_{\Om_1}$.   
The solution depends on the Green's function  
\be
   G_{k, \de  \bom^+}    =  \Big [ - \De   +  \bar \mu_k   +   Q^T_{k,  \de  \bom^+}  \ba   Q_{k,  \de  \bom^+}\Big ]_{\Om_1}^{-1}
\ee
where  
\be    Q_{k,  \de  \bom^+}  \phi  =  \Big (  [Q_1 \phi]_{\de  \Om_1}, \dots,  [Q_k \phi]_{\de  \Om_k} \Big)   
\ee

\begin{lem}  {  \  }
\begin{enumerate}
\item
Given     $   \Phi_{k,  \de  \bom^+} $   The unique minimum  of  
$S(\Om_1,  \Phi_{k, \bom},   \phi    )$  in   $ \Phi_{k, \Om_{k+1}}$  and    $\phi_{\Om_1}$
comes at \be  \label{star}
\phi_{\Om_1} =   \phi_{k, \de   \bom^+}  =  \phi_{k, \de  \bom^+} ( \phi_{\Om^c_1} ,   \Phi_{k,    \de  \bom^+}  )
=    G_{k, \de  \bom^+}   \Big(   Q^T_{k,  \de  \bom^+}   \ba  \Phi_{k,    \de  \bom^+}    +[\De]_{\Om_1, \Om_1^c}\  \phi_{\Om^c_1}  \Big)
\ee
and at   
\be  \Phi_{k, \Om_{k+1}}   =  \Psi_{k, \Om_{k+1}}( \de \bom^+)  =  [Q_k   \phi_{k, \de  \bom^+}]_{\Om_{k+1}}  
\ee
\item 
 We  have the identity
\be   
\label{star2}
 \phi_{k, \de  \bom^+}    =   \phi_{k, \bom} (  \phi_{\Om_1^c},  \Phi_{k, \de  \bom^+ },   \Psi_{k, \Om_{k+1}} ( \de \bom^+)     ) 
\ee
\end{enumerate}
\end{lem}
\bigskip

\re  This is only useful for  $\bar  \mu_k $  has a substantial size    and so can take the place of the missing averaging operator  in
$\Om_{k+1}$ in  $  G_{k, \de  \bom^+}$.
\bigskip

\pr   The  variational equations  for  minimizing   $S(\Om_1,  \Phi_{k, \bom},   \phi    )$  are
\be 
 \begin{split}  \label{star3}
 \Phi_{k, \Om_{k+1}}- Q_k  \phi   = &    0   \\
\Big [ - \De   +  \bar \mu_k   +    Q^T_{k,    \bom}  \ba    Q_{k,    \bom}  \Big ]_{\Om_1} \phi_{\Om_1}
=   &   Q^T_{k,   \bom} \ba        \Phi_{k,   \bom}    +[\De]_{\Om_1, \Om_1^c}\  \phi_{\Om^c_1}
    \\
\end{split}
\end{equation}
Substituting  $\Phi_{k, \Om_{k+1}}  =   Q_k  \phi $  into the second equation  and canceling a  term  $ a_kQ_k^T Q_k\phi$ on each side       it becomes
\be    \Big[ - \De   +  \bar \mu_k   +   Q^T_{k,  \de  \bom^+}  \ba   Q_{k,  \de  \bom^+}  \Big]_{\Om_1}\phi_{\Om_1}
=      Q^T_{k,  \de  \bom^+} \ba       \Phi_{k,    \de  \bom^+}      +[\De]_{\Om_1, \Om_1^c}\  \phi_{\Om^c_1}
\ee
with the solution  $\phi_{\Om_1}  =   \phi_{k, \de  \bom^+}  $  defined by  (\ref{star}).  With this  $\phi$ the minimum in  
 $\Phi_{k, \Om_{k+1}}$  is  $\Psi_{k, \Om_{k+1}  }= [ Q_k  \phi_{k, \de  \bom^+}]_{\Om_{k+1}} $ as claimed.

Before the substitution, 
the solution of the second equation  in (\ref{star3})  is  $\phi_{k, \bom} (   \phi_{\Om_1^c},  \Phi_{k,   \bom } )$.   
At  the minimum      $\Phi_{k, \Om_{k+1}}  =   \Psi_{k, \Om_{k+1}} $      it  becomes  
$ \phi_{k, \bom} (    \phi_{\Om_1^c},  \Phi_{k, \de  \bom^+ },   \Psi_{k, \Om_{k+1}}     ) $.   Hence this is another representation for the minimizer
$ \phi_{k, \de  \bom^+}  $.

\section{a resummation  operation}  \label{B}

Suppose  $\Om,  \La$  are unions of  $M$ cubes  with  $\La \subset  \Om$,  and    we have an expression  
\begin{equation}
\sum_{X \in \cD_k( \bmod   \Om^c),  X \cap  \La \neq  \emptyset} B(X)  
\end{equation}
  with  
  \begin{equation}
  |B(X)| \leq   B_0 e^{- \ka d_M(X, \bmod  \Om^c )}
  \end{equation}
  for some constant  $B_0$.
  We   want  to   write it as  a similar  sum    with  $\Om$  replaced    $\La $  everywhere.  
 Every such  $X$  determines   a    
$Y \in \cD_k( \bmod \   \La^c)$ with  $Y \cap  \La \neq  \emptyset$  by  taking the union with  any  connected component  of $\La^c$  connected to      $X$,  written  $X \to  Y$.
We  define  
\begin{equation}
B'(Y)  =   \sum_{X \in \cD_k( \bmod   \Om^c),    X \cap  \La \neq  \emptyset, X \to  Y     }    B(X)  
\end{equation}
and then 
\begin{equation}
\sum_{X \in \cD_k( \bmod   \Om^c), X \cap  \La \neq  \emptyset } B(X)  =\sum_{Y \in \cD_k( \bmod   \La^c),Y \cap  \La \neq  \emptyset} B'(Y)  
\end{equation}
\bigskip

\begin{lem}    \label{diamond}
\begin{equation}  \label{opal}
| B'(Y) |   \leq   \cO(1)  B_0   e^{-  (\ka- \ka_0-1)  d_M(Y, \bmod  \La^c )}
\end{equation}
\end{lem}
\bigskip

\pr
We   first     claim  that  
\begin{equation}
d_M(Y, \bmod\  \La^c)  \leq   d_M(X, \bmod\  \Om^c) 
\end{equation}
Indeed  let   $\tau$ be a minimal tree joining the cubes in    $X \cap  \Om$  of length  $\ell(\tau) = Md_M(X)$.   Then 
$\tau$  is also a tree joining the cubes in    $Y \cap  \La$  since   $Y \cap  \La= X \cap  \La
\subset  X  \cap  \Om$.   Hence   $Md_M(Y, \bmod\  \La^c)\leq  \ell(\tau)$ and hence the result.

Then we have
\begin{equation}  \label{topaz}
\begin{split}
  | B'(Y ) |  \leq  & \  \sum_{   X \in \cD_k(\bmod  \La^c), X  \cap  ( Y \cap  \La)  \neq  \emptyset         } B_0  e^{     - \ka   d_M(X, \bmod  \Om^c)  }
  \\
   \leq  &\  B_0    e^{    - (\ka- \ka_0)   d_M(Y, \bmod  \La^c)  }
  \sum_{   X \in \cD_k(\bmod  \Om^c),  X  \cap  ( Y \cap  \La)  \neq  \emptyset         }   e^{     -\ka_0   d_M(X, \bmod  \Om^c)  }
  \\
   \leq  &\one B_0    e^{    - (\ka- \ka_0 )   d_M(Y, \bmod  \La^c)  }  |Y \cap   \La|_M  \\
 \end{split}
\end{equation}
Since   $Y \cap  \La \subset  \Om$ the   last  step follows by lemma  E.3 in  part II.  The result now follows  by 
\be
    |Y \cap   \La|_M  \leq   \one (d_M(Y \cap  \La) +1)   =  \one  ( d_M(Y, \bmod\  \La^c)  +1)   \leq  \one  e^{ d_M(Y, \bmod  \La^c)}
\ee

\section{a bound below}  \label{C}

  Let  $\Phi:     \bbT^{0}_{\sN +\sM-k}  \to  \bbR$
     and    $\phi:    \bbT^{-k}_{\sN +\sM-k}\to \bbR$,  and  let    $X$  be a union of unit blocks in  $ \bbT^{-k}_{\sN +\sM-k}$.  
 For the following result  we employ Neumann boundary conditions:  only bonds contained in $X$ contribute.

      \begin{lem} \cite{Bal83b}  \label{sylvan1}
    There is a constant  $c_0 = \one $
     such that    for  $0 \leq  \mu  \leq  1$  
   \be
     \frac12  \| \Phi -  Q_k  \phi \|^2_X   +  \frac 12  \|  \pa \phi \|^2_{X}   +  \frac 12  \mu  \| \phi\|^2_X         \geq    c_0\   \Big(    \|  \pa \Phi \|_{X} ^2   +   \mu  \| \Phi\|^2_X  \Big )
   \ee
  \end{lem}
  \bigskip
  
  \pr    
   We    have  for  $  y \in  X  \cap     \bbT^{0}_{\sN +\sM-k} $   
   \begin{equation}
   |\Phi(y)|  \leq     | \Phi(y)  -  (Q_k \phi)(y)| +  | (Q_k \phi)(y)|
   \end{equation}
  which yields    $\| \Phi \|_X  \leq     \| \Phi  -  Q_k \phi \|_X +  \| Q_k \phi  \|_X $  and hence  
    \begin{equation}
    \|  \Phi \|^2_X  \leq  2  \Big(  \| \Phi -  Q_k  \phi \|^2_X   +   \|   \phi \|^2_X   \Big)
   \end{equation}
 This   gives   half the result.

  We   also    need a bound on  $  \| \pa  \Phi\|^2_X$.
  For    a bond   $<y, y+ e_{\mu}>$ in $ X  \cap     \bbT^{0}_{\sN +\sM-k} $ 
  \begin{equation}  \label{swoon}
  \begin{split}
 | \pa_{\mu} \Phi(y)|  = & | \Phi(y+ e_{\mu}) - \Phi(y)|   \\
 \leq &  | \Phi(y+ e_{\mu})  -  (Q_k \phi)(y+ e_{\mu})| +  | (Q_k \phi)(y+ e_{\mu})- (Q_k \phi)(y)|  +  | (Q_k \phi)(y) -  \Phi(y)|  \\
  \end{split}
  \end{equation}
  The middle term is written as  
  \be
  (Q_k \phi)(y+ e_{\mu})- (Q_k \phi)(y)  =   \int_{\De_y} dx \Big(\phi(x + e_{\mu})  - \phi(x) \Big) 
     =   \int_0^1  dz  \int_{\De_y} dx \   (\pa_{\mu}\phi)(x +z e_{\mu})    \ee
  where  $\De_y$ the unit cube  centered on  $y$  and  $z \in  L^{-k} \bbZ$.
  Therefore  
\be
   |(Q_k \phi)(y+ e_{\mu})- (Q_k \phi)(y) |  \leq         \int_0^1  dz  \   \|  \pa_{\mu}\phi(\cdot   +z e_{\mu})     \| _{\De_y}       
   \leq      \|  \pa_{\mu} \phi   \| _{\De_y \cup  (\De_y + e_{\mu})}  
\ee  
   This  leads  to    
   \be
   \sum_{<y, y+ e_{\mu}>  \in X}     |(Q_k \phi)(y+ e_{\mu})- (Q_k \phi)(y) | ^2 
    \leq     \sum_{<y, y+ e_{\mu}>  \in X}     \|  \pa \phi   \|^2 _{\De_y \cup  (\De_y + e_{\mu})}     \leq   2    \|  \pa   \phi   \|^2_X  
 \ee   
 Using this in (\ref{swoon})  yields  
 \be   
  \| \pa  \Phi\|^2_X  \leq      \one  \Big(   \| \Phi -  Q_k  \phi \|^2_X   +   \|  \pa   \phi \|^2_X  \Big)
  \ee
 to complete the proof.


\begin{thebibliography}{BrDiHu95}

\bibitem{Bal82a} T. Balaban,
 (Higgs)$_{2,3}$ quantum fields in a finite volume  I,
  {\em  Commun. Math. Phys.} 85: 603-636, 1982.

  \bibitem{Bal82b} T. Balaban,  (Higgs)$_{2,3}$ quantum fields in a finite volume  II,
  {\em  Commun. Math. Phys.} 86: 555-594, 1982.

  
\bibitem{Bal83a} T. Balaban,
 (Higgs)$_{2,3}$ quantum fields in a finite volume  III,
  {\em  Commun. Math. Phys.} 88: 411-445, 1983.

  
\bibitem{Bal83b} T. Balaban,
 Regularity and decay of lattice Green's functions,
  {\em  Commun. Math. Phys.} 89: 571-597, 1983.




\bibitem{Bal95}   T. Balaban,  A low temperature expansion for classical N-vector models I,
  {\em  Commun. Math. Phys.} 167: 103-154, 1995.


\bibitem{Bal96a}   T. Balaban,     Variational problems for classical N-vector models, 
  {\em  Commun. Math. Phys.} 175:   607-642, 1996.







\bibitem{Bal96b}   T. Balaban,     Localization expansions I,
  {\em  Commun. Math. Phys.} 182: 33-82, 1996.




\bibitem{Bal96c}   T. Balaban,  A low temperature expansion for classical N-vector models II,
  {\em  Commun. Math. Phys.} 182: 675-721, 1996.

\bibitem{Bal98a}      T. Balaban,   A low temperature expansion for classical N-vector models III,
  {\em  Commun. Math. Phys.} 196: 485-521, 1998.





\bibitem{Bal98b}    T. Balaban,    Renormalization and localization expansions II,
  {\em  Commun. Math. Phys.} 198: 1-45, 1998.


\bibitem{Bal98c}  T. Balaban,   Large field renormalization operation for classical N-vector 
models,   {\em  Commun. Math. Phys.} 198:  493-534, 1998.






\bibitem{BaOc99}    T. Balaban, M. O'Carroll,     Low temperature properties for correlation functions in classcial $N$-vector spin models,   {\em  Commun. Math. Phys.} 199: 493-520, 1999.









\bibitem{BDH95}
D.Brydges, J.Dimock, and T.R. Hurd.
\newblock The short distance behavior of $\phi^4_3$,
\newblock {\em Commun. Math. Phys.} 172: 143--186, 1995.



\bibitem{Dim11}  J. Dimock,   The renormalization group according to Balaban - I.  small fields,   arXiv:  1108.1335.

\bibitem{Dim12}  J. Dimock,   The renormalization group according to Balaban - II.  large  fields,   arXiv: 1212.5562.



\bibitem{FeOs76}  J. Feldman, K. Osterwalder,   The Wightman axioms and the mass gap for
weakly coupled $\phi^4_3$ quantum field theories,  {\em Ann.  Physics} 97: 80-135,  1976.



\bibitem{GlJa73}  J. Glimm,  A. Jaffe,   Positivity of the  $\phi^4_3$ Hamiltonian,
{\em Fortschritte der Physik }  21:  327-376,   1973.









\end{thebibliography}
\end{document}